\documentclass[pra,twocolumn,superscriptaddress,amsmath,amssymb,aps,nofootinbib]{revtex4-2}
\usepackage[utf8]{inputenc}
\usepackage{amsmath,amsthm,amssymb,amsfonts,braket}
\usepackage{placeins}[verbose]
\usepackage{tabularx, longtable, multirow}
\usepackage{graphicx}
\usepackage{dcolumn}
\usepackage{bm,dsfont}
\usepackage[english]{babel}
\usepackage{comment}
\usepackage[dvipsnames]{xcolor}
\usepackage{siunitx}
\usepackage{comment}
\usepackage[normalem]{ulem}
\usepackage[T1]{fontenc}

\usepackage{cancel}
\providecommand{\ignore}[1]{}

\def\i{{\sf i}}

\begin{document}
	
\title{Spin-Textured Neutron Beams with Orbital Angular Momentum}

\author{Quan Le Thien}
\affiliation{Department of Physics, Indiana University, Bloomington IN 47405, USA}
\author{S. McKay}
\affiliation{Department of Physics, Indiana University, Bloomington IN 47405, USA}
\affiliation{Center for Exploration of Energy and Matter, Indiana University, Bloomington, 47408, USA} 
\author{R. Pynn}
\affiliation{Department of Physics, Indiana University, Bloomington IN 47405, USA}
\affiliation{Center for Exploration of Energy and Matter, Indiana University, Bloomington, 47408, USA}
\affiliation{Neutron Sciences Directorate, Oak Ridge National Laboratory, Oak Ridge, TN, 37830, USA}
\affiliation{Quantum Science and Engineering Center, Indiana University, Bloomington, IN 47408, USA}
\author{G. Ortiz}
\email{ortizg@iu.edu}
\affiliation{Department of Physics, Indiana University, Bloomington IN 47405, USA}
\affiliation{Quantum Science and Engineering Center, Indiana University, Bloomington, IN 47408, USA}
	
\date{\today}
	
\begin{abstract}
We present a rigorous theoretical framework underpinning the technique of spin-echo modulated small-angle neutron scattering (SEMSANS), and show how the technique can be extended in order to generate spin-textured neutron beams with orbital angular momentum (OAM) via birefringent neutron spin-polarization devices known as magnetic Wollaston prisms. Neutron OAM beams are mathematically characterized by a ``cork-screw'' phase singularity $e^{i \ell \phi}$ about the propagation axis where $\ell$ is the OAM quantum number. To understand the precise relationship between the emergent OAM state and the variety of spin textures realized by various setups, we have developed a path-integral approach that in the interferometric limit makes a judicious use of magnetic Snell's law. We show that our proposed technique produces a complex two-dimensional pattern of spin-OAM entangled states which may be useful as a probe of quantum magnetic materials.
We compare our path-integral approach to the well-known single-path Larmor precession model and present a pedagogical derivation of magnetic Snell's law of refraction for both massive and massless particles based on Maupertuis's action principle.
\end{abstract}
	
\maketitle
	
\section{Introduction}

\textit{Spin texture} is the emergent property of a physical system in which the system's spin is non-trivially coupled with its other dynamical degrees of freedom, such as the position, momentum, or orbital angular momentum (OAM); such a correlation between the spin and OAM degrees of freedom is called a \textit{spin-orbit coupling}.
These intricate correlations (often leading to entanglement) between the spin and the other degrees of freedom are responsible for a rich variety of thermodynamic phases of matter in which, for example, skyrmions and merons may materialize \cite{spin_texture_matter_Tao}, thus providing a basic platform for future applications in spintronics.

While conventional probes provide indirect signatures of topological excitations, spin-textured beams of particles with specific spin-orbit couplings (e.g., those with definite states of OAM) are strongly desired because they may act as direct probes of the target's topology \cite{Yin_2021}.
Beams with  OAM have been experimentally produced with photons \cite{allen_orbital_1992, Sarenac2018, Woods_2021, Kurzynowski_2010, Kurzynowski_2006}, electrons \cite{uchida_generation_2010, verbeeck_production_2010, eKarimi2013},  positrons \cite{lei_generation_2021}, and atoms \cite{Luski_2021} (see \cite{Allen2020, Yao_Padgett_2011, Rubinsztein-Dunlop_2016, Shen_2019} for some thorough reviews of OAM beams).
The generation of neutron OAM has been reported experimentally and/or proposed theoretically using macroscopic spiral phase plates \cite{clark_controlling_2015}, quadrupolar magnetic fields \cite{Nsofini_2016,SarenacTheory_2018}, room-temperature triangular electromagnetic coils \cite{Sarenac2019}, forked diffraction gratings \cite{Sarenac_2022}, polarized helium-3 \cite{Jach_2022}, aluminium prisms in a nested loop interferometer \cite{Geerits_2022}, and strong static electric fields via the relativistic Schwinger interaction \cite{Geerits2021}.
However, it remains an experimental and technological challenge to definitely  demonstrate the production of OAM in neutron beams: the usual optical methods do not work for a variety of reasons, primarily due to the weak interaction of neutrons with matter \cite{munter}. See \cite{Cappelletti_Vinson_2021, Cappelletti_Jach_Vinson_2018} for discussions on some previous demonstrations of neutron OAM.

In this work, we propose a method of generating spin-textured neutron beams which carry  OAM by using {\it magnetic Wollaston prisms} (MWPs), devices that act as polarizing neutron beam splitters \cite{li2014} (optical Wollaston prisms were used to generate photon OAM beams in \cite{Kurzynowski_2010,Kurzynowski_2006}). With the additional flexibility of tuning the various length scales \cite{shen2019,kuhn2021} associated to these spin textures, our beams have the potential to become useful probes of microscopic correlations in quantum materials.

An OAM state of a neutron is described by a phase $e^{i \ell \phi}$, where $\phi$ is the azimuthal angle about the axis of propagation and $\ell \in \mathbb{Z}$ is the OAM quantum number. This azimuthal phase leads to a number of interesting consequences. Firstly, the azimuthal component of the probability current $J_{\phi}$ is non-zero: for concreteness, in cylindrical coordinates ($r,\phi,z$) we find
\begin{equation}
    J_{\phi} = \frac{\hbar}{m} {\rm Im} \left( {\psi}^* \nabla \psi \right) \cdot \hat{\phi} = \frac{\hbar}{m r} \left[{\rm Im} \left( {f}^* \partial_{\phi} f \right) + \ell |f|^2 \right] ,
\end{equation}
where $m$ is the mass of the neutron, ${\rm Im}\left(\cdot\right)$ denotes the imaginary component, $\psi = f(r,\phi,z) e^{i \ell \phi}$ is the total wave function, and $^*$ signifies complex conjugation. When $f \neq f(\phi)$ as in the case of a paraxial beam (which can be written in Laguerre-Gauss modes \cite{Allen1992}), the azimuthal current is directly proportional to the OAM quantum number.
Therefore, we expect some non-zero scattering signatures into channels that emerge from the interaction between the neutron's OAM and the sample's chiral structures or dynamics; for example, such an off-axis current could cause \textit{super-kicks}, which are enhanced scattering events from processes that are kinematically forbidden for beams without OAM \cite{Afanasev2021}.
See \cite{Barnett_2022} for a discussion of the subtle physical properties of OAM states and the interaction of OAM states with matter.
Secondly, a neutron in a pure OAM state must have intensity singularity along the axis that defines its direction of travel to preserve the single-valueness of the wave function; because of this property the phase $e^{i \ell \phi}$ is generally referred to as a \textit{phase singularity}.

In this paper, we investigate the preparation and propagation of a spin-textured neutron beam, and show how the spin texture relates to the OAM content of the beam.  Neutron beams with tunable OAM offer a unique opportunity to observe quantum interference that is otherwise inaccessible in the traditional treatment of plane-wave scattering \cite{ivanov_elastic_2016}.
In addition, the OAM distinguishable subsystem could in principle be entangled with the other degrees of freedom of the neutron, such as its path, spin, or energy. This expansion of available probe subsystems enhances the idea of exploiting the advantages of quantum metrology and sensing by observing unique scattering signatures from entangled matter, as discussed in previous work for the spin-path entanglement case \cite{irfan_quantum_2021}.

We also show how the uniform magnetic field regions produced by MWPs generate spin-textured neutron beams. Specifically, we propose to use multiple pairs of MWPs in the \textit{spin echo modulated small angle neutron scattering} (SEMSANS) configuration (see Sec.~\ref{sec:MWP}).
Previous work carried out a theoretical analysis of neutron interferometers based on MWPs to test the violation of Bell-type contextual inequalities \cite{lu_operator_2020}, but that analysis is not sufficient to describe the SEMSANS setup: we must extend that result to include situations where the entangled neutron beam can be focused on desired spatial planes by precisely tuning the magnetic fields inside the MWPs. 

Starting from a path-integral representation of the dynamics in Sec.~\ref{sec:MWP}, we approximate the neutron's full time-evolution using the \textit{interferometric limit}, in which we only consider the two dominating spin-correlated classical paths. We explicitly show that this limit preserves unitarity in Appendix \ref{app:unitary}.
Our theoretical framework includes both the kinetic and potential energy contributions to the neutron's overall accumulated phase, which stands in stark contrast with the standard single-path Larmor precession approximation where only the potential energy along a single neutron path contributes to the accumulated phase. Essentially, the  single-path approximation neglects all refractive effects of the neutron trajectory. On the other hand, the path integral in the two-path interferometric limit includes the inequivalent refraction of the two spin states of a spin-1/2 particle when subject to a sharp boundary between regions of magnetic field (for justification and discussion of the single-path approximation, see \cite{Golub_1994}).
This difference in refraction angle is epitomized by the known magnetic Snell's law, which can be derived generically for any particle including neutrons from a relativistic action principle, as shown in Appendix \ref{sec: snell law}.
The kinematics and ray geometry for focused beams in the SEMSANS configuration are described for various configurations of pairs of MWPs in Sec. \ref{sec:focused beam}; higher order contributions are provided in Appendix \ref{higher order}. In Sec. \ref{sec: operator}, we develop the interferometric quantum dynamics of SEMSANS. Central to our derivation is an expansion over the refraction and beam divergence angles, which clarifies how the usual single-path Larmor precession approach can be recovered as the lowest-order expansion of our calculation. More importantly, we highlight the refractive corrections that are missed by the single-path method.
Finally, after developing the mathematical formalism to characterize MWP pairs, we explain in Sec. \ref{sec: result} how to combine the prisms to generate a variety of spin textures and OAM. We show in Sec. \ref{OAMC} that our MWP setup can produce regions on the detector where the $\ell=0,\pm1$ OAM states dominate in the OAM density.

\section{Magnetic Wollaston prisms} \label{sec:MWP}

\begin{figure}[ht] 
    \centering
    \includegraphics[width=0.95\linewidth]{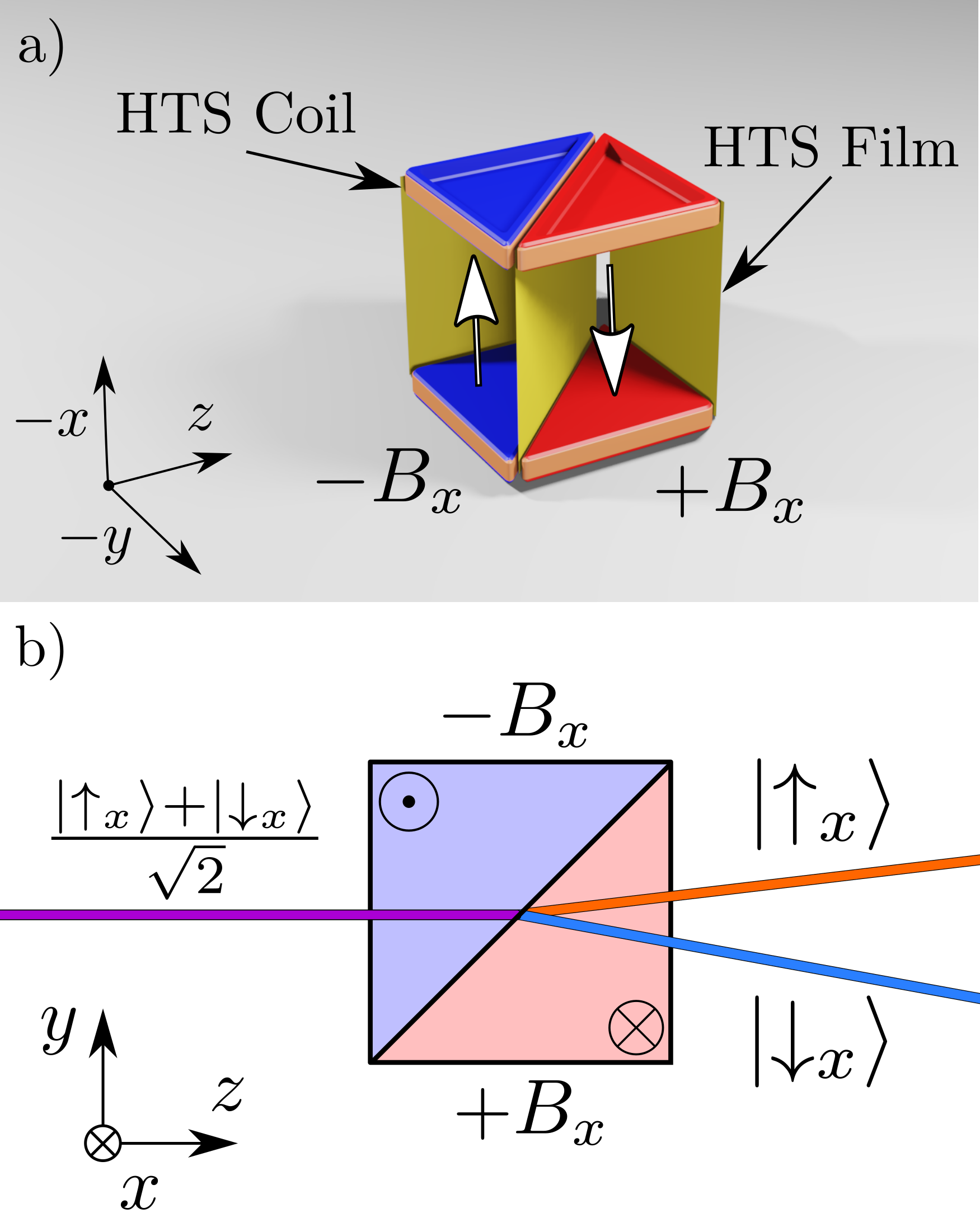}
    \caption{\label{fig:MWP3D} a) Perspective view of a magnetic Wollaston prism (MWP). Two high-temperature superconducting (HTS) films are not shown (front and side films). b) Top-down plan view of a MWP. The incident superposition state $\frac{\ket{\uparrow_x}+\ket{\downarrow_x}}{\sqrt{2}}$ (purple path) is coherently refracted in two separate directions (orange and blue path) at the interface between the two magnetic field ($\pm B_x$) regions. The outgoing neutron is in a mode-entangled (i.e., intraparticle-entangled) state between the path and spin subsystems.}
\end{figure}

\subsection{Working Principles of SEMSANS}

A MWP consists of two uniform triangular prisms containing oppositely directed magnetic fields as shown in Fig. \ref{fig:MWP3D}. High-temperature superconducting (HTS) wire coils wound around soft iron pole pieces generate a magnetic field of up to  150 mT in each triangular region. HTS films encapsulate the device, and a similar film separates the two triangular field regions; these films contain the naturally diverging magnetic field, producing a relatively homogeneous magnetic field with sharp boundaries. Two MWPs separated by an additional rectangular region of magnetic field form a MWP pair. The magnetic field strength and orientation can be independently tuned in each triangular field region as well as in the interposed rectangular region. 
The MWP acts like a polarizing beam splitter as long as the incoming neutron is not polarized parallel or antiparallel to the field direction inside the MWP. That is, each MWP is a spin-path entangler that transversally separates the incoming neutron into two outgoing path states, each labeled by a particular spin state \cite{lu_operator_2020}. This mode-entanglement (i.e., intraparticle-entanglement) has been experimentally demonstrated for neutrons by measurements of Bell-like contextuality inequalities with its concomitant violations \cite{shen2019, kuhn2021, Hasegawa_2010, Klepp_2014, Hasegawa_2003}.

MWPs are often used in \textit{spin echo small angle neutron scattering} (SESANS) and SEMSANS. In SESANS, a MWP pair (the first arm) splits the neutron state into two parallel path states that impinge on the sample; the distance between the mode-entangled spin-path states is called the \textit{entanglement length} $\xi$ (also called the spin echo length). 
The entanglement length when in the SESANS focusing condition (all field magnitudes equal) is given by
\begin{equation}
    \xi_{\mathrm{SE}} = \frac{m |\mu| \lambda^2}{\pi^2 \hbar^2} B |L_1 - L_2| ,
\end{equation}
where $\lambda$ is the wavelength of the neutron; $B$ the magnetic field magnitude in each triangular field region; $|L_1 - L_2|$ the distance between MWPs (see Fig. \ref{fig:parallel non div}); and $\mu = \gamma \mu_N < 0$ the magnetic moment of the neutron, with $\mu_N = e\hbar/(2m_p)$  being the usual nuclear magneton, $m_p$ the mass of the proton, and $\gamma = -1.913$, half the g-factor of a free neutron.
After the sample, another set of prisms (the second arm) recombines the two path states and the depolarization of the measured signal gives the density-density correlation function of the sample \cite{Rekveldt_1996, Rekveldt_2005}.
This configuration was described as a quantum circuit in our previous work \cite{lu_operator_2020} which treated the MWPs as black boxes, without considering the detailed dynamics of each MWP, which is an acceptable approximation for describing the SESANS configuration (see Sec. \ref{sec:interferometric limit}). The SESANS setup is the neutronic analogue of an optical Mach–Zehnder interferometer \cite{BornWolf}.

In SEMSANS, only a single MWP pair is required (the sample is placed after the prism pair).
However, unlike in SESANS, the detailed dynamics cannot be ignored since the second arm is no longer in perfect echo with the first, meaning that the two spin states will spatially interfere, thus generating a pattern of interference fringes on the detector. This spatial interference pattern from each single neutron is on the order of the size of the wavepacket. However, the interference pattern observed at the detector is indeed macroscopic as the few centimeter-width beam is made up of many such mutually incoherent neutrons. 

Usually, the two path states are focused in space at the detector plane, although the states can also be focused at any point after the second prism. The SEMSANS focusing condition is achieved by choosing
\begin{equation} \label{focusing condition}
    B_1 L_1 = B_2 L_2,
\end{equation}
where $B_1, B_2 \ge 0$ are the field magnitudes in the first and second MWPs, and $L_1, L_2$ are the distances between the first and second MWPs to the detector, respectively. The entanglement length in SEMSANS is a function of the distance $L_s$ between the detector and the position of the neutron after exiting the second MWP:
\begin{equation}
    \xi_{\mathrm{SEM}} = \frac{m |\mu| \lambda^2}{\pi^2 \hbar^2} |B_1 - B_2| L_s
\end{equation}
where $|B_1 - B_2|$ is the difference between the magnetic field magnitudes in the first and second prism \cite{Li_2016}. We note that the entanglement length, 
$\xi_{\mathrm{SEM}}~=~(\lambda/{\sf p})L_s$, is inversely proportional to the fringe period $\sf p$ on the detector 
\begin{equation} \label{eq:fringe period}
    {\sf p} = \frac{\pi^2 \hbar^2}{m |\mu| \lambda |B_1 - B_2|}.
\end{equation}
The fringe pattern of the modulated neutron intensity is the fundamental observable in a SEMSANS experiment: the change of the amplitude of the fringe pattern due to scattering from the sample is enhanced when the entanglement length is close to the correlation length of the sample. More specifically, the ratio of the fringe amplitude of the scattered and unscattered beam gives the sample's correlation function \cite{Andersson_2008}.
SEMSANS has already successfully measured the correlation functions of many materials \cite{Strobl_2012, Li_Parnell_Dalgliesh_2019, Li_Parnell_2021}. Both 
SEMSANS and the closely related technique of grating interferometry are examples of a neutronic Talbot-Lau interferometer, which has both x-ray \cite{Pfeiffer_2006} and molecular \cite{Brezger_2002} counterparts. 

In the familiar single-path Larmor precession model, the beam must be \textit{phase-focused} to observe the intensity fringes:  each neutron measured at some pixel on the detector could have taken a slightly different path through the instrument due to the beam divergence and finite-sized source, so each neutron measured at that pixel will have a slightly different Larmor phase, which will result in a decrease of the fringe visibility. We can improve the contrast of the signal (i.e. ``focus'' the measured fringe pattern on the detector) by removing the beam divergence dependence in the Larmor phase (at least to first order) with a certain choice of fields in the MWPs \cite{Li_2016}.
This type of focusing involves an incoherent ensemble of many neutrons. However, a single neutron must also be \textit{geometrically focused} such that the two separated path states overlap at the detector; this type of focusing is intrinsically different, as it involves the rays of the two correlated spin states of a single neutron. Geometric focusing is a single-particle requirement that is similar to the photonic ray-optics notion of focusing. Previously, it was assumed that the geometric focusing condition should agree with the phase focusing condition, but as we will show later, this assumption, which lies at the heart of the single-path Larmor precession approximation, is only true to first order in neutron deflection angle; see Fig. \ref{fig:snell law} and Eq.~\eqref{snell law approx} for the definition of the spin-dependent deflection angle.

Both SESANS and SEMSANS are methods of generating a high-fidelity, structured beam of mode-entangled  neutrons. This work develops the mathematical framework that describes the spin-texturing and OAM state that can be produced using extensions of the SEMSANS technique with focused MWPs.

\subsection{Path-integral in the Interferometric Limit} 
\label{sec:interferometric limit}

A mathematical framework appropriate for SEMSANS must incorporate the possibility that a  neutron may interfere with itself at arbitrary spatial positions, depending on specific experimental parameters. We will apply the usual path-integral formalism to model SEMSANS (for a pedagogical treatment of the path-integral, see \cite{cohen-tannoudji}).
A quantum treatment of SESANS  utilizing a finite-dimensional Hilbert space of path and spin modes was developed previously \cite{lu_operator_2020}, but that treatment is insufficient to describe the spatial self-interference aspect of SEMSANS. 
However, the full quantum mechanical treatment involving an infinite number of paths is still unnecessary due to the smallness of the neutron's transverse intrinsic coherence length compared to the transverse dimensions of the MWPs (roughly $4 \times 4$ cm) and the width of the neutron beam (typically 0.5-4 cm). Under these conditions the phase imparted on a single neutron by the MWPs is approximately constant across the neutron wavepacket.

The transverse and longitudinal intrinsic coherence lengths are parameters that determine the size of an individual neutron's wavepacket (e.g., the full-width half-maximum of a Gaussian wavepacket).
Shull originally reported a lower bound of 21 microns for the transverse intrinsic coherence length \cite{Shull_1969}, which agrees with the result of 24 microns in the measurement of diffraction from phase gratings in the near-normal transmission geometry \cite{Majkrzak_Berk_Maranville_Dura_Jach_2022}. A recent experiment in neutron reflectometry in the specular geometry has reported an intrinsic coherence length of about 1 micron, but as discussed by the authors, this unexpectedly small result is most likely due to the surface curvature of the grating samples \cite{Majkrzak_2014}.
Other experiments that measured the diffraction pattern from a grating \cite{Treimer_Hilger_Strobl_2006}, a single-crystal Bragg prism \cite{Wagh_Abbas_Treimer_2011}, and a Fresnel zone plate \cite{Altissimo_2008} report the transverse intrinsic coherence length to be on the order of 100 microns.

On the other hand, experiments in traditional neutron interferometry have reported much smaller transverse coherence lengths on the order of a few microns or less \cite{Rauch_1996, Pushin_2008}. One would expect that the exact size and shape of the neutron's wavepacket would depend on the specific method of neutron preparation and its interaction with the various optical elements (for example, the mosaicity of the crystal monochromator or the shape and size of the neutron guides). However, this discrepancy between the two sets of experimental data could be resolved by separating the effects of \textit{beam coherence} and \textit{intrinsic coherence} on the visibility of the experimentally observed interference fringes.

As is well-known in classical optics, a totally incoherent extended source of radiation can develop coherence after propagation, an effect that is mathematically described by the van Cittert-Zernike (VCZ) theorem, which in the far-field form of the theorem relates the degree of coherence to the Fourier transform of the intensity \cite{Mandel_Wolf_1995}. The VCZ theorem has also been extended to matter waves, including neutrons \cite{Zarubin_1993, Taylor_1994}.  
While the intrinsic coherence length is defined as the characteristic size of the neutron wavepacket, the beam coherence is a measure of the spontaneous VCZ-like coherence; in a simple single-slit geometry with uniform aperture illumination by a completely incoherent source, the transverse beam coherence length $\beta_{t}$ is defined as $\beta_{t} = d/(k a)$, where $d$ is the distance between the slit and the point of measurement on the axis of propagation, $k$ the magnitude of the neutron wavevector, and $a$ the width of the slit \cite{Keller_1997, Felber_1998}.
Therefore, the intrinsic coherence can be loosely described as a ``quantum'' effect, while the beam coherence could be considered more ``classical'' in nature.
However, this same effect that produces coherence classically could also ``enfeeble'' the much larger intrinsic coherence length associated with the size of the wavepacket, making the experimentally measured coherence length appear much smaller than the intrinsic coherence length of the neutron \cite{Barrachina_Navarrete_Ciappina_2019, Barrachina_Navarrete_Ciappina_2020, Fabre_Navarrete_Sarkadi_Barrachina_2018}.
The general relationship between beam and intrinsic coherence is more complicated, as the degree of measured coherence can actually increase with \textit{increasing} classical uncertainty (i.e., a smaller beam coherence length), a phenomena reminiscent of stochastic resonance \cite{Mariano_2002}.

As an aside, under the assumption of a stationary beam (i.e., the beam is time-independent, so the density matrix commutes with the Hamiltonian), it is not possible in general to experimentally distinguish between a beam of plane waves or a beam of wavepackets if both ensembles overall have the same energy spectra and uncertainties \cite{Stodolsky_1998}.  Results in \cite{Stodolsky_1998} do not apply if the observable that the experimenter chooses to measure does not commute with the momentum operator, or if other information about the initial preparation of the wavepacket beyond the spectra is known \cite{Kiers_1996}. Alternatively, one can consider a time-dependent, non-stationary process, to measure the intrinsic coherence lengths \cite{Golub_Lamoreaux_1992}, or treat the scattering process using ``energy-gated wavepackets'' as described in \cite{Berk_2014}. Therefore, it is possible in principle to decouple the effects of both the intrinsic and beam coherence from an experimental measurement.

Although beam coherence is important experimentally, the beam coherence has no effect on the final results of the following quantum mechanical calculations as these results apply to the mode-entangled states of a single neutron. However, one could calculate the beam coherence of our OAM beam for a given experimental configuration as we report the action of our proposed MWP setup on an incident neutron with arbitrary initial wavelength, position, divergence, and polarization.
The primary limitation on the size of the beam coherence is the neutron flux: $\beta_t$ is inversely proportional to the beam-defining slit width, so a larger beam coherence length requires a longer count time to attain reasonable statistics.

Regardless of the outcome of the experimental discrepancy of the size of the transverse intrinsic coherence length, our following theoretical results would still apply as the calculation only assumes that the intrinsic coherence length is non-zero. The necessity of this requirement is a subtle point: as we will show later, taking into account the wavepacket nature of the neutron is crucial in the path-integral in order to consistently include the contributions from both the kinetic and potential terms in the Hamiltonian in Eq. \eqref{Hamiltonian}.
Our calculation assumes that the incoming state has sufficiently large coherence lengths to observe interference fringes; we discuss these assumptions in more detail in Sec. \ref{sec: operator}.

One can then consider an interferometric limit, where the neutron spin-path states are constrained to the paths geometrically determined by refraction; this limit is analogous to the geometrical optics limit for photon beams.
The interferometric limit is similar to what is called the ``semiclassical ray-tracing'' approach used in \cite{keller_book}, where each spin state is treated as a separate plane wave, both of which propagate independently through the magnetic field configuration.
For a general discussion of the duality between plane waves and ray optics for matter waves including neutrons, see \cite{Klein_Werner_1983, Goldstein_1950}.
The interferometric limit amounts to a two-path approximation in which the spatial subsystem of a single neutron has two available internal spin states; each path corresponds to the classical trajectory taken by the eigenstate of the spin subspace in the magnetic field (i.e., parallel and antiparallel to the field).
This two correlated-path regime is widely used in traditional neutron interferometry where the neutron spin states are separated by centimeters \cite{rauch-werner2015}, while the separation in our beams ranges from nanometers to microns \cite{shen2019,kuhn2021}. 
We also assume pure transmission with no reflection in the interferometric limit; although there is a very small  reflection amplitude at the magnetic-field boundaries, the reflection amplitude is heavily suppressed by the fact that the Zeeman energy is much smaller than the neutron's kinetic energy. We show that the quantum mechanical nature of the neutron is preserved even in the interferometric limit.

\begin{figure*}[ht]
    \centering
    \includegraphics[width=0.95\textwidth]{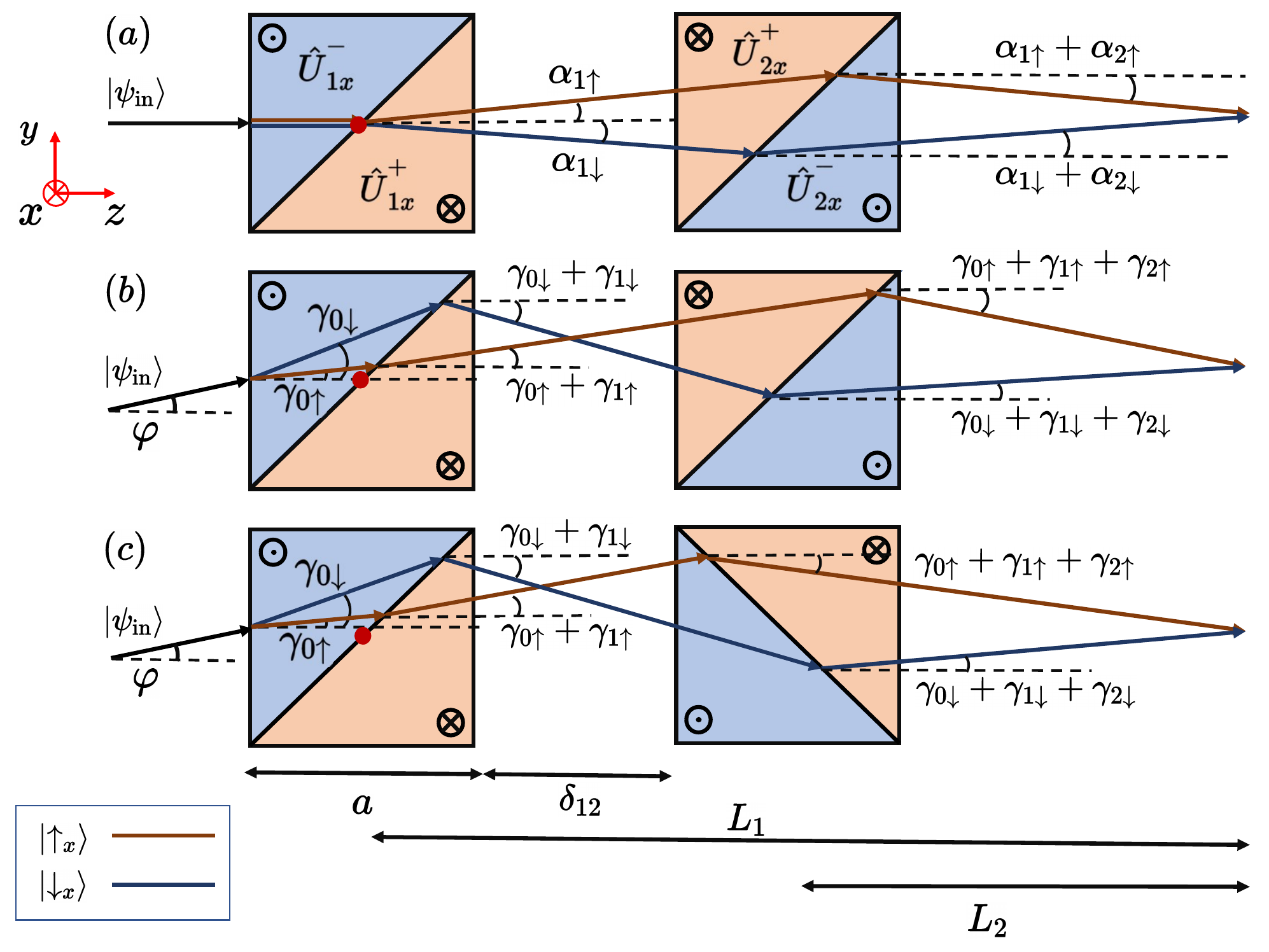}
    \caption{\label{fig:parallel non div}
    Plan view of the neutron paths through a pair of MWPs; each prism is labeled with a subscript $\i=1,2$.
   The rays in each subdiagram represent the semiclassical evolution of a single (representative) incoming neutron ray, characterized by a single plane wave component of the individual neutron's wavepacket. The spatial extent of the neutron wavepacket is determined by the transverse and longitudinal intrinsic coherence lengths (not shown).
    $(a)$ The parallelogram geometry without beam divergence, $(b)$ the parallelogram geometry with beam divergence, and $(c)$ the triangular geometry with beam divergence. The interferometric limit is represented by the fact that we can individually trace the paths of the $\ket{\uparrow_x}$ and $\ket{\downarrow_x}$ spin states (blue and red lines).
    The origin of the coordinate system is taken to be the center of the first MWP as indicated by the red dot in each subfigure. The hypotenuse superconducting film of each MWP is assumed to be at $45^{\circ}$ relative to the side film, and $a$ and $\delta_{12}$ are respectively the edge length of each MWP and the distance between the the outside edges of the two MWPs.
    The distances between the focusing point and centers of the MWPs are denoted $L_1$ and $L_2$. The incoming spin state is defined as $|\psi^s_{\mathrm{in}} \rangle = \cos(\theta_{\sf in}/2) \ket{\uparrow_z} + e^{i \phi_{\sf in}}\sin(\theta_{\sf in}/2) \ket{\downarrow_z}$. To $\mathcal{O}\left[\alpha_\i^2,\alpha_\i\varphi,\varphi^2\right]$, the refraction at the orthogonal MWP boundaries is inconsequential to the final results obtained in Sec. \ref{sec:focused beam} and \ref{sec: operator}; thus for the sake of clarity, these deflection angles are omitted in the diagram.  Finally, we note that the beam divergence $\varphi$ and deflection angles $\alpha_{\i \sigma}$, $\sigma=\uparrow, \downarrow$, in this diagram are greatly magnified compared to experimental values.}
\end{figure*}

Without loss of generality, we assume that the prisms' magnetic fields are either aligned or antialigned along the $x$-axis, and that the neutron travels in the $yz$ plane (see Fig. \ref{fig:parallel non div}). We split the time evolution operator into two stages, corresponding to when the neutron is in each triangular field region. The total time evolution operator for a single MWP is
\begin{equation} \label{operator original}
	\hat{U}^{\sf MWP}_x =  \hat{U}_x^{\pm} \hat{U}_x^{\mp},
\end{equation}
where $\hat{U}_x^\pm$ are time evolution operators in the corresponding triangular field regions of the MWP, and the superscripts designates the field's orientation along the $x$-axis. The Hamiltonians in each region are
\begin{equation}
    \label{Hamiltonian}
    \hat{H}^\pm = \frac{\vec{p}^{\,2}}{2m} - \hat{\mu} \cdot \vec{B}^\pm = \frac{\vec{p}^{\,2}}{2m} \mp \mu \hat{\sigma}_x B_x, 
\end{equation}
where $\vec p$ is the momentum of the neutron, $B_x > 0$ is the field magnitude, $\hat \sigma_x$ is the $x$ Pauli operator with eigenvectors $\ket{\sigma_x}$, and the symbol $\sigma$ is an element of $\{\uparrow, \downarrow\}$. Explicitly, the time evolution operators are
\begin{align}
    \hat{U}_x^\pm =& \exp \left(-\frac{i}{\hbar} \int^{t_o}_{t_i} dt \, \hat{H}^\pm \right),
\end{align}
where $t_i$ is the entrance time (incoming) and $t_o$ is the exit time (outgoing).

We now derive the time evolution operator for a single triangular field region; without loss of generality, we take $\vec{B}^+=B_x \hat x$ with $B_x$ constant and drop the superscript. A general expression for $\hat{U}_x$ can be written in the basis of the incoming position state $\ket{\vec{r}_i}$ and outgoing position states $\ket{\vec{r}_o}$:
\begin{equation} \label{general operator}
    \hat{U}_x = \int_{\mathds{R}^3} d\vec{r}_i d\vec{r}_o \sum_{\sigma = \uparrow ,\downarrow} \ A_{\sigma}(\vec{r}_o,\vec{r}_i)|\sigma_x\rangle \langle\sigma_x|\otimes| \vec{r}_{o}\rangle\langle \vec{r}_i|,
\end{equation}
where the cross terms such as $\ket{\uparrow_x} \bra{\downarrow_x}$ vanish since $|\sigma_x \rangle$ with $\sigma \in \{ \uparrow, \downarrow \}$ are the eigenstates of the Hamiltonian in Eq.~(\ref{Hamiltonian}). Equation~(\ref{general operator}) has the path-integral interpretation that an outgoing neutron at position $\vec{r}_o$ has contributions $A_{\sigma}$ from all possible incoming states at $\vec{r}_i$. The position states $\ket{\vec{r}_{i}}$ and $\ket{\vec{r}_{o}}$ in Eq. \eqref{general operator} are respectively the incoming and outgoing positions of an individual neutron's wavepacket component.
The amplitudes $A_{\sigma}$ are given by the matrix elements
\begin{align}
    A_{\sigma} (\vec{r}_o,\vec{r}_i) &= \langle  \vec{r}_o| \langle \sigma_x |  \hat{U}_x | \sigma_x \rangle | \vec{r}_i \rangle\\
    &=  \langle \vec{r}_o| \exp \left(-\frac{i}{\hbar} \int^{t_o}_{t_i} dt \, \left (\frac{\vec{p}^{\,2}}{2m} - \mu_{\sigma} B_x \right ) \right) | \vec{r}_i \rangle , \nonumber 
\end{align}
where $\mu_{\uparrow} = -|\mu|$ and $\mu_{\downarrow} = |\mu|$. To connect the fully quantum-mechanical expression of $A_{\sigma}$ to the interferometric limit, we express the amplitude using the standard path-integral formalism:
\begin{align}
    A_{\sigma} (\vec{r}_o,\vec{r}_i) = \int_{\vec{r}_i}^{\vec{r}_o} \mathcal{D}r \exp \left( \frac{i}{\hbar} \int_{t_i}^{t_o} dt \, L_{\sigma}(\vec{r}, \dot{\vec{r}})    \right),
\end{align}
where $L_{\sigma}$ is the Lagrangian for each spin state
\begin{align}
    L_{\sigma} (\vec{r}, \dot{\vec{r}} ) = \frac{1}{2}m \dot{\vec{r}}^{\,2} + \mu_{\sigma} B_x,
\end{align}
which is a quadratic function of  $\dot{\vec{r}}$.
The measure $\mathcal{D}r$ denotes that the integral is taken over all possible paths from $\vec{r}_i$ 
to $\vec{r}_o$. The transition amplitude $A_\sigma(\vec{r}_o,\vec{r}_i)$ can be evaluated exactly to yield
\begin{equation}
    \nonumber
    A_{\sigma} (\vec{r}_o,\vec{r}_i) ={\cal N} \exp\left(i\frac{m \left( \vec{r}_o - \vec{r}_i \right)^2}{2\hbar \left( t_o - t_i \right)} - i \Phi_{L,\sigma} (t_o,t_i) \right) ,
    \label{eq: amplitude exact}
\end{equation}
with normalization ${\cal N}=\left (\frac{m}{i 2 \pi \hbar \left( t_o - t_i \right)}\right )^{1/2}$ and a magnetic-field dependent phase
\begin{equation}
    \Phi_{L,\sigma}(t_o,t_i)=-\frac{1}{\hbar}\int_{t_i}^{t_o} dt \, \mu_{\sigma} B_x.
\end{equation}
The magnetic phase can be shown to reduce to the well-known Larmor phase in the single-path approximation (for example, see \cite{Golub_1994}).
The amplitude for any path consists of a phase and a normalization factor weighting the contribution of each path.
Instead of integrating over all possible neutron trajectories from $\vec{r}_i$ to $\vec{r}_o$ in Eq.~\eqref{general operator}, we will employ the  interferometric limit, a type of semiclassical approximation, which only considers the two dominating and correlated paths for each of the two spin states $|\sigma_x\rangle$.
In physical terms, the interferometric limit accounts for neutron refraction (but not reflection) at the boundaries defined by the magnetic field discontinuities. 

In reality, the amplitudes corresponding to the appropriate classical paths from $\vec{r}_i$ to $\vec{r}_{\sigma}(\vec{r}_i)$ will dominate over the others paths in the integral, with $\vec{r}_{\sigma}(\vec{r}_i)$ determined from geometric considerations analogous to ray optics.
This assumption amounts to constraining the path integral to classical paths $\vec{r}_{\sigma} (\vec{r}_i)$ (straight lines in space-time neglecting gravity) for each spin state.
Mathematically, this approximation is equivalent to inserting $ \mathcal{N}^{-1} \delta\left( \vec{r}_o - \vec{r}_{\sigma} \left( \vec r_i\right) \right)$ into Eq.~(\ref{general operator}), where $\mathcal{N}^{-1}$ is the appropriate normalization factor that sets the magnitude of the amplitude along the classical path to unity.
Taking this classical path constraint into account, the time evolution operator of Eq. \eqref{general operator} simplifies to
\begin{align}\hspace*{-0.3cm}
	\hat{U}_x =& \int_{\mathds{R}^3} d\vec{r}_i \sum_{\sigma=\uparrow,\downarrow}  A_{\sigma,{\sf cl}}(\vec{r}_{\sigma},\vec{r}_i)|\sigma_x\rangle \langle \sigma_x|\otimes| \vec{r}_{\sigma}(\vec{r}_i)\rangle\langle \vec{r}_i| ,
    \label{interferometric operator}
\end{align}
with amplitudes along the classical paths $A_{\sigma,{\sf cl}}$ given by
\begin{equation}
\begin{aligned}
    \label{amplitude classical}
     A_{\sigma,{\sf cl}} (\vec{r}_\sigma,\vec{r}_i) =& 
    \exp \left( i \left[ \vec{k}_{\sigma} \cdot \frac{\left(\vec{r}_\sigma - \vec{r}_i \right)}{2}  - \Phi_{L,\sigma}(t_o,t_i)\right] \right) .
\end{aligned}
\end{equation}
Here we have defined the classical wave vector of each neutron spin state as $\hbar \vec{k}_\sigma =m \left(\frac{\vec{r}_\sigma - \vec{r}_i}{t_o-t_i}\right)= m \vec{v}_\sigma$, with $\vec v_\sigma$ being the spin-orientation dependent classical velocity; this simplification is possible because the Zeeman energy is constant inside each triangular region. Notice that the amplitude $A_{\sigma,{\sf cl}}$ comprises both a kinetic phase term (i.e., the $\vec{k}_{\sigma}$ dependent phase) and the magnetic field phase term. For more discussion on the kinetic phase, see Appendix \ref{app:unitary}.

In Eq.~(\ref{amplitude classical}), we made use of the classical path and kinematics to express amplitudes in terms of spatial coordinates $\vec{r}_\sigma$ and $\vec{r}_i$. In the path-integral representation, it is implied that both initial time $t_i$ and final time $t_o$ are identical for the opposite-spin paths.
To determine these phases for the amplitudes in Eq.~(\ref{amplitude classical}), we must account for ray geometry and neutron kinematics (conservation of energy across boundaries) as shown in the next section. Finally, we note that the interferometric limit preserves unitarity; for a proof, see Appendix \ref{app:unitary}. 

\subsection{Magnetic Snell's Law: Refraction}

\begin{figure}[ht]
    \centering
    \includegraphics[width=0.45\textwidth]{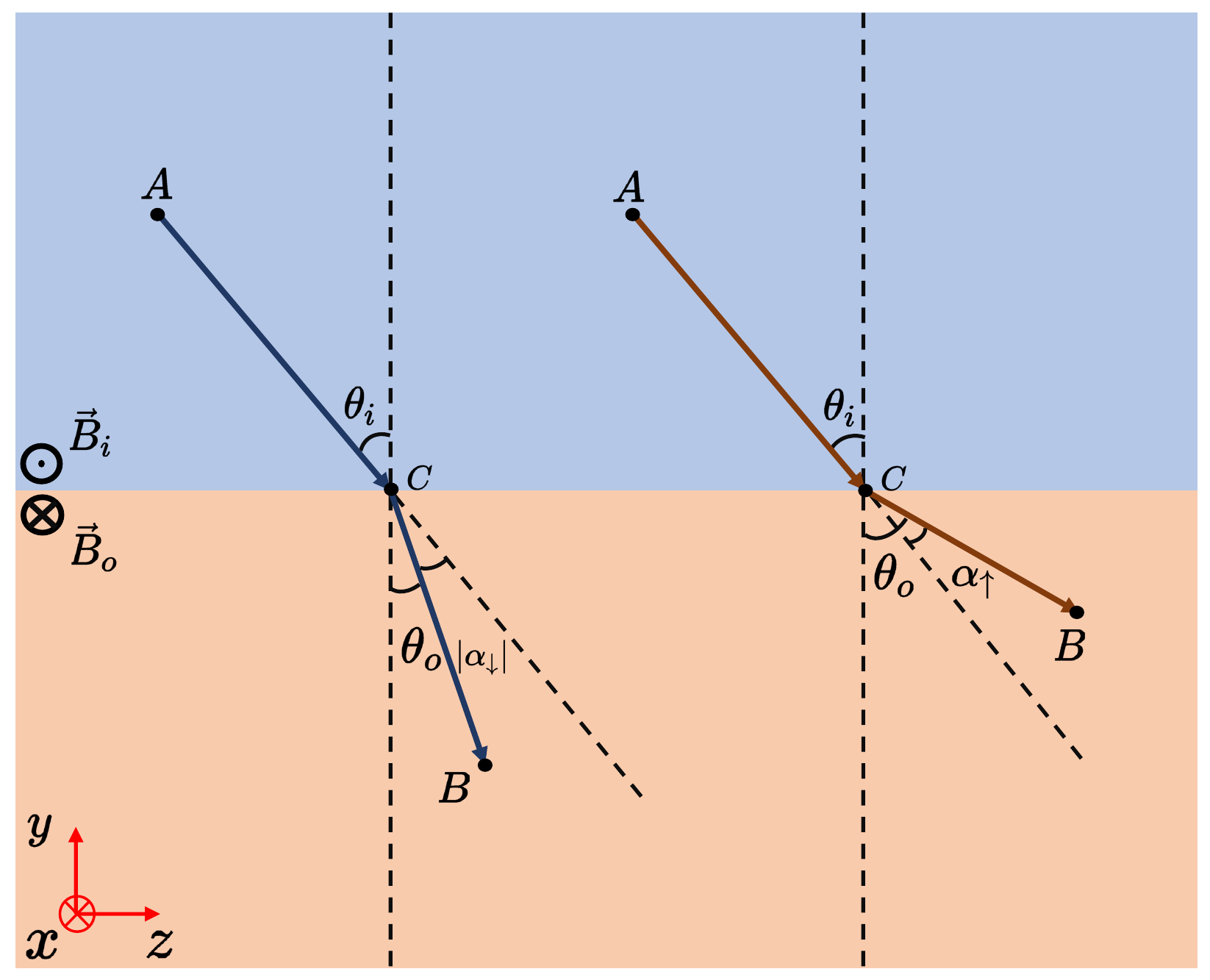}
    \caption{\label{fig:snell law} The schematic for magnetic Snell's law. Two scenarios for the spin states are shown, with the red and blue lines denoting the $\ket{\uparrow_x}$ and $\ket{\downarrow_x}$ states, respectively. Note that the deflection angles $\alpha_\sigma$ carry a sign indicating whether the deflection is towards ($\alpha_\sigma<0$) or away ($\alpha_\sigma>0$) from the normal to the boundary.}
\end{figure}

We now consider how discontinuous field boundaries affect the neutron classical spin paths. As shown in Fig.~\ref{fig:snell law}, when the neutron crosses the boundary of a field region, the corresponding classical paths of the two spin states are refracted by angles related by
\begin{align}
    \label{magnetic snell law}
    \sin \theta_{o\sigma} = \frac{v_{i}}{v_{o\sigma}} \sin \theta_i = \left(1+\frac{2 \mu_\sigma (B_o + B_i)}{m v_{i\sigma}^2}\right)^{-\frac{1}{2}} \sin \theta_i ,
\end{align}
where $\theta_{i}$ and $\theta_{o\sigma}$ are respectively the incoming and outgoing angles of each spin state, $B_o + B_i$ (with $B_o, B_i \ge 0$) is the field discontinuity across the boundary, and $v_{i\sigma}$ and $v_{o\sigma}$ are respectively the incoming and outgoing speed of the corresponding spin state. For simplicity, we take the initial speed and angle for both states to be the same. Following the standard definition, the refraction angles $\theta_{o\sigma}$ are defined relative to the normal to the boundary between the two field regions.

This formula is the magnetic extension of Snell's law in the non-relativistic limit; it can be derived from the variational principle of least action subject to conservation of energy (see Appendix~\ref{sec: snell law}). Due to the smallness of the Zeeman energy compared to the kinetic energy in the current generation of MWPs, one can expand Eq.~(\ref{magnetic snell law}) in terms of the (small) deflection angle $\alpha_\sigma=\theta_{o\sigma}-\theta_i$ from the incoming direction, and so $\sin(\theta_i+\alpha_\sigma)\approx \sin \theta_i + \alpha_\sigma \cos \theta_i$. Therefore, we find that 
\begin{equation} \label{snell law approx}
    \alpha_\sigma \approx - \frac{\mu_\sigma (B_o + B_i)}{m v_{i\sigma}^2} \tan \theta_i.
\end{equation}
%
	
\section{Kinematics and Geometry for Focused Beams in SEMSANS} \label{sec:focused beam}

As a neutron enters a triangular region with a different field strength, its speed and magnetic Zeeman energy both change due to the conservation of total energy (i.e., kinetic plus potential). The neutron's magnetic-field dependent speed is given by
\begin{align}
    \label{eq: speed change}
	v_{\i \sigma}^\pm =& \sqrt{v_{0}^{2} + \frac{2 \mu_\sigma B_\i^\pm}{m}} \approx \left(1 \pm \frac{\mu_\sigma B_\i}{m v_{0}^{2}}\right) v_{0},
\end{align}
where $v_0$ is the initial speed of the neutron before entering the MWP, the subscript $\i$ denotes which MWP is being considered, and the superscript $\pm$ denotes the corresponding field direction where we have used the convention that $B_\i = |B^\pm_\i|\ge 0$.
Notice that Eq.~(\ref{eq: speed change}) implies that the neutron's two spin states have different speeds. The underlying assumption in the derivation of the SEMSANS quantum operator in Sec.~\ref{sec: operator} is that a neutron starting at $t=t_i$ will spread over a finite-size region in space in the longitudinal (transverse) direction; this spread is usually referred to as the longitudinal (transverse) intrinsic coherence length. Hence, it is assumed that there are points within the initial intrinsic quantum coherence volume, corresponding to opposite spin projections, that will simultaneously arrive at the detector and interfere \cite{cohen-tannoudji}.

We will now consider the MWP arrangements displayed in Fig.~\ref{fig:parallel non div}. We refer to the setup shown in  Fig.~\ref{fig:parallel non div}$(a)$ and $(b)$ as the \textit{parallelogram geometry} and Fig.~\ref{fig:parallel non div}$(c)$ as the \textit{triangular geometry}. At the hypotenuse interface inside a single MWP, the neutron is refracted due to the discontinuity in magnetic field direction (see Fig. \ref{fig:parallel non div}$(a)$). The two spin states diverge into an upper and lower path, deviating from the angle $\pi/4$ by $\pm \alpha_{\i \sigma}$.
According to the magnetic Snell's law, Eq.~(\ref{magnetic snell law}), the deflection angles of the two spin states at the first and second interfaces are given respectively by 
\begin{align}
	\alpha_{1\sigma} =& - \frac{\mu_\sigma \left(2B_{1}\right)}{m (v_1^{-})^2} \tan \frac{\pi}{4} \nonumber \\
	\approx& - \frac{2 \mu_\sigma B_{1}}{mv_0^{2}} +  \mathcal{O}\left[\left(\frac{\mu B_1 }{mv_0^2} \right)^{2} \right] \\
	\alpha_{2\sigma} =& - \frac{\mu_\sigma \left(-2B_{2}\right)}{m (v_2^{+})^2} \tan (\frac{\pi}{4}+\alpha_{1\sigma}) \nonumber \\
	\approx&  \frac{2\mu_\sigma B_{2}}{mv_0^{2}} +  \mathcal{O}\left[\left(\frac{\mu  }{m} \right)^{2} \frac{B_1 B_2}{(v_1^{-})^2 (v_2^+)^2}\right],
\end{align}
where for now we are assuming no beam divergence, $\varphi=0$, corresponding to the situation in Fig.~\ref{fig:parallel non div}$(a)$; this assumption implies that $\theta_{i} = \pi/4$ at the first MWP hypotenuse interface. 

In the subsequent calculation, we assume that the deflection angles at the boundaries are small, thus $\tan\alpha_{\i\sigma} \approx \alpha_{\i\sigma}$. Because the only difference between the angles $\alpha_{\i\uparrow}$ and $\alpha_{\i\downarrow}$ is their sign, it is convenient to state results of our calculation in terms of the magnitude of the angle, which we denote as
\begin{equation}
    \alpha_\i = |\alpha_{\i\sigma}| .
\end{equation}

In the parallelogram geometry, when the incoming neutron enters the first MWP with a divergence angle $\varphi\ne 0$ in the $yz$ plane as shown in Fig.~\ref{fig:parallel non div}$(b)$, the two spin paths are deflected with angles 
\begin{equation}
	\gamma_{0\sigma} = \left(1 + \frac{\mu_\sigma B_{1}}{m v_0^2}\right) \varphi =  \left(1 - \frac{\alpha_{1\sigma}}{2}\right) \varphi .
\end{equation}
The deflection angles at the first and second hypotenuse interfaces in the case of a divergent beam now become
\begin{align}
	\gamma_{1\sigma} =& - \frac{ \mu_\sigma \left(2B_{1}\right)}{m (v_1^{-})^2} \tan \left(\frac{\pi}{4}+\gamma_{0\sigma}\right)\\
	\gamma_{2\sigma} =& \frac{ \mu_\sigma \left(2B_{2}\right)}{m (v_2^{+})^2} \tan \left(\frac{\pi}{4} + \gamma_{0\sigma} +\gamma_{1\sigma}\right).
	\label{gammaparallel}
\end{align}
Hence, the angles that the two spin states make with the $z$-axis at the two hypotenuse interfaces are given respectively by $ \gamma_{0\sigma} + \gamma_{1\sigma}$ and $ \gamma_{0\sigma} + \gamma_{1\sigma} + \gamma_{2\sigma}$.

To obtain focusing of the two spin paths in the triangular geometry, the magnetic field  orientations in the second MWP need to be chosen as is indicated in Fig.~\ref{fig:parallel non div}$(c)$. Furthermore, since the refraction at field boundaries depends only on the field discontinuity across the boundaries, the previous Eq. \eqref{gammaparallel} for the deflection angles $\gamma_{2\sigma}$ are changed to
\begin{equation}
	\gamma_{2\sigma} = - \frac{ \mu_\sigma \left(2B_{2}\right)}{m (v^-_{2})^2} \tan \left[\frac{\pi}{4} - \left( \gamma_{0\sigma} +\gamma_{1\sigma}\right)\right].
\end{equation}
%
Now we are ready to derive the geometric focusing condition for the SEMSANS configuration in the interferometric limit for both MWP configurations.

\subsection{Parallelogram Geometry}

We start by considering two MWPs of side length $a$ separated by a distance $\delta_{12}$ with hypotenuse interfaces parallel to each other as is shown in Fig. \ref{fig:parallel non div}$(a)$. 
For parallel incoming rays (i.e., $\varphi=0$), the incident angle $\theta_{i}$ for both spin states as the neutron enters the first MWP is zero; thus no refraction occurs, which leads to
\begin{equation}
    y_{1\sigma} = z_{1\sigma} = y_0,
\end{equation}
where $(y_0,z_0)$ are the coordinates where the neutron enters the first MWP and $(y_{1\sigma},z_{1\sigma})$ for $\sigma \in \{ \uparrow, \downarrow\}$ are the coordinates of the two spin states at the first MWP hypotenuse (see Fig. \ref{fig:parallel non div}).

To determine where the two spin states arrive at the second MWP's hypotenuse interface $(y_{2\sigma},z_{2\sigma})$, we set up the equations of lines with origin at the center of the first MWP. Thus, we need to solve the set of linear equations
\begin{align}
	y_{2\sigma}-y_{1\sigma}=& \tan \alpha_{1\sigma}\left(z_{2\sigma}-y_{1\sigma}\right)\\
	y_{2\sigma} =& z_{2\sigma}-a-\delta_{12}.
\end{align}
Doing so, we find
\begin{align}
	y_{2\sigma} =& \frac{y_0 + \left(a+\delta_{12}- y_0\right) \tan\alpha_{1\sigma}}{1-\tan \alpha_{1\sigma}} \\
	z_{2\sigma} =& \frac{y_0 \left( 1 - \tan\alpha_{1\sigma} \right) +a+\delta_{12}}{1 - \tan\alpha_{1\sigma}} .
\end{align}
We have ignored the refraction as the neutron leaves the first MWP and enters the second MWP because these refractive corrections are of order $\alpha_\i^2$.

To obtain the position where the two spin states' paths focus on the detector, we establish the geometrical equations for the two spin state paths after the hypotenuse interface of the second MWP
\begin{equation} \label{detector equation}
	y_f - y_{2\sigma} = \tan \left(\alpha_{1\sigma} + \alpha_{2\sigma} \right) \left(z_f-z_{2\sigma}\right).
\end{equation}
To second order in $\alpha_\i$, we obtain the following result for the focusing position $(y_f,z_f)$:
\begin{align}
	y_{f} \approx& \ y_0 \\
    \label{focus cond parallel}
	z_f 
	\approx& \ y_0 - \left(a+\delta_{12}\right) \frac{\alpha_2}{\alpha_1 - \alpha_2}  .
\end{align}
This result is equivalent to the focusing condition derived by the single-path Larmor precession approximation given in Eq.~(\ref{focusing condition}) by identifying $z_f - y_0=L_1$ and $a+\delta_{12} = L_1 - L_2$; in other words, the phase and geometric focusing conditions are equivalent to first order in deflection angle. An important physical consequence of Eq.~(\ref{focus cond parallel}) is that the two spin states are spatially focused on a diagonal plane parallel to the two hypotenuse interfaces. Finally, because the neutron beam is only refracted in the $yz$ plane, we have
\begin{equation}
	x_f \approx x_0 .
\end{equation}

We can combine the geometry above and kinematics to calculate the total time $t_{f\sigma}$  the neutron spin state starting at $(x_0,y_0,-a/2)$ with initial speed $v_0$ takes to arrive at the focusing plane. Using the time of flight $t_\sigma = L_\sigma / v_\sigma$, where $L_\sigma$ is the distance travelled by the spin state at a constant speed $v_\sigma$, one obtains to $\mathcal{O}\left[\alpha_\i^2 \right]$
\begin{align}
    \nonumber
    t_{f\sigma} &= \frac{a (\alpha_{1\sigma}-3\alpha_{2\sigma})- 2 \alpha_{2\sigma} \delta_{12}}{2 v_0 (\alpha_{1\sigma}-\alpha_{2\sigma})} \\
    \label{time parallel non div focus}
    & \hspace*{0.5cm} - \frac{y_0 (\alpha_{1\sigma}-\alpha_{2\sigma}) (\alpha_{1\sigma}-\alpha_{2\sigma} -1)}{ v_0 (\alpha_{1\sigma}-\alpha_{2\sigma})} .
\end{align}
%

We now extend our calculation to include beam divergence. Generically, the neutron beam will have divergence angles in both the $yz$ plane (which we call $\varphi$) and the $xz$ plane; however, as shown in previous section, because our setup of MWPs in Fig.~\ref{fig:parallel non div}$(b)$ only changes the neutron trajectory in the $yz$ plane, it is sufficient to only consider the angle $\varphi$.
Calculations of $(y_{\i\sigma},z_{\i\sigma})$ can be carried out straightforwardly, as in the previous section, with the additional consideration of refraction happening at boundaries of the MWPs, since we want corrections to $\mathcal{O}\left[\alpha_\i^2, \alpha_\i \varphi, \varphi^2 \right]$. We find the focusing position $(y_f,z_f)$ to be
%
\begin{align}
	y_f \approx& \ y_0 +  \varphi  \left(y_0 + \frac{a}{2}- (a+\delta_{12}) \frac{\alpha_2 
	}{\alpha_1 -\alpha_2}\right) \\
	z_{f} \approx&\ y_0 - \left(a+\delta_{12}\right) \frac{\alpha_2}{\alpha_1 - \alpha_2} \nonumber \\
	&+ \varphi \left( \frac{a}{2} +2y_0 - \left(a +\delta_{12} \right) \frac{\alpha_2}{\alpha_1 - \alpha_2} \right), \label{eq:zf varphi}
\end{align}
%
where the coordinate $y_0$ of the incoming neutron can be determined from the focusing position:
\begin{equation}
	\label{parallel y0 yf}
	y_0 \approx y_f -  \varphi \left( y_f + \frac{a}{2} - \left(a+\delta_{12}\right) \frac{ B_2 }{B_1-B_2 } \right) .
\end{equation} 
Including beam divergence, the time the neutron takes to arrive at the focusing plane to $\mathcal{O}\left[\alpha_\i^2, \alpha_\i \varphi, \varphi^2 \right]$ is
\begin{align}
    \label{time parallel div focus}
    t_{f\sigma} &= \frac{a (\varphi +1) (\alpha_{1\sigma}-3\alpha_{2\sigma})- 2 \alpha_{2\sigma} (\varphi +1)\delta_{12}}{2 v_0 (\alpha_{1\sigma}-\alpha_{2\sigma})} \nonumber \\
    & \hspace*{0.5cm}+\frac{y_0 (\alpha_{1\sigma}-\alpha_{2\sigma}) (\alpha_{1\sigma}-\alpha_{2\sigma}-2\varphi -1)}{ v_0 (\alpha_{1\sigma}-\alpha_{2\sigma})} .
\end{align}

\subsection{Triangular Geometry}
	
It is straightforward to carry out calculations in the same spirit as in the previous section to obtain the following for the focusing position $(y_f,z_f)$:
\begin{widetext}
\begin{align}
	y_f \approx& \ y_0 + \varphi  \left(\frac{a (\alpha_1-3 \alpha_2)}{2\left(\alpha_1-\alpha_2\right)}-\frac{ \delta_{12} \alpha_2 }{\alpha_1-\alpha_2}+\frac{y_0 \left(\alpha_1+\alpha_2\right)}{\alpha_1-\alpha_2}\right) \\
	z_f \approx& \ y_0  \frac{\alpha_1+\alpha_2}{\alpha_1-\alpha_2} - \frac{\left(a + \delta_{12}\right) \alpha_2}{\alpha_1-\alpha_2} + \varphi \left(\frac{a \left(\alpha_1^2+6 \alpha_1 \alpha_2 - 3 \alpha_2^2\right)}{2 (\alpha_1-\alpha_2)^2}+\frac{\delta_{12} \alpha_2 (3 \alpha_1-\alpha_2)}{(\alpha_1-\alpha
  _2)^2}+\frac{2 y_0 \left(\alpha_1^2-4 \alpha_1 \alpha_2 + \alpha_2^2\right)}{(\alpha_1-\alpha_2)^2}\right) .
\end{align}
\end{widetext}
Interestingly, the focusing plane for the triangular geometry is different than for the parallelogram geometry, as the slope is defined by the relative strength of the magnetic fields and is no longer always at 45$^{\circ}$. We also note that if one neglects the beam divergence $\varphi$ and the initial position of the incoming neutron $y_0$, we obtain the focusing condition in Eq.~(\ref{focusing condition}) similar to the parallelogram case; for a discussion of the difference between the focusing conditions, see Sec. \ref{sec: operator}.
Solving for neutron's incoming position $y_0$ in terms of the focused position $y_f$, we obtain
\begin{align}
    \label{triangular y0 yf}
	y_0 \approx y_f - \varphi  \left(\frac{a (B_1-3 B_2)}{2\left(B_1-B_2\right)}-\frac{ \delta_{12} B_2 }{B_1-B_2}+\frac{y_0 \left(B_1+B_2\right)}{B_1-B_2}\right).
\end{align}
The time the neutron takes to arrive at the focusing plane to $\mathcal{O}\left[\alpha_\i^2, \alpha_\i \varphi, \varphi^2 \right]$ is
\begin{widetext}
\begin{align}
    \nonumber
    t_{f\sigma} &= \frac{a (\alpha_{1\sigma}-3 \alpha_{2\sigma})}{2v_0 \left(\alpha_{1\sigma} - \alpha_{2\sigma} \right)}+\frac{\alpha_{2\sigma} \delta_{12}}{v_0 (\alpha_{2\sigma}-\alpha_{1\sigma})}+\frac{y_0\left(\alpha_{1\sigma}+\alpha_{2\sigma}-\left(\alpha_{1\sigma} - \alpha_{2\sigma}\right)^2\right)}{v_0 \left(\alpha_{1\sigma} -\alpha_{2\sigma} \right)} \\
    \label{time triangular focus}
    & +\varphi  \left(\frac{a \left(\frac{4 \alpha_{1\sigma}^2}{(\alpha_{1\sigma}-\alpha_{2\sigma})^2}-3\right)}{2v_0}+\frac{\alpha_{2\sigma} \delta_{12} (3 \alpha_{1\sigma}-\alpha_{2\sigma})}{v_0(\alpha_{1\sigma}-\alpha_{2\sigma})^2}+\frac{2y_0 \left(\alpha_{1\sigma}^2-4 \alpha_{1\sigma} \alpha_{2\sigma}+\alpha_{2\sigma}^2\right)}{v_0 (\alpha_{1\sigma}-\alpha_{2\sigma})^2}\right) .
\end{align}
\end{widetext}

\section{The Interferometric Quantum Mechanics of SEMSANS} \label{sec: operator}

In this section, we establish the full unitary time evolution operator in the interferometric limit corresponding to the MWP configurations displayed in Fig. \ref{fig:parallel non div}. It is important to note that the process analyzed so far involves the interference of the two coherent spin states of a single neutron. In reality, an important origin of beam divergence is due to thermal fluctuations present in the neutron source. Hence, we need to derive operators for each divergence angle $\varphi$ and take into account the distribution of these divergence angles via the density matrix of the incoming neutron beam.

We can generically cast our operator at the exit of the second MWP in the form
\begin{align}
     \hat{U}_{\nu} &=
    \int_{\mathds{R}^3} d\vec r_0 \sum_{\sigma = \uparrow,\downarrow} \  A_{\sigma,{\sf cl}} \left( \vec r_0 \right) \ket{\sigma_\nu}\bra{\sigma_\nu}  \otimes  \ket{\vec r_{2o\sigma} \left(\vec r_0\right)}\bra{\vec r_0} , \nonumber
\end{align}
%
where $\nu \in \{x,y\}$ and $ |\vec r_{2o\sigma}(\vec r)\rangle \langle \vec r |$ is defined as
\begin{align}
	|\vec r_{2o\sigma}(\vec r)\rangle \langle \vec r | =& |x \rangle \langle x| \otimes   |y_{2o\sigma}(z_{2o}) \rangle \langle y| \otimes | z_{2o} \rangle \langle z |.
\end{align}
The coordinates $\vec{r}_{2o\sigma} = (x, y_{2o\sigma}, z_{2o})$ are those where the spin states exit the second MWP. Notice that $\hat{U}_x$ does not affect the neutron trajectory in the $x$ coordinate as shown in previous section. Similarly, for $\hat{U}_y$, the SEMSANS operator with fields aligned along the $y$ axis, the neutron trajectory in the $y$ coordinate is unaffected. Hence, the actions of $\hat{U}_y$ and $\hat{U}_x$ are independent of each other in terms of their geometrical paths. 

Furthermore, from the previous geometric considerations, we see that the two spin paths also propagate freely in regions without magnetic fields, namely between the first and second MWPs and from the second MWP to the focusing point $\vec{r}_f$. Therefore, there are additional operators $\hat{U}_f$ in these two regions associated with this free propagation, which are given by Eq.~(\ref{interferometric operator}) with $B=0$.
As $\vec r_f$ is the same for both spin states, the overall action of $\hat{U}^{\rm P,T}_x$ on the neutron state arriving at the focusing detector results in a tensor product between its spatial and spin components which can be written as
\begin{align}
    \label{operator general form}
     \hat{U}^{\rm P,T}_{\nu} &=
    \int_{\mathds{R}^3} d\vec r_0 \, \hat{U}^{\rm P,T}_{\nu,{\sf spin}}( \vec r_0 ) \otimes  \ket{\vec r_f \left(\vec r_0\right)}\bra{\vec r_0} , 
\end{align}
where the superscripts $\rm P$ and $\rm T$ refer to the parallelogram and triangular SEMSANS setups, respectively, and 
\begin{align}
     \hat{U}^{\rm P,T}_{\nu,{\sf spin}}( \vec r_0 ) &=\!\!
  \sum_{\sigma = \uparrow,\downarrow} \  A_{\sigma,{\sf cl}} ( \vec r_0 ) \ket{\sigma_\nu}\bra{\sigma_\nu}  .
\end{align}
Hereafter, we will omit writing the spatial components since they will be factored out when focused.
From the geometrical results of previous section and the operator form of the MWP triangular fields in Eq.~(\ref{interferometric operator}), it is straightforward to obtain $\hat{U}^{\sf MWP}_{\i \nu}$ explicitly for the parallelogram setup
\begin{equation}
    \hat{U}^{\sf MWP}_{1\nu}=  \hat{U}^{+}_{1\nu} \hat{U}^{-}_{1\nu}, \quad  \hat{U}^{\sf MWP}_{2\nu}=  \hat{U}^{-}_{2\nu} \hat{U}^{+}_{2\nu} ,
\end{equation}
as well as for the triangular setup
\begin{equation}
    \hat{U}^{\sf MWP}_{1\nu}=  \hat{U}^{+}_{1\nu} \hat{U}^{-}_{1\nu}, \quad  \hat{U}^{\sf MWP}_{2\nu}=  \hat{U}^{+}_{2\nu} \hat{U}^{-}_{2\nu} .
\end{equation}

For the parallelogram setup without beam divergence shown in Fig.~\ref{fig:parallel non div}$(a)$, the operator for a pair of MWPs up to terms $\mathcal{O}\left[\alpha_\i^2\right]$ is
\begin{eqnarray}
\hat{U}_{x}^{\text{P,No Div}} \!\!&=&\! \hat{U}_f(\vec{r}_{2o}, \vec{r}_{f}) \hat{U}^{\sf MWP}_{2x} \hat{U}_f(\vec{r}_{1o}, \vec{r}_{2i}) \hat{U}^{\sf MWP}_{1x} \hat{U}_f\left(\vec{r}_{b\sigma}, \vec{r}_0 \right) \nonumber \\
	&=& e^{i \zeta} \left( \cos \phi^{\text{P,No Div}} + i \sin \phi^{\text{P,No Div}}\ \hat{\sigma}_x \right) , 
\label{operator parallel non div}
\end{eqnarray}
where $\vec{r}_{1o}$ and $\vec{r}_{2i}$ are the coordinates where the neutron exits the first MWP and enters the second MWP, respectively, and $\zeta$ is a global phase.
The original positions $\vec{r}_{b\sigma}$ of the two spin states are assumed to be within the initial quantum coherence volume; in other words, we are assuming that both initial spin states are within the Fresnel zone \cite{cohen-tannoudji}.
We must introduce $\vec{r}_{b\sigma}$ in order to impose the constraint that the two spin states interfering at the focusing plane have the same initial time $t_i$ and final time $t_o$ in the path-integral formalism. Effectively, the free propagation operator $\hat{U}_f \left(\vec{r}_{b\sigma},\vec{r_0} \right)$ evolves the faster spin state during this longitudinal delay.
The resulting phase spatial variation $\phi^{\text{P,No Div}}$ generated by our SEMSANS setup is
\begin{equation}
    \phi^{\text{P,No Div}}_f =  \frac{ 2 |\mu| \left(B_{1}-B_{2}\right) y_{0}}{v_{0} \hbar}.
\end{equation}
Note that the incoming neutron state is a wavepacket, and the relative spatial phase variation $\phi^{\text{P,No Div}}$ is the phase difference between two longitudinally separated plane wave components that contribute to the incoming wavepacket. Particularly, we assume that the distance $\Delta z_{0}$ between the two initial spin states is within the ``longitudinal coherence length'' of the incoming wavepacket, which is given by the longitudinal intrinsic coherence length.  

Similarly, we can also find the operator for the parallelogram and triangular geometries with beam divergence. The general structure of these operators is similar to that of Eq.~(\ref{operator parallel non div}), with the only changes being the overall unobservable phase and the corresponding phase spatial variations, which up to terms $\mathcal{O}\left[\alpha_\i^2, \alpha_\i \varphi, \varphi^2\right]$ are 
\begin{align}
	\label{phase spatial variations incoming position}
	\nonumber
	\phi_f^{\text{P,Div}} =& \frac{2|\mu|\left(B_1-B_2\right)\left(1+\varphi\right)y_0}{v_0 \hbar} \\
	& + \frac{|\mu|\left(B_1 a - B_2 \left(3a+2\delta_{12}\right) \right)\varphi }{v_0 \hbar} \\
	\nonumber
	\phi_{f}^{\text{T,Div}}	=& \frac{2|\mu|\left(B_1-B_2\right)y_0}{v_0 \hbar} + \frac{2|\mu|\left(B_1+B_2\right)y_0 \varphi}{v_0 \hbar} \\
	& + \frac{|\mu| \left(B_1 a - B_2 \left(3a+2\delta_{12}\right)\right) \varphi }{v_0 \hbar} .
\end{align}

For non-divergent beams, we know from geometric consideration that $y_0 \approx y_f$. Moreover, for divergent beams, our unitary operators can be expressed entirely in terms of the focused positions $y_f$ by making use of Eqs.~(\ref{parallel y0 yf}) and (\ref{triangular y0 yf}):
\begin{eqnarray}
	\phi^{\text{P,Div}} (y_f) &=& \phi^{\text{T,Div}} (y_f)= \frac{2|\mu|\left(B_1-B_2\right)}{v_0 \hbar}  y_f \nonumber \\ 
	&=& \phi^{\text{P,No Div}} (y_f) .
	\label{phase spatial variation yf}
\end{eqnarray}
Surprisingly, the spatial variations of phase induced by the SEMSANS operator are the same for both geometries at the focusing plane: the only difference between the geometries is the angle of the focusing plane. This spatial variation is simply proportional to the $y$ coordinate from the center of the focusing plane; we should then expect a robust interference pattern on the focusing plane since there is no blurring effect coming from averaging over the beam divergence $\varphi$.
Here, the initial longitudinal spatial separation 
\begin{align}
    \Delta z_0 =  v_0 \left( t_{f\uparrow} - t_{f\downarrow} \right) = 2 y_f \left( \alpha_2 - \alpha_1 \right)
    \label{long-coh-length}
\end{align}
is assumed to be smaller than the neutron's longitudinal intrinsic coherence length (notice there is no initial separation required at the center of the focusing plane).
One may be able to estimate the longitudinal coherence length by measuring the dampening of the interference fringes.

\begin{figure}[ht]
    \centering
    \includegraphics[width=0.47\textwidth]{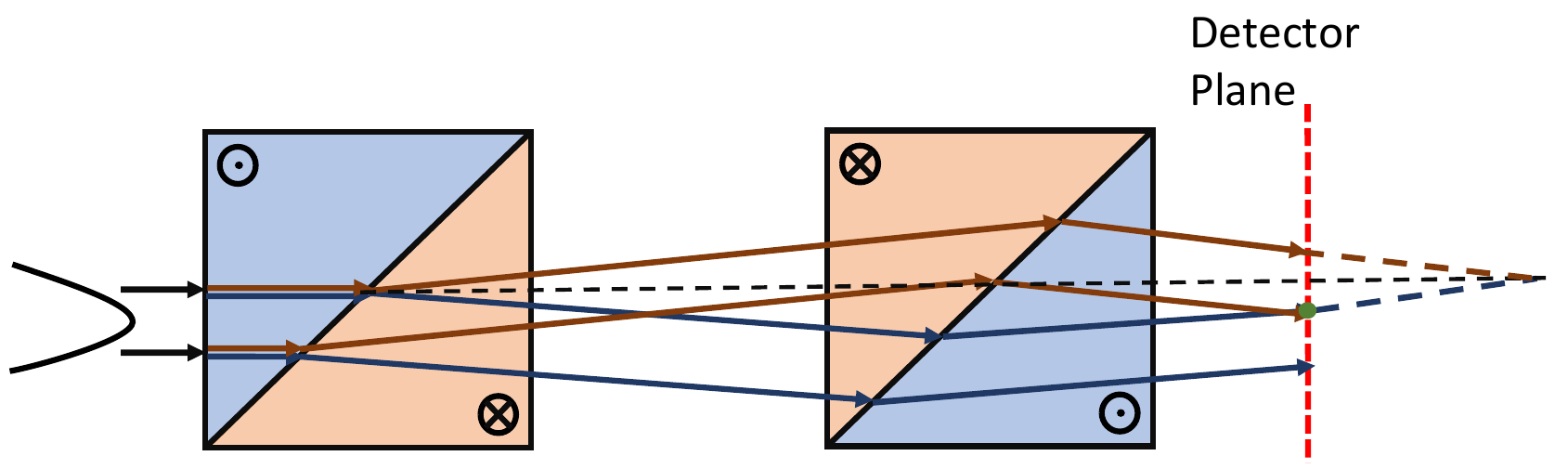}
    \caption{\label{fig:off-focus} Diagram of the usual detection scheme, where the red ray representing the state $\ket{\uparrow_x}$ interferes at a point on the detector plane (red dashed line) with the blue ray representing the state $\ket{\downarrow_x}$ which starts at a different initial position. Both sets of red and blue rays originate from a \textit{single neutron} with a finite intrinsic coherence volume. Traditionally, the detector plane in experimental setups is parallel to the $xy$ plane. Two rays with same spin states never interfere with each other because they are always parallel to and thus will not intersect.}
\end{figure}
	
So far, we have only considered neutron detection at the focusing plane defined as the plane where the two rays intersect after propagating through a pair of MWPs. It is important to account for the general case where the detector position is arbitrary (see Fig. \ref{fig:off-focus}). Because the corresponding measurement would then not be taken at the focusing plane, the neutron state at the point of detection is not a result of interference between the two spin states originating from the same initial point $(x_0,y_0)$ as considered above. 
%
To formalize this idea, notice that the focusing condition of the two spin states is cemented in the previous section via the operator $\hat{U}_f(\vec{r}_{o2\sigma}, \vec{r}_f)$.
We now introduce the new operator used to described the free propagation after the second MWP:
\begin{align}
    \hat{U}_d = \hat{U}_f (\vec{r}_{2o}, \vec{r}) ,
\end{align}
where $\vec r=(x,y,z)$ is an arbitrary position. Since a MWP with a field oriented in the $\pm x$ direction does not alter the trajectory in the $x$ direction, we will omit writing the $x$ coordinate.
In the usual experimental setup, the detector is situated at a plane parallel to the exterior edge of the MWPs a distance $z$ away from the center of the first MWP. In order to determine the two spin paths contributing to the interference at the detector, we need to solve for $\vec r_{0\sigma}$ in the two equations that describe the two neutron spin paths past the second MWP:
\begin{align}
    y -y_{2\sigma}(y_{0\sigma}) = \tan\left( \gamma_{0\sigma} + \gamma_{1\sigma} + \gamma_{2\sigma} \right) \left[z - z_{2\sigma}(y_{0\sigma}) \right].
\end{align}
Formally, the operator can be easily written down similar to Eq.~(\ref{operator general form}), but with the modified position of the detector:
\begin{eqnarray}\hspace*{-0.5cm}
     \hat{U}_{x} &=&
     \!\!\sum_{\sigma = \uparrow,\downarrow}\int_{\mathds{R}^3} d\vec{r}_{0\sigma}  \ A_{\sigma} \left( \vec{r}_{0\sigma} \right) \ket{\sigma_x}\bra{\sigma_x}  \otimes  \ket{\vec{r} }\bra{\vec{r}_{0\sigma}} ,
     \label{operator general form off focus}
\end{eqnarray}
where $\vec r_{0\sigma}$ are the initial neutron positions which can be obtain by inverting Eq.~(\ref{detector equation}) for the desired geometry.

The structure of these defocused operators is still similar to Eq.~(\ref{operator parallel non div}). This result reflects our interferometric limit consideration for the point of detection where there are two spin paths intersecting and thus, our operator always lives on the space of a two-level spin system. For the  parallelogram case without beam divergence, we get
\begin{align}
    \label{unitary operator final position parallel no div}
    \phi^{\text{P, No Div}} &= \frac{2 |\mu| \left(B_{1}-B_{2}\right) y}{v_{0} \hbar}.
\end{align}
Notice that due to the constraint on the classical path, we have expressed the spatially varying phase $\phi^{\text{P, No Div}}$ in terms of the general final position of the neutron $\vec{r}$. 

For an initial neutron with divergence angle $\varphi$, the spatial phase variations in the parallelogram geometry $\phi^{\text{P, Div}}$ and in the triangular geometry $\phi^{\text{T, Div}}$ are given by
\begin{align} 
    \nonumber
    \phi^{\text{P, Div}} =& \frac{2|\mu|\left(B_1-B_2\right)\left(1+\varphi\right)y}{v_0 \hbar} \\
    \label{unitary operator final position parallel}
	& - \frac{2|\mu| \left( B_1 L_1 - B_2 L_2 \right) \varphi}{v_0 \hbar}  \\
	\nonumber
	\phi^{\text{T,Div}}	=& \frac{2|\mu|\left(B_1-B_2\right)y}{v_0 \hbar}  + \frac{2|\mu|\left(B_1+B_2\right)y \varphi}{v_0 \hbar} \\
	\label{unitary operator final position triangular}
	&  -  \frac{2|\mu| \left( B_1 L_1 - B_2 L_2 \right) \varphi }{v_0\hbar} ,
\end{align}
where we have used the following equation that relates the arbitrarily chosen detector position $z$ to the previously defined distances $L_1$ and $L_2$:
\begin{equation}
    B_1 L_1 - B_2 L_2 = z \left(B_1 - B_2\right) + \left( a + \delta_{12} \right) B_2 .
\end{equation}

There are two important differences between the expressions \eqref{unitary operator final position parallel}-\eqref{unitary operator final position triangular} and Eq. \eqref{phase spatial variation yf}. On one hand, in Eq. \eqref{unitary operator final position parallel}, there is an extra term proportional to the divergence angle $\varphi$, independent of the $y$ coordinate.
Because rays with different beam divergence are mutually incoherent, this variation in $\varphi$ results in an overall blurring of interference pattern. However, for both geometries, we can cancel the purely $\varphi$-dependent term by applying the focusing condition of Eq. \eqref{focusing condition}. 
Interestingly, as discussed previously in Sec. \ref{sec:MWP}, this phase focusing requirement is equivalent to the geometrical focusing condition only when $\varphi=0$.
However, the geometric and phase focusing conditions are different when considering $\varphi \neq 0$. The difference between the proper detector position according to the geometric and phase focusing conditions is proportional to $\varphi$. With a realistic value of $\varphi \approx 1^{\circ}$, the numerical correction  to the phase from the $\varphi$ term in Eq. \eqref{eq:zf varphi} would be of order $10^{-2}$, which corresponds to a difference in focusing plane of roughly 1 mm.

On the other hand, to produce the observable interference pattern that is due to the spatially varying phase described in Eqs.~(\ref{unitary operator final position parallel})-(\ref{unitary operator final position triangular}), the two opposite-spin initial points in the intrinsic coherence volume must be displaced not only longitudinally, as in \eqref{long-coh-length}, but also transversally.
Again, this requirement implies that we are assuming that the incoming neutron state is a wavepacket whose size is determined by the transverse and longitudinal intrinsic coherence lengths. The two spin states that will eventually interfere at the detector must originate from the same wavepacket (i.e., within the same intrinsic coherence volume), as the neutrons that form the macroscopic beam are all mutually incoherent.
The required initial transverse displacement $\Delta y_0$ between the two spin states interfering at an arbitrary point $\vec{r}=(x,y,z)$ is given by
\begin{align}
    \Delta y_0 &= y_{0\uparrow} - y_{0\downarrow} \nonumber \\
    &= 2 \left[ \left( y-z \right) \left(\alpha_1 - \alpha_2\right) - \left( a+\delta_{12}\right) \alpha_2 \right] .
\end{align}
Hence, one can in principle estimate both the longitudinal and transverse coherence intrinsic lengths of the neutron.



\section{Generating Spin Textures and Orbital Angular Momentum} \label{sec: result}

\begin{figure*}[ht]
    \centering
    \includegraphics[width=0.95\textwidth]{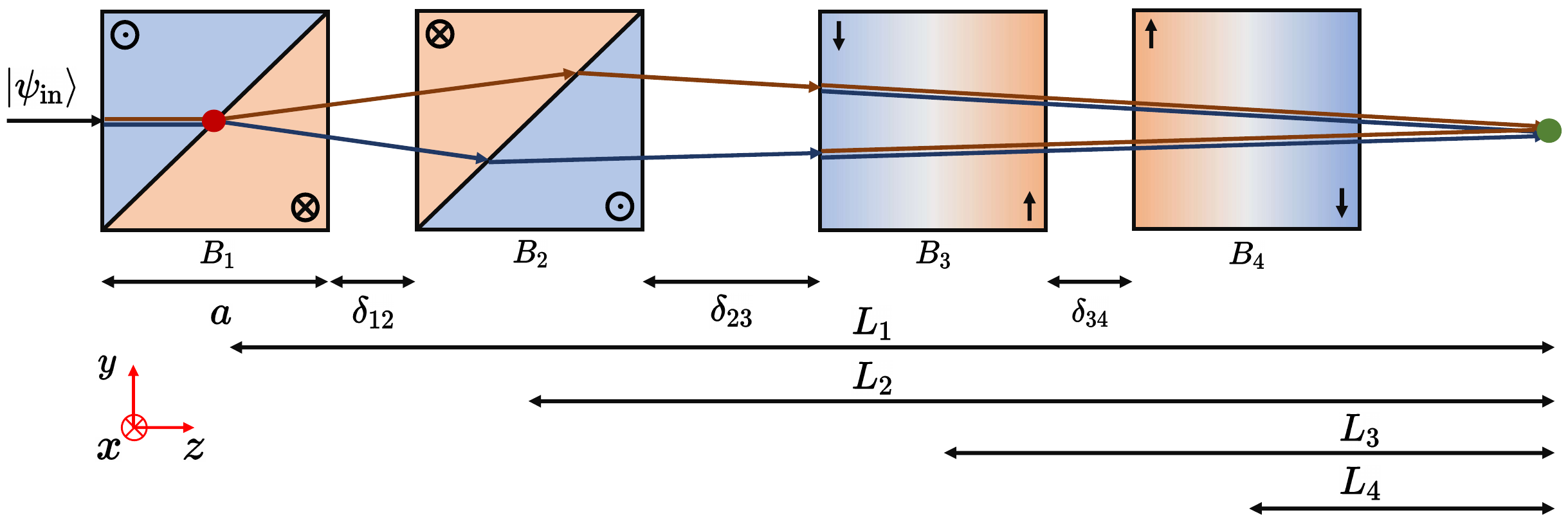}
    \caption{\label{fig:spin texture set up}
    Schematic of an array of two pairs of MWPs described by $\hat{U}_{\rm pair}$, with each pair in the parallelogram SEMSANS configuration. These rays show the action of this configuration on a single representative neutron ray; the incoming ray represents a single plane wave component of an individual neutron’s wavepacket. The spatial extent of the neutron wavepacket is determined by the transverse and longitudinal intrinsic coherence lengths (not shown).
    The magnetic fields $B_\i$, with $\i=1,2,3,4$, are chosen according to the focusing condition of each pair, namely $B_1 L_1= B_2 L_2$ and $B_3 L_3=B_4 L_4$. The red and green points denote the coordinate origin and the focusing point of the original ray, respectively. Before entering the third MWP, the red (blue) arrow denotes the state $\ket{\uparrow_x}$ ($\ket{\downarrow_x}$). As they enter the third MWP, since the magnetic field direction is now $-\hat{y}$, the rays are split again into $\ket{\uparrow_y}$ (red) and $\ket{\downarrow_y}$ (blue) in the $xz$ plane, as indicated by the double line in the last MWP pair.}
\end{figure*}

We now discuss the experimental protocol required to impart various spin textures and OAM densities to the incident neutron beam using pairs of MWPs in the SEMSANS configuration, an example of which is diagrammed in Fig. \ref{fig:spin texture set up}.
First, we rephrase our previous results for arbitrary magnetic field direction. In the interferometric limit, the general structure of the (spin part of the) unitary operator representing a pair of MWPs with both field setups in the SEMSANS configurations has the spatial phase variations form of Eqs.~(\ref{unitary operator final position parallel}) and (\ref{unitary operator final position triangular}). We can generalize this form to the case where the magnetic fields in the pair of MWPs point along an arbitrary direction $\hat{n}$. Up to an overall phase, we find
\begin{equation}
     \hat{U}^{\rm P,T}_{\hat{n}} = \cos \phi^{\rm P,T}\left(n_\perp \right) + i \sin \phi^{\rm  P,T} \left( n_\perp \right) \hat{\sigma}_{\hat{n}} ,
\end{equation}
where $\hat{n}_\perp = \hat{v} \times \hat{n}$ with $\hat{v}$ being the beam propagation direction, and $n_\perp$ is the coordinate measured along the $\hat{n}_\perp$-direction.
We assume that the detector is a vertical plane and placed at the position such that $B_1 L_1 = B_2 L_2$. Thus, the phase spatial variations can be generalized in a similar way: to $\mathcal{O}\left[ \alpha_\i^2, \alpha_\i \varphi^{\;}_{\hat{n}}, \varphi_{\hat{n}}^2 \right]$, we obtain
\begin{equation}
    \phi^{\rm P,T}_{f}\left(n_\perp \right) = \kappa^{\rm P,T}_{\hat n} n_{\perp} ,
\end{equation}
where $\kappa^{\rm P,T}_{\hat n}$ is found from Eq.~(\ref{unitary operator final position parallel}) and (\ref{unitary operator final position triangular}) for both the parallel and triangular geometry to be
\begin{align}
    \kappa^{\text{P}}_{\hat n} =& \frac{2|\mu| \Delta B_{\hat n} \left(1 + \varphi\right)}{v_0 \hbar} \\
    \kappa^{\text{T}}_{\hat n} =& \frac{2|\mu| \Delta B_{\hat n}}{v_0 \hbar}  + \frac{2|\mu| B_{\hat n} \, \varphi}{v_0 \hbar} ,
\end{align}
where $\Delta B_{\hat n}=B_1-B_2\le 0$ or $\Delta B_{\hat n}=B_3-B_4\le 0$ is the difference in the magnitude of the magnetic field in a single MWP pair, and $B_{\hat n} =  B_1+B_2$ or $B_{\hat n} = B_3 + B_4$.
As an aside, when concatenating two pairs of MWPs in the SEMSANS configurations with orthogonal magnetic field orientations, one needs to use the general expressions for the phase spatial variation shown in Eqs. \eqref{unitary operator final position parallel}-\eqref{unitary operator final position triangular}. This requirement is due to each pair of MWPs having a different focusing plane; therefore, there is no common focusing plane.

We now consider the specific case where the field orientations of the two pairs of MWPs are perpendicular to each other in the $xy$ plane (see Fig. \ref{fig:spin texture set up}). Without any loss of generality, let us take the beam direction to be in the $z$ direction ($\hat{v}=\hat{z}$) and the first MWP pair to have the field orientated in the $x$ direction ($\hat{n}_1=\hat{x}$, $n_{1\perp}=y$), which implies that the field direction in the second pair is in the $y$ direction ($\hat{n}_{2} = \hat{y}$, $n_{2\perp} = -x$).
The minus sign in $n_{2\perp}$ is necessary due to the cross product that defines $\hat{n}_{2\perp}$.
The effective (i.e., spin part only) unitary operator, in the interferometric limit, can then be represented by the two by two matrix 

\begin{widetext}
\begin{equation} \label{operator pair}
   \hat{U}_{\text{pair},{\sf spin}} = \hat{U}_{y} \hat{U}_{x} = \begin{pmatrix}
 \cos \kappa_x x \cos \kappa_y y - i \sin \kappa_x x \sin \kappa_y y &  - \sin \kappa_x x \cos \kappa_y y  + i   \cos \kappa_x x \sin \kappa_y y \\
  \sin \kappa_x x \cos \kappa_y y + i \cos \kappa_x x \sin \kappa_y y  & \cos \kappa_x x \cos \kappa_y y + i \sin \kappa_x x \sin \kappa_y y \\
\end{pmatrix} ,
\end{equation}
\end{widetext}
where $\kappa_x$ and $\kappa_y$ denote, respectively, the phase gradient imparted by the second ($B_3,B_4$ fields) and first ($B_1,B_2$ fields) pair of MWPs, and we have omitted the superscripts $\rm P,T$.
For the rest of this section, we will assume that both pairs of MWPs share the same $\Delta B_{\hat n}$ leading to $\kappa_x=\kappa_y=\kappa$. This choice of $\kappa_x $ and $\kappa_y$ is only possible for the parallelogram geometry because for the triangular case, one must fix not only the difference between the magnetic fields but also their sum. This choice for $\kappa$ can be satisfied by a new geometric focusing condition that extends Eq.~\eqref{focusing condition} to apply to two pairs of MWPs, which is given by the following set of equations:
\begin{subequations} \label{eq:ch focus}
\begin{align}
    B_2 &= B_1 \frac{L_1}{L_2} \\
    B_3 &= B_1 \frac{L_4(L_1 - L_2)}{L_2(L_3 - L_4)} \\
    B_4 &= B_1 \frac{L_3(L_1 - L_2)}{L_2(L_3 - L_4)}
\end{align}
\end{subequations}
where $B_1$ is chosen to achieve the desired spin texture period and the various distances $L_\i$ for $\i =1,2,3,4$ are the distances between the midpoint of each MWP and the geometric focusing plane as shown in Fig. \ref{fig:spin texture set up}. The spin texture period in both the $x$ and $y$ directions is still given by Eq.~\eqref{eq:fringe period}.

\subsection{Spin texture} \label{sec:st}

\begin{figure*}[p]
    \centering
    \includegraphics[width=0.925\linewidth]{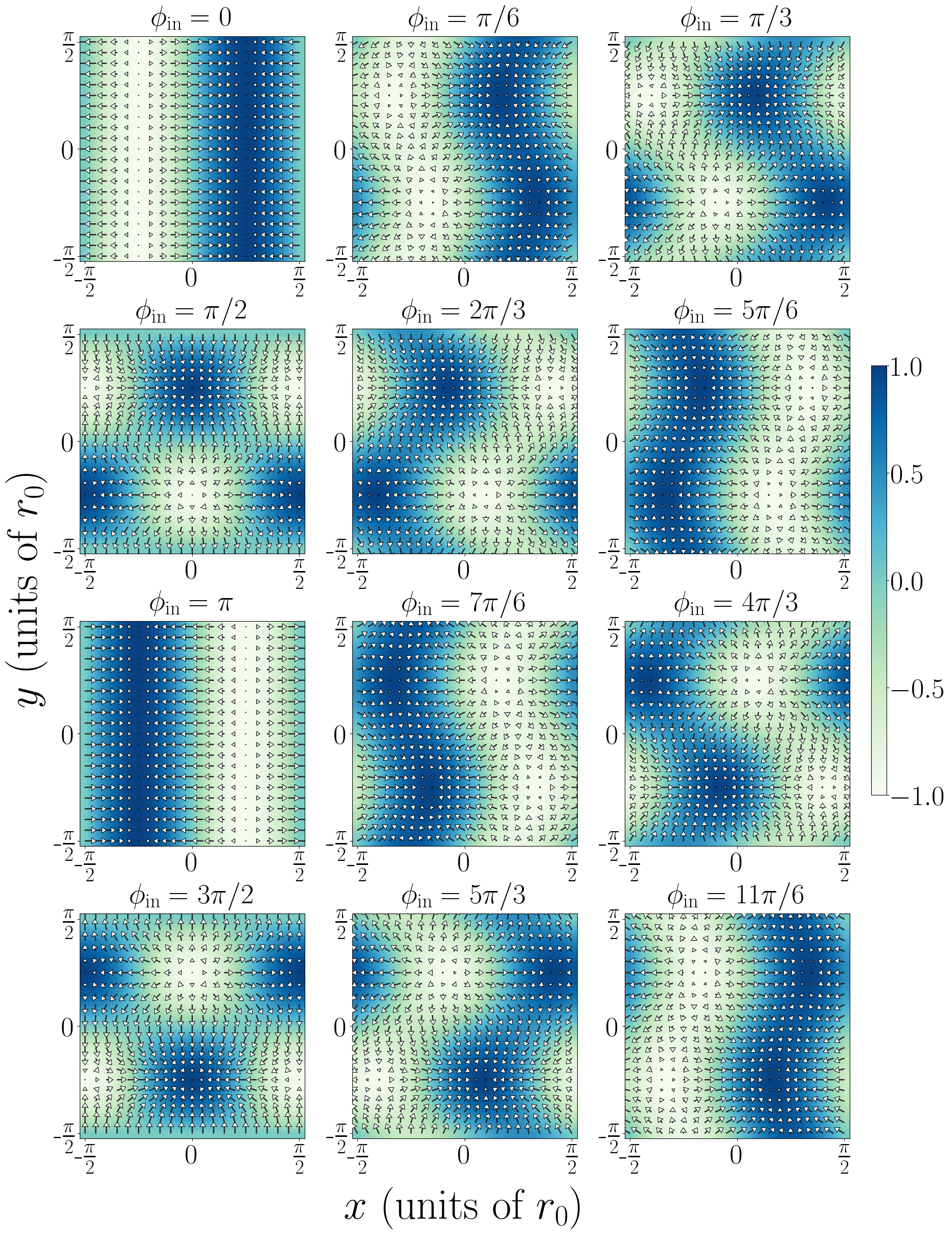}
    \caption{\label{fig:spin texture phis}
    Checkerboard spin textures $\langle \vec \sigma(x,y) \rangle $ plotted in units of $r_0 = v_0 \hbar / (2 |\mu| \Delta B)$, with a minimum $r_0$ of roughly 50 microns (see Sec. \ref{sec:st}). Note that $r_0$ is inversely proportional to both the neutron wavelength and the magnetic field magnitude. The azimuthal initial spin state angle $\phi_{\mathrm{in}}$ is varied while $\theta_{\mathrm{in}} = \pi /2$ is held constant, which ensures that the incident neutron polarization is in the $xy$ plane. The color represents the magnitude of the  longitudinal component $\hat{z}$ of the polarization, and arrows its transverse component ($xy$ plane); the length of the arrows is scaled to the magnitude of the transverse projection. Only the unit cell is shown.}
\end{figure*}

It is enlightening to consider the effect of $\hat{U}_\text{pair}$, Eq.~(\ref{operator pair}), on an arbitrary incident neutron spin state. The action of such an operator on an incoming neutron pure state creates intricate spin textures that manifest in the measurement of the neutron spin polarization.
For a generic initial polarization state given by $\ket{\psi^s_{\mathrm{in}}} = \cos(\theta_{\mathrm{in}}/2)\, \ket{\uparrow_z} + e^{i\phi_{\mathrm{in}}} \sin(\theta_{\mathrm{in}}/2) \, \ket{\downarrow_z}$, which can be represented by the column vector
\begin{equation}
    \ket{\psi^s_{\mathrm{in}}} = \begin{pmatrix}
    \cos\frac{\theta_{\mathrm{in}}}{2} \\
    e^{i\phi_{\mathrm{in}}} \sin\frac{\theta_{\mathrm{in}}}{2}
    \end{pmatrix},
\end{equation}
one obtains the following experimentally observable pattern:
\begin{widetext}
\begin{subequations}\label{spin texture}
\begin{align}
     \langle \psi^s |\hat{\sigma}_x |\psi^s \rangle =&  \sin \theta_{\mathrm{in}}  \cos \phi_{\mathrm{in}}  \cos 2 \kappa x + \sin 2 \kappa x \left(\cos \theta_{\mathrm{in}}  \cos 2 \kappa y - \sin \theta_{\mathrm{in}}  \sin \phi_{\mathrm{in}}  \sin 2 \kappa y\right) \\
    \langle \psi^s |\hat{\sigma}_y |\psi^s \rangle =& \sin \theta_{\mathrm{in}} \sin \phi_{\mathrm{in}}  \cos 2 \kappa y + \cos \theta_{\mathrm{in}}  \sin 2 \kappa y \\
     \langle \psi^s |\hat{\sigma}_z |\psi^s \rangle =& - \sin \theta_{\mathrm{in}}  \cos \phi_{\mathrm{in}}  \sin 2 \kappa x + \cos 2 \kappa x \left( \cos \theta_{\mathrm{in}}  \cos2 \kappa y - \sin \theta_{\mathrm{in}} \sin \phi_{\mathrm{in}} \sin 2 \kappa y\right) ,
\end{align}
\end{subequations}
\end{widetext}
where we defined $|\psi^s \rangle=\hat{U}_{\text{pair},{\sf spin}}\, \ket{\psi^s_{\mathrm{in}}}$. See Fig. \ref{fig:spin texture phis} for examples of some spin textures.
Due to the intrinsic periodicities in the longitudinal component of the polarization pattern, a state generated at the detector such as that of $\phi_{\rm in}=\pi/2$ in Fig. \ref{fig:spin texture phis} is called a \textit{checkerboard state}. 

If one is to use these spin-textured beams to probe a target whose magnetic correlation length is smaller than the one defined by the spin texture period, the fundamental scattering would not differ from that of an untextured neutron beam.
However, in the opposite limit when the period of the spin texture is equal to or smaller than the magnetic correlation length, one must use the full quantum-mechanical description after the action of $\hat{U}_{\text{pair}}$ in Eq.~(\ref{operator pair}) to compute the scattering cross section. 
These spin textured beams may be useful for measuring mesoscopic magnetically ordered systems, such as skyrmions; for example, we would expect a scattering resonance when the period of the spin texture equals the period of the skyrmion, analogous to the neutron wavelength in diffraction.

The magnetic fields of presently available MWPs are limited to about 150 mT by the maximum current of around 50 A that can be carried by their superconducting coils. The period of the spin textures that can be produced depends on the magnitude of the magnetic fields as well as on the distances between prisms and the neutron wavelength (see Eq. \eqref{eq:fringe period}).
SEMSANS experiments to date have produced intensity oscillations with a minimum period of about 150 microns \cite{Dadisman_2019}. If we choose $L_1$ = 1.3 m, $L_2$ = 0.9 m, $L_3$ = 0.7 m, and $L_4$ = 0.3 m in Fig. \ref{fig:spin texture set up} and Eq. \eqref{eq:ch focus}, and a maximum magnetic field of 150 mT in the second prism (which has the largest field due to the focusing condition), the period of the spin textures shown in  Fig. \ref{fig:spin texture phis}, which scales inversely with neutron wavelength, is 145 microns for a neutron wavelength of 1 nm. The spin texture period can also be reduced by a factor of two by doubling the number of MWPs used, although such a setup would increase the length of the beamline required.

\subsection{Momentum-Spin Entanglement}

In this section and the following, we will focus on the nature of the quantum correlations generated by the action of our operator $\hat{U}_{\text{pair}}$. To better understand the nature of the entanglement realized by $\hat{U}_{\text{pair}}$, it is instructive to consider again both the spin and spatial components, and to express the neutron state in momentum space.
Since our operator does not affect the propagation of the neutron along the $z$ direction after the neutron has exited the last MWP, we only focus on the transverse $xy$ plane by making use of the completeness relation for $|\vec k_\perp \rangle$ where $\vec k_\perp=(k_x,k_y)$, with spatial coordinate representation $\langle \vec r_\perp| \vec k_{\perp} \rangle=\frac{1}{2 \pi \hbar} \exp(i \vec k_{\perp} \cdot \, \vec r_{\perp})$.
Expressing the operator in the momentum representation yields
\begin{align}
\nonumber
    \hat{U}_{\mathrm{pair}} &= \int_{\mathds{R}^3} d\vec r_0 d\vec k_\perp d\vec k_{0\perp} \hat{U}_{\mathrm{pair},{\sf spin}}\left(\vec r\right) |\vec k_\perp\rangle \langle \vec k_\perp| \vec r\rangle\\
    \nonumber
    & \hspace*{4.4cm}\times \langle \vec r_0| \vec k_{0\perp} \rangle \langle \vec k_{0\perp} | \\
    &=  \int_{\mathds{R}} dz_0 d\vec k_\perp d\vec k_{0\perp} \hat U_{\mathrm{pair},{\sf spin}}(\vec{k}_\perp-\vec{k}_{0\perp}) |\vec k_\perp\rangle \langle \vec k_{0\perp} | \nonumber\\
    & \hspace*{4.3cm} \otimes |z \rangle \langle z_0 | , 
    \label{Upair}
\end{align}
where we made use of $\vec r_{\perp} \approx \vec r_{0\perp}$ and $\ket{\vec{r} \,} = \ket{\vec{r}_{\perp}} \otimes \ket{z}$,
and the integrals in $\vec k_\perp$ and $\vec k_{0\perp}$ are over ${\mathds{R}^2}$. We must return to writing the full operator $\hat{U}_{\rm pair}$ with the explicit position-dependence for the sake of clarity when obtaining the Fourier transform of $\hat{U}_{\rm pair}$.
Thus, the momentum space representation of $\hat{U}_{\mathrm{pair},{\sf spin}}$ can be represented by the matrix 
\begin{widetext}
\begin{align}
\label{operator pair momentum}
    &\hat U_{\mathrm{pair},{\sf spin}}(\vec{k}_\perp-\vec{k}_{0\perp}) = \frac{1}{(2\pi \hbar)^2}\int_{\mathds{R}^2} d\vec r_\perp \hat{U}_{\mathrm{pair},{\sf spin}}\left(\vec r_\perp\right) e^{-i \left(\vec{k}_\perp - \vec{k}_{0\perp}\right)\cdot \vec{r}_{\perp}} \\
    &= \frac{1}{2\sqrt{2}} 
\begin{pmatrix}
   e^{i\frac{\pi}{4}} \left( \delta_{+x} \delta_{+y}  + \delta _{-x} \delta_{-y} \right)+ e^{-i\frac{\pi}{4}} \left(  \delta_{-x} \delta _{+y}+   \delta_{+x} \delta_{-y} \right) & e^{i\frac{\pi}{4}}  \left( \delta _{-x} \delta_{-y}  -  \delta_{+x} \delta_{+y} \right) + e^{-i\frac{\pi}{4}} \left(\delta _{+x} \delta_{-y} - \delta _{-x} \delta_{+y} \right) \\
    e^{-i\frac{\pi}{4}}  \left(  \delta_{-x} \delta_{-y} - \delta _{+x} \delta_{+y} \right) +e^{i\frac{\pi}{4}} \left( \delta _{+x} \delta_{-y} -  \delta_{-x} \delta_{+y} \right)  & e^{-i\frac{\pi}{4}} \left(   \delta_{+x} \delta_{+y} + \delta_{-x} \delta_{-y} \right) + e^{i\frac{\pi}{4}} \left(\delta _{-x} \delta_{+y}+  \delta _{+x} \delta _{-y} \right)  \nonumber
\end{pmatrix} ,
\end{align}
\end{widetext}
where we have used the short-hand notation $\delta_{\pm \nu} = \delta\left[ \kappa \pm \left( k_\nu - k_{0\nu} \right) \right]$ for $\nu \in \{x,y\}$ for the Dirac delta function.
Because $\vec{k}_0$ and $\vec{k}$ can be interpreted respectively as the incoming and outgoing wavevectors, the parameter $\kappa$ can be interpreted as the transverse wavevector transfer.

Next, we consider the action of the operator $\hat U_{\mathrm{pair}}$ given in Eq.~(\ref{Upair}) on 
an incoming plane wave polarized along its direction of motion, such that the incoming state has no transverse momentum, i.e., $\vec{k}_{0\perp} = (0,0)$.
The incoming state can then be written as $| \Psi_{\rm in}\rangle= e^{-i E_0 t_i / \hbar} \ket{\uparrow_z} \otimes |{k}_{0z} \rangle$, where $t_i$ is some initial time, with $E_0$ being the initial total energy of the neutron. Therefore, we have 
\begin{widetext}
\begin{align}
    \nonumber
    \hat{U}_{\mathrm{pair}{\sf}} | \Psi_{\rm in}\rangle
    &=\frac{1}{2\sqrt{2}}\begin{pmatrix}
        e^{i\frac{\pi}{4}} \left( |\vec \kappa^+_\perp \rangle  + |-\vec \kappa^+_\perp \rangle  \right) + e^{-i\frac{\pi}{4}} \left( |\vec \kappa^-_\perp \rangle  + |-\vec \kappa^-_\perp \rangle \right)   \\
        e^{-i\frac{\pi}{4}} \left( |\vec \kappa^+_\perp \rangle  - |-\vec \kappa^+_\perp \rangle \right) - e^{i\frac{\pi}{4}} \left( |\vec \kappa^-_\perp \rangle  - |-\vec \kappa^-_\perp \rangle \right) 
    \end{pmatrix}  \int_{\mathds{R}} dz_0   \exp\left(i \frac{k_{0z} \left(z - z_0\right)}{2} \right) \frac{1}{\sqrt{2\pi \hbar}} \exp \left(i \frac{ k_{0z} z_0}{2}\right) |z \rangle \\
    &= \frac{1}{2\sqrt{2}}\begin{pmatrix}
        e^{i\frac{\pi}{4}} \left( |\vec \kappa^+_\perp \rangle  + |-\vec \kappa^+_\perp \rangle  \right) + e^{-i\frac{\pi}{4}} \left( |\vec \kappa^-_\perp \rangle  + |-\vec \kappa^-_\perp \rangle \right)   \\
        e^{-i\frac{\pi}{4}} \left( |\vec \kappa^+_\perp \rangle  - |-\vec \kappa^+_\perp \rangle \right) - e^{i\frac{\pi}{4}} \left( |\vec \kappa^-_\perp \rangle  - |-\vec \kappa^-_\perp \rangle \right) 
    \end{pmatrix}  \int_{\mathds{R}} dz   \exp\left(i \frac{k_{0z} z}{2} \right) \frac{1}{\sqrt{2\pi \hbar}}  |z \rangle ,
\end{align}
\end{widetext}
where we have changed the integration variable from $z_0$ to $z$ due to constraints from the classical path, while  $\vec \kappa^\pm_\perp=(\kappa_x,\pm \kappa_y)$ and in our setup we typically consider $\kappa_x=\kappa_y=\kappa$. Because we are using semiclassical kinematics to express all time dependence in terms of spatial coordinates, the explicit time dependence is encoded in the wavevector and spatial coordinates.
In this form, we can see that $\hat{U}_{\rm pair}$ imparts transverse momenta to the initial state, which means that we can no longer write the outgoing state of the neutron as an unentangled state with respect to the tensor product decomposition $\mathcal{H}_s \otimes \mathcal{H}_{\vec{k}}$ of the spin and momentum subsystems. Hence, the action of $\hat{U}_{\text{pair}}$ can be thought of as an {\it entangler} of spin and the transverse momentum of the distinguishable subsystems.
This point is crucial in understanding the novel scattering signatures of the OAM beams generated by our array of pairs of MWPs: even if the incoming state has no transverse momentum, the resulting outgoing beam displays quantum correlations between the transverse momenta and spin.

\subsection{Orbital Angular Momentum}
\label{OAMC}

\begin{figure*}[p]
    \centering
    \includegraphics[width=0.925\linewidth]{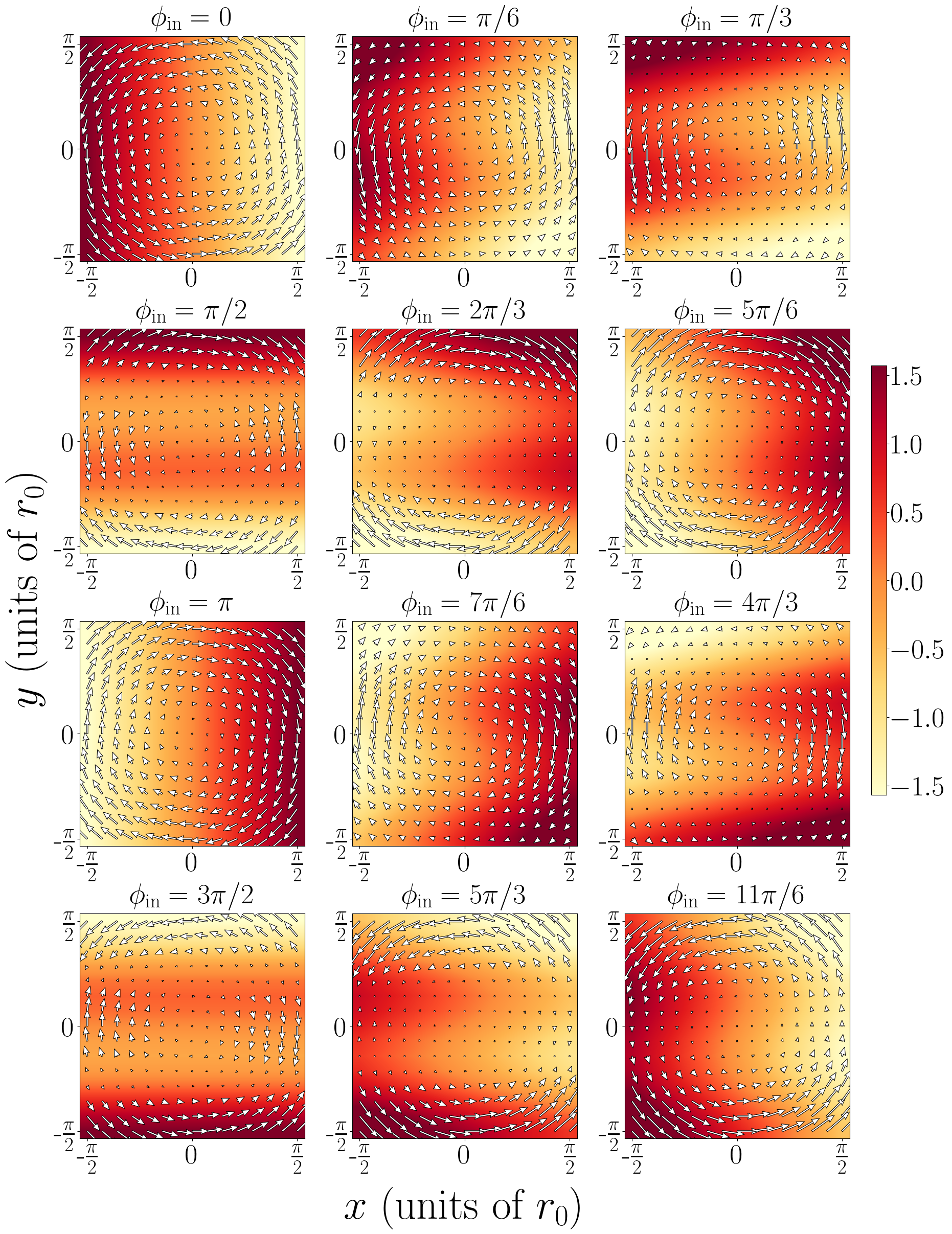}
    \caption{ \label{fig:oam texture phis}
    Checkerboard OAM textures $\langle \vec L(x,y) \rangle$ plotted in units of $r_0 = v_0 \hbar / (2 |\mu| \Delta B)$, with a minimum $r_0$ of roughly 50 microns (see Sec. \ref{sec:st}). Note that $r_0$ is inversely proportional to both the neutron wavelength and the magnetic field magnitude. The azimuthal initial spin state angle $\phi_{\mathrm{in}}$ is varied while $\theta_{\mathrm{in}} = \pi /2$ is held constant, which ensures that the incident neutron polarization is in the $xy$ plane. The color corresponds to the longitudinal component $\hat{z}$ of the OAM density and arrows its transverse component ($xy$ plane); the length of the arrows is scaled to the magnitude of the transverse projection. We only include the OAM terms imparted by our setup, namely the terms proportional to $\kappa$ in Eqs. \eqref{eq:OAM density}.}
\end{figure*}

The inherent periodicity in the spatial coordinates $x$ and $y$ is manifest in Eq.~(\ref{operator pair}). We now show how the generated spin textures are coupled to the orbital motion of the neutron; specifically, there are some special points at the detector plane where the operator $\hat{U}_{\mathrm{pair}}$ locally imparts quantized OAM to the beam. Hence, our proposed setup also generates a rich and complex structure of spin-orbit entangled states. 

Generally, the coordinates of these special points take the form
\begin{equation}
    (x_m,y_n) = \left(\frac{m\pi}{\kappa},\frac{n\pi}{\kappa} \right) ,
\end{equation}
where $m$ and $n$ are either integers or half-odd integers. It is convenient to switch to polar coordinates $(r,\phi)$, where $x-x_m=r \cos \phi$ and $y-y_n=r \sin \phi$ are defined with respect to the origin $(x_m,y_n)$, and expand $\hat{U}_{\text{pair}}$ around these points to $\mathcal{O} \left[\kappa^2 r^2\right]$. We also omit the overall global phase for the sake of simplicity.
The expansion $\forall (m,n) \in \mathbb{Z}$ yields
\begin{equation}
    \label{operator oam}
    \lim_{\substack{ x \rightarrow x_m \\ y \rightarrow y_n}}\hat{U}_{\text{pair},{\sf spin}} = 1 + \kappa r \left(  \hat{l}_+ \hat{\sigma}_- - \hat{l}_- \hat{\sigma}_+ \right) ,
\end{equation}
where $\hat{\sigma}_\pm= (\hat{\sigma}_x\pm\hat{\sigma}_y)/2$ are the usual spin-1/2 ladder operators and $\hat{l}_\pm = e^{\pm i \phi}$ are the OAM ladder operators.
%
%
When $m$ and $n$ are half-odd integers (i.e., $m+n \in \mathds{Z}$), then the action of $\hat{U}_{\text{pair},{\sf spin}}$ is given by
\begin{align}
    \label{operator oam case 2}
     \lim_{\substack{ x \rightarrow x_m \\ y \rightarrow y_n}}\hat{U}_{\text{pair},{\sf spin}} = \hat{\sigma}_z + \kappa r \left( \hat{l}_+ \hat{\sigma}_+ + \hat{l}_- \hat{\sigma}_-  \right).
\end{align}
%
%
Finally, whenever $m \in \mathds{Z}$ and $n$ is a half-odd integer or vice-versa (i.e.,  $m+n$ is a half-odd integer), we obtain 
\begin{align}
    \lim_{\substack{ x \rightarrow x_m \\ y \rightarrow y_n}}\hat{U}_{\text{pair},{\sf spin}} = i\hat{\sigma}_y + \frac{\kappa r}{2} \left[ ( \hat{l}_+ + \hat{l}_- \right) + \left( \hat{l}_+ - \hat{l}_- ) \hat{\sigma}_z  \right] .
\end{align}
All three of these families of special points on the focusing plane will host a spin-orbit state with OAM quantum number $\ell = -1,0,1$, with the specific OAM state determined by the initial neutron spin polarization. Similar expansions were found in previous work \cite{Sarenac2018,SarenacTheory_2018,Sarenac2019}.

To connect the OAM content of the beam to the spin texture, one can define the OAM density $\vec{L}$ of the  final state $\ket{\psi_{\sf d}}=\hat{U}_{\text{pair}}\ket{\psi_{\mathrm{in}}}$ as 
\begin{equation}
    {\vec{L}}(\vec r)  = \psi_{\sf d}^*(\vec r)  {\vec{L}} \psi_{\sf d}(\vec r),
\end{equation}
where the OAM operator $\vec{L}=\left(\vec{r} - z_{\sf d}\right) \times \vec{p}$ is defined with respect to an origin shifted along the $z$ axis to be at the center of the general detector plane. Here, we have defined the initial state as a plane wave that propagates along the $z$ direction as
\begin{eqnarray}
    \ket{\psi_{\mathrm{in}}} &=&e^{-i E_0 t_i / \hbar} \ket{\psi^s_{\mathrm{in}}} \otimes  \ket{k_{0z}} \\
    &=& e^{-i E_0 t_i / \hbar}\left( \cos(\theta_{\mathrm{in}}/2)\, \ket{\uparrow_z} + e^{i\phi_{\mathrm{in}}} \sin(\theta_{\mathrm{in}}/2) \, \ket{\downarrow_z} \right) \nonumber \\
    &&\hspace*{2.0cm} \otimes  \ket{k_{0z}}, \nonumber
\end{eqnarray}
where, again, $t_i$ is some initial time, $E_0$ the initial energy of the neutron, and $k_{0z}$ the initial wavevector of the neutron.
Calculating the OAM density, we find
%
\begin{widetext}
\begin{subequations} \label{eq:OAM density}
\begin{align}
    {L}_x(\vec{r}) =& k_{0z} y + \frac{\kappa y}{1+\varphi} \left( \cos\theta_{\rm in} \sin 2\kappa y + \sin\theta_{\rm in} \left( \cos \phi_{\rm in} + \sin \phi_{\rm in} \cos 2 \kappa y \right) \right) \\
    {L}_y(\vec r) =& - k_{0z} x - \frac{\kappa x}{1+\varphi} \left( \cos\theta_{\rm in} \sin 2\kappa y + \sin\theta_{\rm in} \left( \cos \phi_{\rm in} + \sin \phi_{\rm in} \cos 2 \kappa y \right) \right)  \\
     {L}_z(\vec r) =& \kappa \left (y \cos \theta_{\text{in}} \sin 2 \kappa y + \sin \theta_{\text{in}} (y \sin \phi_{\text{in}} \cos 2 \kappa y + x \cos \phi_{\text{in}} )\right) .
\end{align}
\end{subequations}
\end{widetext}
We plot a few examples of some OAM textures in Fig. \ref{fig:oam texture phis}.
\section{Outlook and conclusions} \label{conclusion}

While neutron beams are successfully utilized as indirect probes of many exotic phases of matter, such as quantum spin liquids and magnetic skyrmion lattices \cite{li2019, thanh2022}, or as test particles to investigate fundamental physical phenomena including gravity and dark matter \cite{snow2022}, a creative endeavor lies in producing arbitrary neutron states.
In this work, we have presented an experimental protocol to generate neutron beams with spin textures that couple to the neutron's orbital motion and impart OAM to the beam. The setup involves two pairs of MWPs whose magnetic fields and relative positions and orientations can be chosen to generate a variety of complex structures and, therefore, could extend the analysis of quantum magnetic materials with exotic spin structures, such as chiral magnets \cite{seki,tkachov}. The range of length scales that can be probed by the current generation of MWPs extend down to around 100 microns, which is a limit determined by the maximal magnetic field that can be contained by the superconducting films.

We have also presented a systematic theoretical investigation from quantum mechanical first principles of a SEMSANS setup of MWPs. Starting from a path-integral representation of the unitary time evolution operator, a simplification is introduced by invoking the interferometric limit where the refraction of the neutron at magnetic field boundaries is taken into account by defining two opposite-spin correlated trajectories that interfere at the detector. Our approach is reminiscent of the connection between wave and geometric optics.
An important consequence of the refractive effects on the two opposite-spin paths is the appearance of geometric focusing of these two states. Specifically, we have highlighted that there is an inherent assumption that the longitudinal and transverse coherence coherence lengths are larger than the spatial separation of the neutron's two spin states for such an interferometric limit to remain valid.
Furthermore, we have shown that in addition to the usual magnetic-field dependent phase considered previously, there is a phase contribution coming from the kinetic-energy. However, under the assumption that the final position $\vec r_f$ and time $t_f$ of the neutron's opposite-spin rays coincide at the detection plane given that those rays started at the same initial time $t_i$ at different initial positions $\vec r_{i\sigma}$ (within the quantum coherence volume of the neutron), the kinetic contribution to the phase cancels out.
In this way, our formalism should allow the measurement of the transverse and longitudinal intrinsic coherence lengths of the neutron by measuring the distance (relative to the center of the detection plane) at which the interference fringes fade away.
Finally, we concluded that the phase focusing condition described by Eq.~(\ref{focusing condition}), derived from the single-path Larmor precession model, agrees with our geometric focusing condition to lowest order in $\alpha_\i$.

To describe the OAM states generated by MWPs, we derived the quantum operator of our proposed SEMSANS setup, where the magnetic field directions in the two pairs of MWPs are perpendicular to one another. This setup creates a spin-textured neutron beam where there is a lattice of points with respect to which the MWPs impart a rich variety of spin-OAM entanglement; these structures are a result of the interference between the two opposite-spin paths with momenta $\vec{k}_\uparrow$ and $\vec{k}_\downarrow$.
One important implication of this result is that the paraxial approximation usually used in neutron scattering may not be used for the analysis of OAM when using MWPs \cite{Barnett_2022}.
To compute the refractive corrections we utilized the magnetic Snell's law, an interesting universal derivation of which (based on a relativistic action principle) can be found in Appendix \ref{sec: snell law}. This derivation is universal in the sense that it is applicable to both massless and massive particles, relativistic or not.
We also explicitly showed in the momentum representation that our configuration of MWPs act as a transverse momentum and spin entangler, which intuitively explains why these neutron beams carry OAM, as well as why the constituent neutrons are in spin-orbit entangled states.

An analog of the unitary operator $\hat{U}_{\text{pair}}$ for the case $|\psi_{\mathrm{in}}\rangle = e^{-i \omega t_i}\ket{\uparrow_z}\otimes \ket{k_{0z}}$ (e.g., the state with $\theta_{\mathrm{in}},\phi_{\mathrm{in}}=0$) was experimentally realized using triangular coils  (LOV prisms) \cite{Sarenac2019}. However, in that work, the spin-texture period is on the order of a few centimeters.
The calculation of the OAM was done in a quantum version of the single-path Larmor precession, where no refraction was assumed. While that single-path model was sufficient to analyze the generated spin texture, to calculate the OAM states, it is required to take refraction into account. One can do this by either using a wavepacket model (for example, see \cite{keller_book}), or as we did via the path-integral formulation.
As mentioned above, a key feature of our calculation that extends previous work \cite{Sarenac2018, Sarenac2019} is the kinetic phase contribution to the phase spatial variations to first-order in deflection angles $\alpha_\i$. Most importantly, the experimental and technical challenges and future prospects of LOV prisms compared to MWPs are completely different. 

Another surprising yet interesting result is that the refractive effects encoded in $\alpha_\i$ are not observable in the magnetic-field dependent phase up to $\mathcal{O} \left[\alpha_\i^2, \alpha_\i \varphi, \varphi^2\right]$, because all of these refractive terms can be factored out as an overall global phase.
The absence of $\alpha_\i$ arises from the fact that the refractive corrections to the two spin states' paths are both proportional to $\pm \alpha_\i$, meaning that they only differ by a sign.
Hence, any linear correction $\mathcal{O} \left[\alpha_\i^2, \alpha_\i \varphi, \varphi^2\right]$ to the magnetic-field dependent phase that each spin state accumulates is always exactly equal but opposite. However, due to their opposite spins, the magnetic-field dependent phases are accumulated in opposite direction; therefore, both states acquire exactly the same correction.
Because all of these corrections are to the phases of the coefficients $A_{\sigma}$ in Eq.~(\ref{operator general form}), they can be factored out as an overall global phase since the corrections are the same for the two opposite-spin paths. 
The final phase observed (i.e., the total relative phase between the two states) in the measurement of the polarization at the detection plane in the parallelogram configuration is $\Phi_L$, which can be written as an expansion of the deflection angles, which takes the general form
$\Phi_L = \Phi_L^{(1)}+\Phi_L^{(2)}+\ldots$, where $\Phi_L^{(n)}$ is the n\textsuperscript{th} order refractive correction. To first order, we have
\begin{align}
\nonumber
\Phi^{(1)}_L&= \frac{4|\mu|\left(B_1-B_2\right)\left(1+\varphi\right)y}{v_0 \hbar} \\
	& \hspace*{0.5cm}- \frac{4|\mu| \left( B_1 L_1 - B_2 L_2 \right) \varphi}{v_0 \hbar},
\end{align}
which is indeed equivalent to the phase predicted by the single-path Larmor precession model.

In order to obtain a non-trivial refractive correction to the magnetic-field dependent phases of the two opposite-spin paths, one needs to consider the refractive correction to second order. The second order refractive effects will give corrections to both the geometric focusing condition and the phase-gradient.
The $\mathcal{O} \left[\alpha_\i^3,...\right]$ corrections to the phase spatial variations for both geometries are of the general form
\begin{align}
    \nonumber
    \Phi_L^{(2)} &= \frac{|\mu|}{2v_0 \hbar} \left( A_1 \alpha_{1}^2 + A_{12} \alpha_1 \alpha_2 + A_2 \alpha_2^2 + A_{\varphi} \varphi^2 \right. \\
    & \hspace*{1.5cm}\left. + A_{1\varphi} \alpha_1 \varphi + A_{2\varphi} \alpha_2 \varphi  \right) ,
    \label{eq: higher order form}
\end{align}
where $A_1$, $A_{12}$, $A_2$, $A_\varphi$, $A_{1\varphi}$ and $ A_{2\varphi}$ are coefficients given in Appendix ~\ref{higher order} that depend on the magnetic field strengths, the position and size of the MWPs, and the initial state of the neutron.
We notice some interesting features. First of all, contrary to the lowest-order $\mathcal{O} \left[\alpha_\i^2, \alpha_\i \varphi, \varphi^2\right]$ calculation where the additional phase is proportional to the beam divergence $\varphi$, we find terms that are intrinsically due to the refractive effect of the beam even in the absence of the beam divergence.
Secondly, $ A_{1\varphi}$ and $A_{2\varphi}$ are contributed purely from the kinetic phase, and no cross-term $\alpha_\i \varphi$ appears in the magnetic field-dependent phase for the same reason that the first-order refractive correction vanishes. Therefore, to properly obtain the higher order refractive corrections, one must consider the kinetic phase.
Last of all, we find that the correction to the phase spatial variation now depends not only in the distance  $a+\delta_{12}$ between centers of the MWPs, but also depends on the physical size of the MWP in a complicated way.

\FloatBarrier
\begin{acknowledgements}
We thank David V. Baxter, Dennis Krause, Stephen Kuhn and Mike Snow for useful discussions. The IU Quantum Science and Engineering Center is supported by the Office of the IU Bloomington Vice Provost for Research through its Emerging Areas of Research program. We acknowledge support from the US Department of Commerce through cooperative agreement number 70NANB15H259. The development of magnetic Wollaston prisms was funded by the US Department of Energy through its STTR program (grant number DE-SC0009584).
\end{acknowledgements}

\FloatBarrier
\appendix

\section{The Kinetic Phase and Unitarity in the Interferometric Limit} \label{app:unitary}

We now discuss in more detail why the free propagator is given by $\exp \left( i \vec{k}_{\sigma} \cdot \frac{\left(\vec{r}_\sigma - \vec{r}_i \right)}{2}\right)$ in the interferometric limit.
This result is deeply rooted in our interferometric limit where space and time can be related to each other via the classical trajectories. For the sake of simplicity, consider the exact free propagation of an initial plane wave
\begin{equation}
    \Psi \left(\vec{r}_i,t_i\right) = 1/(2\pi\hbar)^\frac{3}{2} \exp(i( \vec{k} \cdot \vec{r}_i - \omega t_i)),
\end{equation}
which travels from spacetime coordinates $\left( \vec{r}_i, t_i \right)$ to $\left( \vec{r}_f, t_f \right)$. The wave function after propagation is given by 
\begin{widetext}
\begin{align}
    \Psi \left(\vec{r}_f,t_f\right) =& \int_{\mathds{R}^3} d\vec{r}_i \ \sqrt{\frac{m}{i 2 \pi \hbar \left( t_f - t_i \right)}}  \exp\left(i\frac{m \left( \vec{r}_f - \vec{r}_i \right)^2}{2\hbar \left( t_f - t_i \right)} \right)  \Psi \left(\vec{r}_i,t_i\right)  =  \frac{1}{(2\pi\hbar)^\frac{3}{2}} \exp\left(i  \left(\vec{k} \cdot \vec{r}_f - \omega t_f \right) \right),
\end{align}
\end{widetext}
where $\omega=\hbar k^2/2m$ is the dispersion relation for a free, non-relativistic,  particle.
In our interferometric approximation, we replace the above convolution integral by a multiplicative phase related to the classical action. Such a phase factor in the case of a plane wave is 
\begin{equation}
\nonumber
    \exp\left(i \left[ \vec{k} \cdot \left(\vec{r}_f -\vec r_i\right) - \omega \left( t_f - t_i \right) \right] \right),
\end{equation}
which can be further simplified  by recognizing that the strictly classical trajectory of the interferometric limit allows us to write 
\begin{equation}
    \omega (t_f-t_i)  = \frac{1}{2} \frac{m \vec{v}}{\hbar} \cdot \vec{v}\, (t_f-t_i) = \frac{1}{2} \vec{k} \cdot  (\vec{r}_f-\vec{r}_i).
\end{equation}
Hence, we obtain exactly the so-called kinetic phase.

Next, we explicitly show that even after taking the interferometric limit, 
the resulting operator shown in Eq.~(\ref{interferometric operator}) remains unitary
\begin{widetext}
\begin{align}
    \nonumber
	\left( \hat{U}_\nu^{\pm} \right)^{\dagger} \hat{U}_\nu^\pm  =& \int_{\mathds{R}^3} d \vec r_i d \vec r_i^{\, \prime} \, \sum_{\sigma, \sigma^\prime=\uparrow,\downarrow} A_{\sigma^\prime}^{\pm*}(\vec r_i^{\, \prime})A_{\!\sigma}^\pm(\vec r_i)|\sigma_\nu^\prime\rangle \langle\sigma_\nu^\prime \, \!| \sigma_\nu\rangle\left\langle\sigma_\nu \, \!| \otimes| \vec r_i^{\, \prime}\right\rangle\left\langle \vec r_{o\sigma^\prime}\left(\vec r_i^{\, \prime}\right) | \vec r_{o\sigma}\left(\vec r_i\right)\right\rangle\left\langle \vec r_i\right|\\
	\nonumber
    =& \int_{\mathds{R}^3} d \vec r_i d \vec r_i^{\, \prime} \, \sum_{\sigma, \sigma^\prime=\uparrow,\downarrow} A_{\sigma^\prime}^{\pm*}(\vec r_i^{\, \prime})A_{\!\sigma}^\pm(\vec r_i)|\sigma_\nu^\prime\rangle \delta_{\sigma\sigma^\prime}\left\langle\sigma_\nu \, \!| \otimes| \vec r_i^{\, \prime}\right\rangle \delta\left(\vec r_{o\sigma^\prime}\left(\vec r_i^{\, \prime}\right) - \vec r_{o\sigma}\left(\vec r_i\right) \right)  \left\langle \vec r_i\right|\\
    \nonumber
    =& \int_{\mathds{R}^3} d \vec r_i d \vec r_i^{\, \prime} \, \sum_{\sigma=\uparrow,\downarrow} A_{\sigma}^{\pm*}(\vec r_i^{\, \prime})A_{\!\sigma}^\pm(\vec r_i)|\sigma_\nu\rangle \left\langle\sigma_\nu \, \!| \otimes| \vec r_i^{\, \prime}\right\rangle \delta\left(\vec r_i^{\, \prime} - \vec r_i \right)  \left\langle \vec r_i\right|\\
	=&  \int_{\mathds{R}^3} d \vec r_i \ \sum_{\sigma=\uparrow,\downarrow} \left|A_\sigma^\pm \left(\vec{r}_i\right)\right|^2 \ket{\sigma_\nu}\bra{\sigma_\nu} \otimes \left|\vec r_i\right\rangle\langle \vec r_i| =\mathds{1}, \qquad \nu \in \{x,y\}
\end{align}
\end{widetext}
where we make use of $\left| A_\sigma^\pm \right| = 1$ in the geometrical ray limit such that $\vec r_{o\uparrow ,\downarrow}\left(\vec r_i\right)$ is a bijective function, and thus we have $\delta\left(\vec r_{o\sigma^\prime}\left(\vec r_i^{\, \prime}\right) - \vec r_{o\sigma}\left(\vec r_i\right) \right) = \delta \left( \vec r_i-\vec r_i^{\, \prime} \right)$. Therefore, $\hat{U}_\nu^\pm$ is unitary. Finally, as the total time evolution operator of the prism $\hat{U}^{\sf MWP}_\nu$ is a product of 2 unitary operators, it must also be unitary.

\section{Derivation of Magnetic Snell's law} \label{sec: snell law}

The law of the refraction of light was discovered by the dutch mathematician and astronomer 
Willebrord Snellius (Snell) in 1621. Its justification has a colorful history \cite{Rojo}. Curiously, 
Descartes (1637) and Fermat (1657) developed two conflicting explanations. 
Descartes based his derivation on what is known today as the conservation of momentum in the direction parallel to the boundary separating the two media. For incident (refracted) angle $\theta_i$ ($\theta_o$) and speed of light $c_i$ ($c_o$), he obtained
\begin{align}
    \label{snell law neutron}
    c_i \sin\theta_i = c_{o} \sin\theta_o ,
\end{align}
which qualitatively disagrees with Snell's result. Meanwhile, Fermat obtained his result 
\begin{equation}
    \label{snell law light}
    \frac{\sin\theta_i}{c_i} = \frac{\sin\theta_o}{c_o} ,
\end{equation}
by assuming the principle of least time. It turns out that Fermat obtained the correct result for light refraction.

Interestingly, it is also known that Snell's law for non-relativistic neutron optics follows Decartes' original result, 
with $c_i \, (c_o) \rightarrow v_{i\sigma} \, (v_{o \sigma})$. Therefore, Snell's law for neutron refraction cannot be derived from Fermat's principle of least time. From a theoretical point of view, there should be a unified principle yielding both of these behaviors in the appropriate limits. In this section, we present a derivation of the magnetic Snell's law based on a stationary action principle which unifies both refraction laws for light and neutrons when the two media do not move with respect to each other. As will be shown, the difference arises from the distinction between massless and massive particles (in the non-relativistic limit).

We start by considering the action for a relativistic particle in a potential $V(\vec{r})$ characterizing the medium 
\begin{equation}
    S ( \vec{r},\dot{\vec{r}}) =  \int_{t_1}^{t_2} \!\! dt \, 
    (-\frac{mc^2}{\gamma} - V(\vec{r}))=  \int_{t_1}^{t_2} \!\! dt \,L(\vec{r},\dot{\vec{r}}) ,
\end{equation}
where $\gamma=1/\sqrt{1-\frac{v^2}{c^2}}$, $\vec{v}$ is the velocity of the particle and $c$ the speed of light. For refraction, 
$V(\vec{r})$ is constant in each medium with a discontinuous jump across the boundary. In Hamilton's variational principle, the action $S$ is a functional of both the coordinate $\vec{r}$ and its first time derivative $\dot{\vec{r}}$. Its variation leads to the usual Euler-Lagrange equations of motion. However, for particles undergoing pure refraction, their energy is assumed to be conserved as they cross the boundary. Due to this extra constraint it is more convenient to use the action defined in Maupertuis's action principle
\begin{equation}
    S( \vec{r} ) = \int_A^B \vec{p} \cdot d\vec{r} ,
\end{equation}
where $\vec{p}=\partial L/\partial\dot{\vec{r}}$ is the usual canonical momentum. Since our potential $V(\vec{r})$ depends only on the spatial coordinate (and perhaps on the spin of the particle), in the optical potential limit, $\vec{p}$ is constant in each region and parallel to the displacement $d\vec{r}$, resulting in the simplification
\begin{equation}
    S ( \vec{r}) = \int_A^B p \, ds ,
\end{equation}
where $p$ and $ds$ are the magnitudes of the canonical momentum and displacement, respectively. Since  Maupertuis's action principle uses the constant energy constraint to eliminate  $\dot{\vec{r}}$, one can express the magnitude of the canonical momentum $p$ in terms of the particle's energy as
\begin{equation}
    p = m v \gamma=\frac{\sqrt{(E - V(\vec{r}))^{2}-m^{2} c^{4}}}{c} .
\end{equation}

For our experimental setup in Fig.~\ref{fig:snell law}, the total classical action for the neutron $|\sigma_x \rangle$-state moving from $A$ to $B$ is
\begin{equation}
    S_{{\sf tot},\sigma} = S_{AC,\sigma} + S_{BC,\sigma} ,
\end{equation}
where $S_{AC,\sigma}$ and $S_{BC,\sigma}$ are the classical actions in the two field regions 
\begin{align}
    S_{AC,\sigma}=& \frac{\sqrt{(E - V_{i\sigma})^{2}-m^{2} c_{i}^{4}}}{c_{i}} \ |\vec{r}_C-\vec{r}_{A}| \\
    S_{CB,\sigma}=& \frac{\sqrt{(E - V_{o\sigma})^{2}-m^{2} c_{o}^{4}}}{c_{o}} \ |\vec{r}_B-\vec{r}_{C} | ,
\end{align}
with coordinates $\vec{r}_{A,B,C}$ corresponding to the points indicated in Fig. ~\ref{fig:snell law}.
Minimizing the variations of $S_{{\sf tot},\sigma}$ with respect to $z_C$, where the neutron crosses the boundary (whose normal is the $y$ axis in Fig.~\ref{fig:snell law}), leads to
\begin{align}
\nonumber
    \frac{\partial S_{{\sf tot},\sigma}}{\partial z_{C}}&= \frac{\sqrt{(E - V_{i\sigma})^{2}-m^{2} c_{i}^{4}}}{c_{i}} \ \frac{z_{C}-z_{A}}{ |\vec{r}_{C}-\vec{r}_{A}|} \\
    & - \frac{\sqrt{ (E - V_{o\sigma} )^{2}-m^{2} c_{o}^{4}}}{c_{o}} \  \frac{z_{B}-z_{C}}{ |\vec{r}_{B}-\vec{r}_{C} |}= 0.
\end{align}
Using trigonometric relationships between $\theta_i$, $\theta_o$, $v_{i\sigma}$ and $v_{o\sigma}$, it is straightforward to obtain Snell's law for relativistic neutrons
\begin{equation}
    \label{snell law neutron relativistic}
 \frac{p_{i}}{p_{o}}=   \frac{c_o}{c_{i}} \frac{\sqrt{(E - V_{i\sigma})^{2}-m^{2} c_{i}^{4}}}{\sqrt{(E - V_{o\sigma})^{2}-m^{2} c_{o}^{4}}} = 
      \frac{\sin\theta_o}{\sin\theta_i} ,
\end{equation}
%
where $p_a=m v_{a\sigma}/\sqrt{1-\frac{v_{a\sigma}^2}{c_a^2}}$ with $a = i,o$. 
Moreover, since most of the current experiments with neutrons are carried out in the 
non-relativistic limit, the momentum simplifies to $p_a \approx m v_{a\sigma}$, where 
\begin{align}
    v_{o\sigma} =  \sqrt{ v_{i\sigma}^2 + \frac{2 \mu_\sigma (B_o + B_i)}{m}} ,
\end{align}
due to conservation of energy. Then, for massive non-relativistic particles we obtain the familiar 
\begin{align}
    \label{snell law non-relativistic massive}
    v_{i\sigma} \sin\theta_i = v_{o\sigma} \sin\theta_o .
\end{align}

The case of light, that is, relativistic massless particles called photons, is subject 
to the same variational 
principle with the same end result, Eq. \eqref{snell law neutron relativistic}. However, for photons 
the dispersion relation is given by $p_a=\hbar \omega_a/c_a$, where $\hbar\omega_a=E_a$ is its energy. 
From conservation of energy 
\begin{align}
    \label{snell law relativistic massless}
    c_{o} \sin\theta_i = c_{i} \sin\theta_o ,
\end{align}
and one obtains Snell's law for light refraction, Eq.~(\ref{snell law light}). 


%
%
%
%

%
%

\section{Higher-Order Correction Coefficients} \label{higher order}

We now list the second-order corrections due to the deflection angles $\alpha_1$ and $\alpha_2$ and the divergence angle $\varphi$. The $A_1$, $A_{12}$, $A_2$ and $A_\varphi$ coefficients in Eq.~\eqref{eq: higher order form} for the parallelogram geometry are given by the following equations:
\begin{align}
    A_1^{\text{P,Div}} =&  \frac{B_1}{2} \left( 5y + 4z - 3a\right) - 2B_2 \left( a + \delta_{12} -z \right)  \nonumber\\
    \nonumber
    A_{12}^{\text{P,Div}} =& 4 B_1 \left( 3z - 4y - 3a - 3\delta_{12} \right) \nonumber\\
    & \hspace*{2cm} + 2B_2\left( 5a + 6 \delta_{12} + 8y - 6z\right) \nonumber\\
    A_2^{\text{P,Div}} =&  B_1 \left( 2z - 3a - 2\delta_{12}\right)- \frac{B_2}{2} \left( 7a + 4\delta_{12} -5y-4z \right)\nonumber\\
    A_\varphi^{\text{P,Div}} =& 2B_1 \left( 3y-2z \right) + 2B_2 \left( 2z - 2a - 2\delta_{12} - 3y \right)\nonumber \\
    A^{\text{P,Div}}_{1\varphi} =& 2 m v_0 \left(z-y\right) \nonumber \\ A^{\text{P,Div}}_{2\varphi} =& 2 m v_0 \left(a + \delta_{12} + y-z\right).  
\end{align}

For the triangular geometry, the second order coefficients are given by the following equations:
\begin{align}
    A_1^{\text{T,Div}} &=  \frac{B_1}{2} \left( 5y + 4z - 3a\right) + 2B_2 \left( 5a + 5\delta_{12} -6y -5z \right) \nonumber\\
    A_{12}^{\text{T,Div}} &= 4B_1 \left( a + \delta_{12} - 2y - z\right) +2B_2 \left( 8y +6z - 5a + 6 \delta_{12} \right)  \nonumber\\
    A_2^{\text{T,Div}} &=  B_1 \left( 3a + 2 \delta_{12} - 2z\right)- \frac{B_2}{2} \left( 4z -5y - 7a - 4\delta_{12} \right)\nonumber\\
    A_\varphi^{\text{T,Div}} &= 2B_1 \left( 3y-2z \right) + 2B_2 \left( 2a  + 2\delta_{12} - 3y - 2z \right) \nonumber \\
    A^{\text{T,Div}}_{1\varphi} &= 2 m v_0 \left(z-y\right) \nonumber\\
    A^{\text{T,Div}}_{2\varphi} &= 2 m v_0 \left(a + \delta_{12} - y - z\right).   
\end{align}

\FloatBarrier
\bibliography{sources.bib}

\begin{thebibliography}{83}%
\makeatletter
\providecommand \@ifxundefined [1]{%
 \@ifx{#1\undefined}
}%
\providecommand \@ifnum [1]{%
 \ifnum #1\expandafter \@firstoftwo
 \else \expandafter \@secondoftwo
 \fi
}%
\providecommand \@ifx [1]{%
 \ifx #1\expandafter \@firstoftwo
 \else \expandafter \@secondoftwo
 \fi
}%
\providecommand \natexlab [1]{#1}%
\providecommand \enquote  [1]{``#1''}%
\providecommand \bibnamefont  [1]{#1}%
\providecommand \bibfnamefont [1]{#1}%
\providecommand \citenamefont [1]{#1}%
\providecommand \href@noop [0]{\@secondoftwo}%
\providecommand \href [0]{\begingroup \@sanitize@url \@href}%
\providecommand \@href[1]{\@@startlink{#1}\@@href}%
\providecommand \@@href[1]{\endgroup#1\@@endlink}%
\providecommand \@sanitize@url [0]{\catcode `\\12\catcode `\$12\catcode
  `\&12\catcode `\#12\catcode `\^12\catcode `\_12\catcode `\%12\relax}%
\providecommand \@@startlink[1]{}%
\providecommand \@@endlink[0]{}%
\providecommand \url  [0]{\begingroup\@sanitize@url \@url }%
\providecommand \@url [1]{\endgroup\@href {#1}{\urlprefix }}%
\providecommand \urlprefix  [0]{URL }%
\providecommand \Eprint [0]{\href }%
\providecommand \doibase [0]{https://doi.org/}%
\providecommand \selectlanguage [0]{\@gobble}%
\providecommand \bibinfo  [0]{\@secondoftwo}%
\providecommand \bibfield  [0]{\@secondoftwo}%
\providecommand \translation [1]{[#1]}%
\providecommand \BibitemOpen [0]{}%
\providecommand \bibitemStop [0]{}%
\providecommand \bibitemNoStop [0]{.\EOS\space}%
\providecommand \EOS [0]{\spacefactor3000\relax}%
\providecommand \BibitemShut  [1]{\csname bibitem#1\endcsname}%
\let\auto@bib@innerbib\@empty
\bibitem [{\citenamefont {Tao}\ and\ \citenamefont
  {Tsymbal}(2018)}]{spin_texture_matter_Tao}%
  \BibitemOpen
  \bibfield  {author} {\bibinfo {author} {\bibfnamefont {L.~L.}\ \bibnamefont
  {Tao}}\ and\ \bibinfo {author} {\bibfnamefont {E.~Y.}\ \bibnamefont
  {Tsymbal}},\ }\href@noop {} {\bibfield  {journal} {\bibinfo  {journal} {Nat.
  Commun.}\ }\textbf {\bibinfo {volume} {9}} (\bibinfo {year}
  {2018})}\BibitemShut {NoStop}%
\bibitem [{\citenamefont {Yin}\ \emph {et~al.}(2021)\citenamefont {Yin},
  \citenamefont {Pan},\ and\ \citenamefont {Zahid~Hasan}}]{Yin_2021}%
  \BibitemOpen
  \bibfield  {author} {\bibinfo {author} {\bibfnamefont {J.-X.}\ \bibnamefont
  {Yin}}, \bibinfo {author} {\bibfnamefont {S.~H.}\ \bibnamefont {Pan}},\ and\
  \bibinfo {author} {\bibfnamefont {M.}~\bibnamefont {Zahid~Hasan}},\ }\href
  {https://doi.org/10.1038/s42254-021-00293-7} {\bibfield  {journal} {\bibinfo
  {journal} {Nat. Rev. Phys.}\ }\textbf {\bibinfo {volume} {3}},\ \bibinfo
  {pages} {249} (\bibinfo {year} {2021})}\BibitemShut {NoStop}%
\bibitem [{\citenamefont {Allen}\ \emph
  {et~al.}(1992{\natexlab{a}})\citenamefont {Allen}, \citenamefont
  {Beijersbergen}, \citenamefont {Spreeuw},\ and\ \citenamefont
  {Woerdman}}]{allen_orbital_1992}%
  \BibitemOpen
  \bibfield  {author} {\bibinfo {author} {\bibfnamefont {L.}~\bibnamefont
  {Allen}}, \bibinfo {author} {\bibfnamefont {M.~W.}\ \bibnamefont
  {Beijersbergen}}, \bibinfo {author} {\bibfnamefont {R.~J.~C.}\ \bibnamefont
  {Spreeuw}},\ and\ \bibinfo {author} {\bibfnamefont {J.~P.}\ \bibnamefont
  {Woerdman}},\ }\href {https://doi.org/10.1103/PhysRevA.45.8185} {\bibfield
  {journal} {\bibinfo  {journal} {Phys. Rev. A}\ }\textbf {\bibinfo {volume}
  {45}},\ \bibinfo {pages} {8185} (\bibinfo {year}
  {1992}{\natexlab{a}})}\BibitemShut {NoStop}%
\bibitem [{\citenamefont {Sarenac}\ \emph
  {et~al.}(2018{\natexlab{a}})\citenamefont {Sarenac}, \citenamefont {Cory},
  \citenamefont {Nsofini}, \citenamefont {Hincks}, \citenamefont {Miguel},
  \citenamefont {Arif}, \citenamefont {Clark}, \citenamefont {Huber},\ and\
  \citenamefont {Pushin}}]{Sarenac2018}%
  \BibitemOpen
  \bibfield  {author} {\bibinfo {author} {\bibfnamefont {D.}~\bibnamefont
  {Sarenac}}, \bibinfo {author} {\bibfnamefont {D.~G.}\ \bibnamefont {Cory}},
  \bibinfo {author} {\bibfnamefont {J.}~\bibnamefont {Nsofini}}, \bibinfo
  {author} {\bibfnamefont {I.}~\bibnamefont {Hincks}}, \bibinfo {author}
  {\bibfnamefont {P.}~\bibnamefont {Miguel}}, \bibinfo {author} {\bibfnamefont
  {M.}~\bibnamefont {Arif}}, \bibinfo {author} {\bibfnamefont {C.~W.}\
  \bibnamefont {Clark}}, \bibinfo {author} {\bibfnamefont {M.~G.}\ \bibnamefont
  {Huber}},\ and\ \bibinfo {author} {\bibfnamefont {D.~A.}\ \bibnamefont
  {Pushin}},\ }\href {https://doi.org/10.1103/PhysRevLett.121.183602}
  {\bibfield  {journal} {\bibinfo  {journal} {Phys. Rev. Lett.}\ }\textbf
  {\bibinfo {volume} {121}},\ \bibinfo {pages} {183602} (\bibinfo {year}
  {2018}{\natexlab{a}})}\BibitemShut {NoStop}%
\bibitem [{\citenamefont {Woods}\ \emph {et~al.}(2021)\citenamefont {Woods},
  \citenamefont {Chen}, \citenamefont {Chopdekar}, \citenamefont {Farmer},
  \citenamefont {Mazzoli}, \citenamefont {Koch}, \citenamefont {Tremsin},
  \citenamefont {Hu}, \citenamefont {Scholl}, \citenamefont {Kevan},
  \citenamefont {Wilkins}, \citenamefont {Kwok}, \citenamefont {De~Long},
  \citenamefont {Roy},\ and\ \citenamefont {Hastings}}]{Woods_2021}%
  \BibitemOpen
  \bibfield  {author} {\bibinfo {author} {\bibfnamefont {J.~S.}\ \bibnamefont
  {Woods}}, \bibinfo {author} {\bibfnamefont {X.~M.}\ \bibnamefont {Chen}},
  \bibinfo {author} {\bibfnamefont {R.~V.}\ \bibnamefont {Chopdekar}}, \bibinfo
  {author} {\bibfnamefont {B.}~\bibnamefont {Farmer}}, \bibinfo {author}
  {\bibfnamefont {C.}~\bibnamefont {Mazzoli}}, \bibinfo {author} {\bibfnamefont
  {R.}~\bibnamefont {Koch}}, \bibinfo {author} {\bibfnamefont {A.~S.}\
  \bibnamefont {Tremsin}}, \bibinfo {author} {\bibfnamefont {W.}~\bibnamefont
  {Hu}}, \bibinfo {author} {\bibfnamefont {A.}~\bibnamefont {Scholl}}, \bibinfo
  {author} {\bibfnamefont {S.}~\bibnamefont {Kevan}}, \bibinfo {author}
  {\bibfnamefont {S.}~\bibnamefont {Wilkins}}, \bibinfo {author} {\bibfnamefont
  {W.-K.}\ \bibnamefont {Kwok}}, \bibinfo {author} {\bibfnamefont {L.~E.}\
  \bibnamefont {De~Long}}, \bibinfo {author} {\bibfnamefont {S.}~\bibnamefont
  {Roy}},\ and\ \bibinfo {author} {\bibfnamefont {J.~T.}\ \bibnamefont
  {Hastings}},\ }\href {https://doi.org/10.1103/PhysRevLett.126.117201}
  {\bibfield  {journal} {\bibinfo  {journal} {Phys. Rev. Lett.}\ }\textbf
  {\bibinfo {volume} {126}},\ \bibinfo {pages} {117201} (\bibinfo {year}
  {2021})}\BibitemShut {NoStop}%
\bibitem [{\citenamefont {Kurzynowski}\ \emph {et~al.}(2010)\citenamefont
  {Kurzynowski}, \citenamefont {Woźniak},\ and\ \citenamefont
  {Borwińska}}]{Kurzynowski_2010}%
  \BibitemOpen
  \bibfield  {author} {\bibinfo {author} {\bibfnamefont {P.}~\bibnamefont
  {Kurzynowski}}, \bibinfo {author} {\bibfnamefont {W.~A.}\ \bibnamefont
  {Woźniak}},\ and\ \bibinfo {author} {\bibfnamefont {M.}~\bibnamefont
  {Borwińska}},\ }\href {https://doi.org/10.1088/2040-8978/12/3/035406}
  {\bibfield  {journal} {\bibinfo  {journal} {J. Opt.}\ }\textbf {\bibinfo
  {volume} {12}},\ \bibinfo {pages} {035406} (\bibinfo {year}
  {2010})}\BibitemShut {NoStop}%
\bibitem [{\citenamefont {Kurzynowski}\ \emph {et~al.}(2006)\citenamefont
  {Kurzynowski}, \citenamefont {Woźniak},\ and\ \citenamefont
  {Frączek}}]{Kurzynowski_2006}%
  \BibitemOpen
  \bibfield  {author} {\bibinfo {author} {\bibfnamefont {P.}~\bibnamefont
  {Kurzynowski}}, \bibinfo {author} {\bibfnamefont {W.~A.}\ \bibnamefont
  {Woźniak}},\ and\ \bibinfo {author} {\bibfnamefont {E.}~\bibnamefont
  {Frączek}},\ }\href {https://doi.org/10.1364/AO.45.007898} {\bibfield
  {journal} {\bibinfo  {journal} {Appl. Opt.}\ }\textbf {\bibinfo {volume}
  {45}},\ \bibinfo {pages} {7898} (\bibinfo {year} {2006})}\BibitemShut
  {NoStop}%
\bibitem [{\citenamefont {Uchida}\ and\ \citenamefont
  {Tonomura}(2010)}]{uchida_generation_2010}%
  \BibitemOpen
  \bibfield  {author} {\bibinfo {author} {\bibfnamefont {M.}~\bibnamefont
  {Uchida}}\ and\ \bibinfo {author} {\bibfnamefont {A.}~\bibnamefont
  {Tonomura}},\ }\href {https://doi.org/10.1038/nature08904} {\bibfield
  {journal} {\bibinfo  {journal} {Nature}\ }\textbf {\bibinfo {volume} {464}},\
  \bibinfo {pages} {737} (\bibinfo {year} {2010})}\BibitemShut {NoStop}%
\bibitem [{\citenamefont {Verbeeck}\ \emph {et~al.}(2010)\citenamefont
  {Verbeeck}, \citenamefont {Tian},\ and\ \citenamefont
  {Schattschneider}}]{verbeeck_production_2010}%
  \BibitemOpen
  \bibfield  {author} {\bibinfo {author} {\bibfnamefont {J.}~\bibnamefont
  {Verbeeck}}, \bibinfo {author} {\bibfnamefont {H.}~\bibnamefont {Tian}},\
  and\ \bibinfo {author} {\bibfnamefont {P.}~\bibnamefont {Schattschneider}},\
  }\href {https://doi.org/10.1038/nature09366} {\bibfield  {journal} {\bibinfo
  {journal} {Nature}\ }\textbf {\bibinfo {volume} {467}},\ \bibinfo {pages}
  {301} (\bibinfo {year} {2010})}\BibitemShut {NoStop}%
\bibitem [{\citenamefont {Karimi}\ \emph {et~al.}(2014)\citenamefont {Karimi},
  \citenamefont {Grillo}, \citenamefont {Boyd},\ and\ \citenamefont
  {Santamato}}]{eKarimi2013}%
  \BibitemOpen
  \bibfield  {author} {\bibinfo {author} {\bibfnamefont {E.}~\bibnamefont
  {Karimi}}, \bibinfo {author} {\bibfnamefont {V.}~\bibnamefont {Grillo}},
  \bibinfo {author} {\bibfnamefont {R.~W.}\ \bibnamefont {Boyd}},\ and\
  \bibinfo {author} {\bibfnamefont {E.}~\bibnamefont {Santamato}},\ }\href
  {https://doi.org/https://doi.org/10.1016/j.ultramic.2013.12.002} {\bibfield
  {journal} {\bibinfo  {journal} {Ultramicroscopy}\ }\textbf {\bibinfo {volume}
  {138}},\ \bibinfo {pages} {22} (\bibinfo {year} {2014})}\BibitemShut
  {NoStop}%
\bibitem [{\citenamefont {Lei}\ \emph {et~al.}(2021)\citenamefont {Lei},
  \citenamefont {Bu}, \citenamefont {Wang}, \citenamefont {Shen},\ and\
  \citenamefont {Ji}}]{lei_generation_2021}%
  \BibitemOpen
  \bibfield  {author} {\bibinfo {author} {\bibfnamefont {S.}~\bibnamefont
  {Lei}}, \bibinfo {author} {\bibfnamefont {Z.}~\bibnamefont {Bu}}, \bibinfo
  {author} {\bibfnamefont {W.}~\bibnamefont {Wang}}, \bibinfo {author}
  {\bibfnamefont {B.}~\bibnamefont {Shen}},\ and\ \bibinfo {author}
  {\bibfnamefont {L.}~\bibnamefont {Ji}},\ }\href
  {https://doi.org/10.1103/PhysRevD.104.076025} {\bibfield  {journal} {\bibinfo
   {journal} {Phys. Rev. D}\ }\textbf {\bibinfo {volume} {104}},\ \bibinfo
  {pages} {076025} (\bibinfo {year} {2021})}\BibitemShut {NoStop}%
\bibitem [{\citenamefont {Luski}\ \emph {et~al.}(2021)\citenamefont {Luski},
  \citenamefont {Segev}, \citenamefont {David}, \citenamefont {Bitton},
  \citenamefont {Nadler}, \citenamefont {Barnea}, \citenamefont {Gorlach},
  \citenamefont {Cheshnovsky}, \citenamefont {Kaminer},\ and\ \citenamefont
  {Narevicius}}]{Luski_2021}%
  \BibitemOpen
  \bibfield  {author} {\bibinfo {author} {\bibfnamefont {A.}~\bibnamefont
  {Luski}}, \bibinfo {author} {\bibfnamefont {Y.}~\bibnamefont {Segev}},
  \bibinfo {author} {\bibfnamefont {R.}~\bibnamefont {David}}, \bibinfo
  {author} {\bibfnamefont {O.}~\bibnamefont {Bitton}}, \bibinfo {author}
  {\bibfnamefont {H.}~\bibnamefont {Nadler}}, \bibinfo {author} {\bibfnamefont
  {A.~R.}\ \bibnamefont {Barnea}}, \bibinfo {author} {\bibfnamefont
  {A.}~\bibnamefont {Gorlach}}, \bibinfo {author} {\bibfnamefont
  {O.}~\bibnamefont {Cheshnovsky}}, \bibinfo {author} {\bibfnamefont
  {I.}~\bibnamefont {Kaminer}},\ and\ \bibinfo {author} {\bibfnamefont
  {E.}~\bibnamefont {Narevicius}},\ }\href
  {https://doi.org/10.1126/science.abj2451} {\bibfield  {journal} {\bibinfo
  {journal} {Science}\ }\textbf {\bibinfo {volume} {373}},\ \bibinfo {pages}
  {1105} (\bibinfo {year} {2021})}\BibitemShut {NoStop}%
\bibitem [{\citenamefont {Allen}\ \emph {et~al.}(2020)\citenamefont {Allen},
  \citenamefont {Barnett},\ and\ \citenamefont {Padgett}}]{Allen2020}%
  \BibitemOpen
  \bibfield  {author} {\bibinfo {author} {\bibfnamefont {L.}~\bibnamefont
  {Allen}}, \bibinfo {author} {\bibfnamefont {S.~M.}\ \bibnamefont {Barnett}},\
  and\ \bibinfo {author} {\bibfnamefont {M.~J.}\ \bibnamefont {Padgett}},\
  }\href@noop {} {\emph {\bibinfo {title} {Optical Angular Momentum}}}\
  (\bibinfo  {publisher} {CRC Press},\ \bibinfo {address} {London, England},\
  \bibinfo {year} {2020})\BibitemShut {NoStop}%
\bibitem [{\citenamefont {Yao}\ and\ \citenamefont
  {Padgett}(2011)}]{Yao_Padgett_2011}%
  \BibitemOpen
  \bibfield  {author} {\bibinfo {author} {\bibfnamefont {A.~M.}\ \bibnamefont
  {Yao}}\ and\ \bibinfo {author} {\bibfnamefont {M.~J.}\ \bibnamefont
  {Padgett}},\ }\href {https://doi.org/10.1364/AOP.3.000161} {\bibfield
  {journal} {\bibinfo  {journal} {Adv. Opt. Photonics}\ }\textbf {\bibinfo
  {volume} {3}},\ \bibinfo {pages} {161} (\bibinfo {year} {2011})}\BibitemShut
  {NoStop}%
\bibitem [{\citenamefont {Rubinsztein-Dunlop}\ \emph
  {et~al.}(2016)\citenamefont {Rubinsztein-Dunlop}, \citenamefont {Forbes},
  \citenamefont {Berry}, \citenamefont {Dennis}, \citenamefont {Andrews},
  \citenamefont {Mansuripur}, \citenamefont {Denz}, \citenamefont {Alpmann},
  \citenamefont {Banzer}, \citenamefont {Bauer}, \citenamefont {Karimi},
  \citenamefont {Marrucci}, \citenamefont {Padgett}, \citenamefont
  {Ritsch-Marte}, \citenamefont {Litchinitser}, \citenamefont {Bigelow},
  \citenamefont {Rosales-Guzm{\'{a}}n}, \citenamefont {Belmonte}, \citenamefont
  {Torres}, \citenamefont {Neely}, \citenamefont {Baker}, \citenamefont
  {Gordon}, \citenamefont {Stilgoe}, \citenamefont {Romero}, \citenamefont
  {White}, \citenamefont {Fickler}, \citenamefont {Willner}, \citenamefont
  {Xie}, \citenamefont {McMorran},\ and\ \citenamefont
  {Weiner}}]{Rubinsztein-Dunlop_2016}%
  \BibitemOpen
  \bibfield  {author} {\bibinfo {author} {\bibfnamefont {H.}~\bibnamefont
  {Rubinsztein-Dunlop}}, \bibinfo {author} {\bibfnamefont {A.}~\bibnamefont
  {Forbes}}, \bibinfo {author} {\bibfnamefont {M.~V.}\ \bibnamefont {Berry}},
  \bibinfo {author} {\bibfnamefont {M.~R.}\ \bibnamefont {Dennis}}, \bibinfo
  {author} {\bibfnamefont {D.~L.}\ \bibnamefont {Andrews}}, \bibinfo {author}
  {\bibfnamefont {M.}~\bibnamefont {Mansuripur}}, \bibinfo {author}
  {\bibfnamefont {C.}~\bibnamefont {Denz}}, \bibinfo {author} {\bibfnamefont
  {C.}~\bibnamefont {Alpmann}}, \bibinfo {author} {\bibfnamefont
  {P.}~\bibnamefont {Banzer}}, \bibinfo {author} {\bibfnamefont
  {T.}~\bibnamefont {Bauer}}, \bibinfo {author} {\bibfnamefont
  {E.}~\bibnamefont {Karimi}}, \bibinfo {author} {\bibfnamefont
  {L.}~\bibnamefont {Marrucci}}, \bibinfo {author} {\bibfnamefont
  {M.}~\bibnamefont {Padgett}}, \bibinfo {author} {\bibfnamefont
  {M.}~\bibnamefont {Ritsch-Marte}}, \bibinfo {author} {\bibfnamefont {N.~M.}\
  \bibnamefont {Litchinitser}}, \bibinfo {author} {\bibfnamefont {N.~P.}\
  \bibnamefont {Bigelow}}, \bibinfo {author} {\bibfnamefont {C.}~\bibnamefont
  {Rosales-Guzm{\'{a}}n}}, \bibinfo {author} {\bibfnamefont {A.}~\bibnamefont
  {Belmonte}}, \bibinfo {author} {\bibfnamefont {J.~P.}\ \bibnamefont
  {Torres}}, \bibinfo {author} {\bibfnamefont {T.~W.}\ \bibnamefont {Neely}},
  \bibinfo {author} {\bibfnamefont {M.}~\bibnamefont {Baker}}, \bibinfo
  {author} {\bibfnamefont {R.}~\bibnamefont {Gordon}}, \bibinfo {author}
  {\bibfnamefont {A.~B.}\ \bibnamefont {Stilgoe}}, \bibinfo {author}
  {\bibfnamefont {J.}~\bibnamefont {Romero}}, \bibinfo {author} {\bibfnamefont
  {A.~G.}\ \bibnamefont {White}}, \bibinfo {author} {\bibfnamefont
  {R.}~\bibnamefont {Fickler}}, \bibinfo {author} {\bibfnamefont {A.~E.}\
  \bibnamefont {Willner}}, \bibinfo {author} {\bibfnamefont {G.}~\bibnamefont
  {Xie}}, \bibinfo {author} {\bibfnamefont {B.}~\bibnamefont {McMorran}},\ and\
  \bibinfo {author} {\bibfnamefont {A.~M.}\ \bibnamefont {Weiner}},\
  }\href@noop {} {\bibfield  {journal} {\bibinfo  {journal} {J. Opt.}\ }\textbf
  {\bibinfo {volume} {19}},\ \bibinfo {pages} {013001} (\bibinfo {year}
  {2016})}\BibitemShut {NoStop}%
\bibitem [{\citenamefont {Shen}\ \emph {et~al.}(2019)\citenamefont {Shen},
  \citenamefont {Wang}, \citenamefont {Xie}, \citenamefont {Min}, \citenamefont
  {Fu}, \citenamefont {Liu}, \citenamefont {Gong},\ and\ \citenamefont
  {Yuan}}]{Shen_2019}%
  \BibitemOpen
  \bibfield  {author} {\bibinfo {author} {\bibfnamefont {Y.}~\bibnamefont
  {Shen}}, \bibinfo {author} {\bibfnamefont {X.}~\bibnamefont {Wang}}, \bibinfo
  {author} {\bibfnamefont {Z.}~\bibnamefont {Xie}}, \bibinfo {author}
  {\bibfnamefont {C.}~\bibnamefont {Min}}, \bibinfo {author} {\bibfnamefont
  {X.}~\bibnamefont {Fu}}, \bibinfo {author} {\bibfnamefont {Q.}~\bibnamefont
  {Liu}}, \bibinfo {author} {\bibfnamefont {M.}~\bibnamefont {Gong}},\ and\
  \bibinfo {author} {\bibfnamefont {X.}~\bibnamefont {Yuan}},\ }\href
  {https://doi.org/10.1038/s41377-019-0194-2} {\bibfield  {journal} {\bibinfo
  {journal} {Light Sci. Appl.}\ }\textbf {\bibinfo {volume} {8}},\ \bibinfo
  {pages} {90} (\bibinfo {year} {2019})}\BibitemShut {NoStop}%
\bibitem [{\citenamefont {Clark}\ \emph {et~al.}(2015)\citenamefont {Clark},
  \citenamefont {Barankov}, \citenamefont {Huber}, \citenamefont {Arif},
  \citenamefont {Cory},\ and\ \citenamefont {Pushin}}]{clark_controlling_2015}%
  \BibitemOpen
  \bibfield  {author} {\bibinfo {author} {\bibfnamefont {C.~W.}\ \bibnamefont
  {Clark}}, \bibinfo {author} {\bibfnamefont {R.}~\bibnamefont {Barankov}},
  \bibinfo {author} {\bibfnamefont {M.~G.}\ \bibnamefont {Huber}}, \bibinfo
  {author} {\bibfnamefont {M.}~\bibnamefont {Arif}}, \bibinfo {author}
  {\bibfnamefont {D.~G.}\ \bibnamefont {Cory}},\ and\ \bibinfo {author}
  {\bibfnamefont {D.~A.}\ \bibnamefont {Pushin}},\ }\href
  {https://doi.org/10.1038/nature15265} {\bibfield  {journal} {\bibinfo
  {journal} {Nature}\ }\textbf {\bibinfo {volume} {525}},\ \bibinfo {pages}
  {504} (\bibinfo {year} {2015})}\BibitemShut {NoStop}%
\bibitem [{\citenamefont {Nsofini}\ \emph {et~al.}(2016)\citenamefont
  {Nsofini}, \citenamefont {Sarenac}, \citenamefont {Wood}, \citenamefont
  {Cory}, \citenamefont {Arif}, \citenamefont {Clark}, \citenamefont {Huber},\
  and\ \citenamefont {Pushin}}]{Nsofini_2016}%
  \BibitemOpen
  \bibfield  {author} {\bibinfo {author} {\bibfnamefont {J.}~\bibnamefont
  {Nsofini}}, \bibinfo {author} {\bibfnamefont {D.}~\bibnamefont {Sarenac}},
  \bibinfo {author} {\bibfnamefont {C.~J.}\ \bibnamefont {Wood}}, \bibinfo
  {author} {\bibfnamefont {D.~G.}\ \bibnamefont {Cory}}, \bibinfo {author}
  {\bibfnamefont {M.}~\bibnamefont {Arif}}, \bibinfo {author} {\bibfnamefont
  {C.~W.}\ \bibnamefont {Clark}}, \bibinfo {author} {\bibfnamefont {M.~G.}\
  \bibnamefont {Huber}},\ and\ \bibinfo {author} {\bibfnamefont {D.~A.}\
  \bibnamefont {Pushin}},\ }\href {https://doi.org/10.1103/PhysRevA.94.013605}
  {\bibfield  {journal} {\bibinfo  {journal} {Phys. Rev. A}\ }\textbf {\bibinfo
  {volume} {94}},\ \bibinfo {pages} {013605} (\bibinfo {year}
  {2016})}\BibitemShut {NoStop}%
\bibitem [{\citenamefont {Sarenac}\ \emph
  {et~al.}(2018{\natexlab{b}})\citenamefont {Sarenac}, \citenamefont {Nsofini},
  \citenamefont {Hincks}, \citenamefont {Arif}, \citenamefont {Clark},
  \citenamefont {Cory}, \citenamefont {Huber},\ and\ \citenamefont
  {Pushin}}]{SarenacTheory_2018}%
  \BibitemOpen
  \bibfield  {author} {\bibinfo {author} {\bibfnamefont {D.}~\bibnamefont
  {Sarenac}}, \bibinfo {author} {\bibfnamefont {J.}~\bibnamefont {Nsofini}},
  \bibinfo {author} {\bibfnamefont {I.}~\bibnamefont {Hincks}}, \bibinfo
  {author} {\bibfnamefont {M.}~\bibnamefont {Arif}}, \bibinfo {author}
  {\bibfnamefont {C.~W.}\ \bibnamefont {Clark}}, \bibinfo {author}
  {\bibfnamefont {D.~G.}\ \bibnamefont {Cory}}, \bibinfo {author}
  {\bibfnamefont {M.~G.}\ \bibnamefont {Huber}},\ and\ \bibinfo {author}
  {\bibfnamefont {D.~A.}\ \bibnamefont {Pushin}},\ }\href
  {https://doi.org/10.1088/1367-2630/aae3ac} {\bibfield  {journal} {\bibinfo
  {journal} {New J. Phys.}\ }\textbf {\bibinfo {volume} {20}},\ \bibinfo
  {pages} {103012} (\bibinfo {year} {2018}{\natexlab{b}})}\BibitemShut
  {NoStop}%
\bibitem [{\citenamefont {Sarenac}\ \emph {et~al.}(2019)\citenamefont
  {Sarenac}, \citenamefont {Kapahi}, \citenamefont {Chen}, \citenamefont
  {Clark}, \citenamefont {Cory}, \citenamefont {Huber}, \citenamefont
  {Taminiau}, \citenamefont {Zhernenkov},\ and\ \citenamefont
  {Pushin}}]{Sarenac2019}%
  \BibitemOpen
  \bibfield  {author} {\bibinfo {author} {\bibfnamefont {D.}~\bibnamefont
  {Sarenac}}, \bibinfo {author} {\bibfnamefont {C.}~\bibnamefont {Kapahi}},
  \bibinfo {author} {\bibfnamefont {W.}~\bibnamefont {Chen}}, \bibinfo {author}
  {\bibfnamefont {C.~W.}\ \bibnamefont {Clark}}, \bibinfo {author}
  {\bibfnamefont {D.~G.}\ \bibnamefont {Cory}}, \bibinfo {author}
  {\bibfnamefont {M.~G.}\ \bibnamefont {Huber}}, \bibinfo {author}
  {\bibfnamefont {I.}~\bibnamefont {Taminiau}}, \bibinfo {author}
  {\bibfnamefont {K.}~\bibnamefont {Zhernenkov}},\ and\ \bibinfo {author}
  {\bibfnamefont {D.~A.}\ \bibnamefont {Pushin}},\ }\href
  {https://doi.org/10.1073/pnas.1906861116} {\bibfield  {journal} {\bibinfo
  {journal} {Proc. Natl. Acad. Sci. U.S.A}\ }\textbf {\bibinfo {volume}
  {116}},\ \bibinfo {pages} {20328} (\bibinfo {year} {2019})}\BibitemShut
  {NoStop}%
\bibitem [{\citenamefont {Sarenac}\ \emph {et~al.}(2022)\citenamefont
  {Sarenac}, \citenamefont {Henderson}, \citenamefont {Ekinci}, \citenamefont
  {Clark}, \citenamefont {Cory}, \citenamefont {Debeer-Schmitt}, \citenamefont
  {Huber}, \citenamefont {Kapahi},\ and\ \citenamefont
  {Pushin}}]{Sarenac_2022}%
  \BibitemOpen
  \bibfield  {author} {\bibinfo {author} {\bibfnamefont {D.}~\bibnamefont
  {Sarenac}}, \bibinfo {author} {\bibfnamefont {M.~E.}\ \bibnamefont
  {Henderson}}, \bibinfo {author} {\bibfnamefont {H.}~\bibnamefont {Ekinci}},
  \bibinfo {author} {\bibfnamefont {C.~W.}\ \bibnamefont {Clark}}, \bibinfo
  {author} {\bibfnamefont {D.~G.}\ \bibnamefont {Cory}}, \bibinfo {author}
  {\bibfnamefont {L.}~\bibnamefont {Debeer-Schmitt}}, \bibinfo {author}
  {\bibfnamefont {M.~G.}\ \bibnamefont {Huber}}, \bibinfo {author}
  {\bibfnamefont {C.}~\bibnamefont {Kapahi}},\ and\ \bibinfo {author}
  {\bibfnamefont {D.~A.}\ \bibnamefont {Pushin}},\ }\href@noop {} {\bibinfo
  {title} {Experimental realization of neutron helical waves}} (\bibinfo {year}
  {2022}),\ \Eprint {https://arxiv.org/abs/arXiv:2205.06263} {arXiv:2205.06263}
  \BibitemShut {NoStop}%
\bibitem [{\citenamefont {Jach}\ and\ \citenamefont
  {Vinson}(2022)}]{Jach_2022}%
  \BibitemOpen
  \bibfield  {author} {\bibinfo {author} {\bibfnamefont {T.}~\bibnamefont
  {Jach}}\ and\ \bibinfo {author} {\bibfnamefont {J.}~\bibnamefont {Vinson}},\
  }\href {https://doi.org/10.1103/PhysRevC.105.L061601} {\bibfield  {journal}
  {\bibinfo  {journal} {Phys. Rev. C}\ }\textbf {\bibinfo {volume} {105}},\
  \bibinfo {pages} {L061601} (\bibinfo {year} {2022})}\BibitemShut {NoStop}%
\bibitem [{\citenamefont {Geerits}\ \emph {et~al.}(2022)\citenamefont
  {Geerits}, \citenamefont {Lemmel}, \citenamefont {Berger},\ and\
  \citenamefont {Sponar}}]{Geerits_2022}%
  \BibitemOpen
  \bibfield  {author} {\bibinfo {author} {\bibfnamefont {N.}~\bibnamefont
  {Geerits}}, \bibinfo {author} {\bibfnamefont {H.}~\bibnamefont {Lemmel}},
  \bibinfo {author} {\bibfnamefont {A.-S.}\ \bibnamefont {Berger}},\ and\
  \bibinfo {author} {\bibfnamefont {S.}~\bibnamefont {Sponar}},\ }\href@noop {}
  {\bibinfo {title} {Phase vortex lattices in neutron interferometry}}
  (\bibinfo {year} {2022}),\ \Eprint {https://arxiv.org/abs/1911.07974}
  {arXiv:1911.07974} \BibitemShut {NoStop}%
\bibitem [{\citenamefont {Geerits}\ and\ \citenamefont
  {Sponar}(2021)}]{Geerits2021}%
  \BibitemOpen
  \bibfield  {author} {\bibinfo {author} {\bibfnamefont {N.}~\bibnamefont
  {Geerits}}\ and\ \bibinfo {author} {\bibfnamefont {S.}~\bibnamefont
  {Sponar}},\ }\href {https://doi.org/10.1103/PhysRevA.103.022205} {\bibfield
  {journal} {\bibinfo  {journal} {Phys. Rev. A}\ }\textbf {\bibinfo {volume}
  {103}},\ \bibinfo {pages} {022205} (\bibinfo {year} {2021})}\BibitemShut
  {NoStop}%
\bibitem [{\citenamefont {Munter}(2021)}]{munter}%
  \BibitemOpen
  \bibfield  {author} {\bibinfo {author} {\bibfnamefont {A.}~\bibnamefont
  {Munter}},\ }\href@noop {} {\bibinfo {title} {Neutron scattering lengths and
  cross sections}},\ \bibinfo {howpublished}
  {{https://www.ncnr.nist.gov/resources/n-lengths/}} (\bibinfo {year}
  {2021})\BibitemShut {NoStop}%
\bibitem [{\citenamefont {Cappelletti}\ and\ \citenamefont
  {Vinson}(2021)}]{Cappelletti_Vinson_2021}%
  \BibitemOpen
  \bibfield  {author} {\bibinfo {author} {\bibfnamefont {R.~L.}\ \bibnamefont
  {Cappelletti}}\ and\ \bibinfo {author} {\bibfnamefont {J.}~\bibnamefont
  {Vinson}},\ }\href {https://doi.org/10.1002/pssb.202000257} {\bibfield
  {journal} {\bibinfo  {journal} {Phys. Status Solidi B}\ }\textbf {\bibinfo
  {volume} {258}},\ \bibinfo {pages} {2000257} (\bibinfo {year}
  {2021})}\BibitemShut {NoStop}%
\bibitem [{\citenamefont {Cappelletti}\ \emph {et~al.}(2018)\citenamefont
  {Cappelletti}, \citenamefont {Jach},\ and\ \citenamefont
  {Vinson}}]{Cappelletti_Jach_Vinson_2018}%
  \BibitemOpen
  \bibfield  {author} {\bibinfo {author} {\bibfnamefont {R.~L.}\ \bibnamefont
  {Cappelletti}}, \bibinfo {author} {\bibfnamefont {T.}~\bibnamefont {Jach}},\
  and\ \bibinfo {author} {\bibfnamefont {J.}~\bibnamefont {Vinson}},\ }\href
  {https://doi.org/10.1103/PhysRevLett.120.090402} {\bibfield  {journal}
  {\bibinfo  {journal} {Phys. Rev. Lett.}\ }\textbf {\bibinfo {volume} {120}},\
  \bibinfo {pages} {090402} (\bibinfo {year} {2018})}\BibitemShut {NoStop}%
\bibitem [{\citenamefont {Li}\ \emph {et~al.}(2014)\citenamefont {Li},
  \citenamefont {Parnell}, \citenamefont {Hamilton}, \citenamefont
  {Maranville}, \citenamefont {Wang}, \citenamefont {Semerad}, \citenamefont
  {Baxter}, \citenamefont {Cremer},\ and\ \citenamefont {Pynn}}]{li2014}%
  \BibitemOpen
  \bibfield  {author} {\bibinfo {author} {\bibfnamefont {F.}~\bibnamefont
  {Li}}, \bibinfo {author} {\bibfnamefont {S.~R.}\ \bibnamefont {Parnell}},
  \bibinfo {author} {\bibfnamefont {W.~A.}\ \bibnamefont {Hamilton}}, \bibinfo
  {author} {\bibfnamefont {B.~B.}\ \bibnamefont {Maranville}}, \bibinfo
  {author} {\bibfnamefont {T.}~\bibnamefont {Wang}}, \bibinfo {author}
  {\bibfnamefont {R.}~\bibnamefont {Semerad}}, \bibinfo {author} {\bibfnamefont
  {D.~V.}\ \bibnamefont {Baxter}}, \bibinfo {author} {\bibfnamefont {J.~T.}\
  \bibnamefont {Cremer}},\ and\ \bibinfo {author} {\bibfnamefont
  {R.}~\bibnamefont {Pynn}},\ }\href@noop {} {\bibfield  {journal} {\bibinfo
  {journal} {Rev. Sci. Instr.}\ }\textbf {\bibinfo {volume} {85}},\ \bibinfo
  {pages} {053303} (\bibinfo {year} {2014})}\BibitemShut {NoStop}%
\bibitem [{\citenamefont {Shen}\ \emph {et~al.}(2020)\citenamefont {Shen},
  \citenamefont {Kuhn}, \citenamefont {Dalgliesh}, \citenamefont {de~Haan},
  \citenamefont {Geerits}, \citenamefont {Irfan}, \citenamefont {Li},
  \citenamefont {Lu}, \citenamefont {Parnell}, \citenamefont {Plomp},
  \citenamefont {van Well}, \citenamefont {Washington}, \citenamefont {Baxter},
  \citenamefont {Ortiz}, \citenamefont {Snow},\ and\ \citenamefont
  {Pynn}}]{shen2019}%
  \BibitemOpen
  \bibfield  {author} {\bibinfo {author} {\bibfnamefont {J.}~\bibnamefont
  {Shen}}, \bibinfo {author} {\bibfnamefont {S.~J.}\ \bibnamefont {Kuhn}},
  \bibinfo {author} {\bibfnamefont {R.~M.}\ \bibnamefont {Dalgliesh}}, \bibinfo
  {author} {\bibfnamefont {V.~O.}\ \bibnamefont {de~Haan}}, \bibinfo {author}
  {\bibfnamefont {N.}~\bibnamefont {Geerits}}, \bibinfo {author} {\bibfnamefont
  {A.~A.~M.}\ \bibnamefont {Irfan}}, \bibinfo {author} {\bibfnamefont
  {F.}~\bibnamefont {Li}}, \bibinfo {author} {\bibfnamefont {S.}~\bibnamefont
  {Lu}}, \bibinfo {author} {\bibfnamefont {S.~R.}\ \bibnamefont {Parnell}},
  \bibinfo {author} {\bibfnamefont {J.}~\bibnamefont {Plomp}}, \bibinfo
  {author} {\bibfnamefont {A.~A.}\ \bibnamefont {van Well}}, \bibinfo {author}
  {\bibfnamefont {A.}~\bibnamefont {Washington}}, \bibinfo {author}
  {\bibfnamefont {D.~V.}\ \bibnamefont {Baxter}}, \bibinfo {author}
  {\bibfnamefont {G.}~\bibnamefont {Ortiz}}, \bibinfo {author} {\bibfnamefont
  {W.~M.}\ \bibnamefont {Snow}},\ and\ \bibinfo {author} {\bibfnamefont
  {R.}~\bibnamefont {Pynn}},\ }\href
  {https://doi.org/https://doi.org/10.1038/s41467-020-14741-y} {\bibfield
  {journal} {\bibinfo  {journal} {Nat. Commun.}\ }\textbf {\bibinfo {volume}
  {11}},\ \bibinfo {pages} {930} (\bibinfo {year} {2020})}\BibitemShut
  {NoStop}%
\bibitem [{\citenamefont {Kuhn}\ \emph {et~al.}(2021)\citenamefont {Kuhn},
  \citenamefont {McKay}, \citenamefont {Shen}, \citenamefont {Geerits},
  \citenamefont {Dalgliesh}, \citenamefont {Dees}, \citenamefont {Irfan},
  \citenamefont {Li}, \citenamefont {Lu}, \citenamefont {Vangelista},
  \citenamefont {Baxter}, \citenamefont {Ortiz}, \citenamefont {Parnell},
  \citenamefont {Snow},\ and\ \citenamefont {Pynn}}]{kuhn2021}%
  \BibitemOpen
  \bibfield  {author} {\bibinfo {author} {\bibfnamefont {S.~J.}\ \bibnamefont
  {Kuhn}}, \bibinfo {author} {\bibfnamefont {S.}~\bibnamefont {McKay}},
  \bibinfo {author} {\bibfnamefont {J.}~\bibnamefont {Shen}}, \bibinfo {author}
  {\bibfnamefont {N.}~\bibnamefont {Geerits}}, \bibinfo {author} {\bibfnamefont
  {R.~M.}\ \bibnamefont {Dalgliesh}}, \bibinfo {author} {\bibfnamefont
  {E.}~\bibnamefont {Dees}}, \bibinfo {author} {\bibfnamefont {A.~A.~M.}\
  \bibnamefont {Irfan}}, \bibinfo {author} {\bibfnamefont {F.}~\bibnamefont
  {Li}}, \bibinfo {author} {\bibfnamefont {S.}~\bibnamefont {Lu}}, \bibinfo
  {author} {\bibfnamefont {V.}~\bibnamefont {Vangelista}}, \bibinfo {author}
  {\bibfnamefont {D.~V.}\ \bibnamefont {Baxter}}, \bibinfo {author}
  {\bibfnamefont {G.}~\bibnamefont {Ortiz}}, \bibinfo {author} {\bibfnamefont
  {S.~R.}\ \bibnamefont {Parnell}}, \bibinfo {author} {\bibfnamefont {W.~M.}\
  \bibnamefont {Snow}},\ and\ \bibinfo {author} {\bibfnamefont
  {R.}~\bibnamefont {Pynn}},\ }\href@noop {} {\bibfield  {journal} {\bibinfo
  {journal} {Phys. Rev. Research}\ }\textbf {\bibinfo {volume} {3}},\ \bibinfo
  {pages} {023227} (\bibinfo {year} {2021})}\BibitemShut {NoStop}%
\bibitem [{\citenamefont {Allen}\ \emph
  {et~al.}(1992{\natexlab{b}})\citenamefont {Allen}, \citenamefont
  {Beijersbergen}, \citenamefont {Spreeuw},\ and\ \citenamefont
  {Woerdman}}]{Allen1992}%
  \BibitemOpen
  \bibfield  {author} {\bibinfo {author} {\bibfnamefont {L.}~\bibnamefont
  {Allen}}, \bibinfo {author} {\bibfnamefont {M.~W.}\ \bibnamefont
  {Beijersbergen}}, \bibinfo {author} {\bibfnamefont {R.~J.~C.}\ \bibnamefont
  {Spreeuw}},\ and\ \bibinfo {author} {\bibfnamefont {J.~P.}\ \bibnamefont
  {Woerdman}},\ }\href {https://doi.org/10.1103/PhysRevA.45.8185} {\bibfield
  {journal} {\bibinfo  {journal} {Phys. Rev. A}\ }\textbf {\bibinfo {volume}
  {45}},\ \bibinfo {pages} {8185} (\bibinfo {year}
  {1992}{\natexlab{b}})}\BibitemShut {NoStop}%
\bibitem [{\citenamefont {Afanasev}\ \emph {et~al.}(2021)\citenamefont
  {Afanasev}, \citenamefont {Carlson},\ and\ \citenamefont
  {Mukherjee}}]{Afanasev2021}%
  \BibitemOpen
  \bibfield  {author} {\bibinfo {author} {\bibfnamefont {A.}~\bibnamefont
  {Afanasev}}, \bibinfo {author} {\bibfnamefont {C.~E.}\ \bibnamefont
  {Carlson}},\ and\ \bibinfo {author} {\bibfnamefont {A.}~\bibnamefont
  {Mukherjee}},\ }\href {https://doi.org/10.1103/PhysRevResearch.3.023097}
  {\bibfield  {journal} {\bibinfo  {journal} {Phys. Rev. Research}\ }\textbf
  {\bibinfo {volume} {3}},\ \bibinfo {pages} {023097} (\bibinfo {year}
  {2021})}\BibitemShut {NoStop}%
\bibitem [{\citenamefont {Barnett}\ \emph {et~al.}(2022)\citenamefont
  {Barnett}, \citenamefont {Speirits},\ and\ \citenamefont
  {Babiker}}]{Barnett_2022}%
  \BibitemOpen
  \bibfield  {author} {\bibinfo {author} {\bibfnamefont {S.~M.}\ \bibnamefont
  {Barnett}}, \bibinfo {author} {\bibfnamefont {F.~C.}\ \bibnamefont
  {Speirits}},\ and\ \bibinfo {author} {\bibfnamefont {M.}~\bibnamefont
  {Babiker}},\ }\href {https://doi.org/10.1088/1751-8121/ac6bd1} {\bibfield
  {journal} {\bibinfo  {journal} {J. Phys. A Math. Theor.}\ }\textbf {\bibinfo
  {volume} {55}},\ \bibinfo {pages} {234008} (\bibinfo {year}
  {2022})}\BibitemShut {NoStop}%
\bibitem [{\citenamefont {Ivanov}\ \emph {et~al.}(2016)\citenamefont {Ivanov},
  \citenamefont {Seipt}, \citenamefont {Surzhykov},\ and\ \citenamefont
  {Fritzsche}}]{ivanov_elastic_2016}%
  \BibitemOpen
  \bibfield  {author} {\bibinfo {author} {\bibfnamefont {I.}~\bibnamefont
  {Ivanov}}, \bibinfo {author} {\bibfnamefont {D.}~\bibnamefont {Seipt}},
  \bibinfo {author} {\bibfnamefont {A.}~\bibnamefont {Surzhykov}},\ and\
  \bibinfo {author} {\bibfnamefont {S.}~\bibnamefont {Fritzsche}},\ }\href
  {https://doi.org/10.1103/PhysRevD.94.076001} {\bibfield  {journal} {\bibinfo
  {journal} {Phys. Rev. D}\ }\textbf {\bibinfo {volume} {94}},\ \bibinfo
  {pages} {076001} (\bibinfo {year} {2016})}\BibitemShut {NoStop}%
\bibitem [{\citenamefont {Irfan}\ \emph {et~al.}(2021)\citenamefont {Irfan},
  \citenamefont {Blackstone}, \citenamefont {Pynn},\ and\ \citenamefont
  {Ortiz}}]{irfan_quantum_2021}%
  \BibitemOpen
  \bibfield  {author} {\bibinfo {author} {\bibfnamefont {A.~A.~M.}\
  \bibnamefont {Irfan}}, \bibinfo {author} {\bibfnamefont {P.}~\bibnamefont
  {Blackstone}}, \bibinfo {author} {\bibfnamefont {R.}~\bibnamefont {Pynn}},\
  and\ \bibinfo {author} {\bibfnamefont {G.}~\bibnamefont {Ortiz}},\ }\href
  {https://doi.org/10.1088/1367-2630/ac12e0} {\bibfield  {journal} {\bibinfo
  {journal} {New J. Phys.}\ }\textbf {\bibinfo {volume} {23}},\ \bibinfo
  {pages} {083022} (\bibinfo {year} {2021})}\BibitemShut {NoStop}%
\bibitem [{\citenamefont {Lu}\ \emph {et~al.}(2020)\citenamefont {Lu},
  \citenamefont {Irfan}, \citenamefont {Shen}, \citenamefont {Kuhn},
  \citenamefont {Snow}, \citenamefont {Baxter}, \citenamefont {Pynn},\ and\
  \citenamefont {Ortiz}}]{lu_operator_2020}%
  \BibitemOpen
  \bibfield  {author} {\bibinfo {author} {\bibfnamefont {S.}~\bibnamefont
  {Lu}}, \bibinfo {author} {\bibfnamefont {A.~A.~M.}\ \bibnamefont {Irfan}},
  \bibinfo {author} {\bibfnamefont {J.}~\bibnamefont {Shen}}, \bibinfo {author}
  {\bibfnamefont {S.~J.}\ \bibnamefont {Kuhn}}, \bibinfo {author}
  {\bibfnamefont {W.~M.}\ \bibnamefont {Snow}}, \bibinfo {author}
  {\bibfnamefont {D.~V.}\ \bibnamefont {Baxter}}, \bibinfo {author}
  {\bibfnamefont {R.}~\bibnamefont {Pynn}},\ and\ \bibinfo {author}
  {\bibfnamefont {G.}~\bibnamefont {Ortiz}},\ }\href
  {https://doi.org/10.1103/PhysRevA.101.042318} {\bibfield  {journal} {\bibinfo
   {journal} {Phys. Rev. A}\ }\textbf {\bibinfo {volume} {101}},\ \bibinfo
  {pages} {042318} (\bibinfo {year} {2020})}\BibitemShut {NoStop}%
\bibitem [{\citenamefont {Golub}\ \emph {et~al.}(1994)\citenamefont {Golub},
  \citenamefont {Gähler},\ and\ \citenamefont {Keller}}]{Golub_1994}%
  \BibitemOpen
  \bibfield  {author} {\bibinfo {author} {\bibfnamefont {R.}~\bibnamefont
  {Golub}}, \bibinfo {author} {\bibfnamefont {R.}~\bibnamefont {Gähler}},\
  and\ \bibinfo {author} {\bibfnamefont {T.}~\bibnamefont {Keller}},\ }\href
  {https://doi.org/10.1119/1.17459} {\bibfield  {journal} {\bibinfo  {journal}
  {Am. J. Phys.}\ }\textbf {\bibinfo {volume} {62}},\ \bibinfo {pages} {779}
  (\bibinfo {year} {1994})}\BibitemShut {NoStop}%
\bibitem [{\citenamefont {Hasegawa}\ \emph {et~al.}(2010)\citenamefont
  {Hasegawa}, \citenamefont {Loidl}, \citenamefont {Badurek}, \citenamefont
  {Durstberger-Rennhofer}, \citenamefont {Sponar},\ and\ \citenamefont
  {Rauch}}]{Hasegawa_2010}%
  \BibitemOpen
  \bibfield  {author} {\bibinfo {author} {\bibfnamefont {Y.}~\bibnamefont
  {Hasegawa}}, \bibinfo {author} {\bibfnamefont {R.}~\bibnamefont {Loidl}},
  \bibinfo {author} {\bibfnamefont {G.}~\bibnamefont {Badurek}}, \bibinfo
  {author} {\bibfnamefont {K.}~\bibnamefont {Durstberger-Rennhofer}}, \bibinfo
  {author} {\bibfnamefont {S.}~\bibnamefont {Sponar}},\ and\ \bibinfo {author}
  {\bibfnamefont {H.}~\bibnamefont {Rauch}},\ }\href
  {https://doi.org/10.1103/PhysRevA.81.032121} {\bibfield  {journal} {\bibinfo
  {journal} {Phys. Rev. A}\ }\textbf {\bibinfo {volume} {81}},\ \bibinfo
  {pages} {032121} (\bibinfo {year} {2010})}\BibitemShut {NoStop}%
\bibitem [{\citenamefont {Klepp}\ \emph {et~al.}(2014)\citenamefont {Klepp},
  \citenamefont {Sponar},\ and\ \citenamefont {Hasegawa}}]{Klepp_2014}%
  \BibitemOpen
  \bibfield  {author} {\bibinfo {author} {\bibfnamefont {J.}~\bibnamefont
  {Klepp}}, \bibinfo {author} {\bibfnamefont {S.}~\bibnamefont {Sponar}},\ and\
  \bibinfo {author} {\bibfnamefont {Y.}~\bibnamefont {Hasegawa}},\ }\href
  {https://doi.org/10.1093/ptep/ptu085} {\bibfield  {journal} {\bibinfo
  {journal} {Prog. Theor. Exp. Phys.}\ }\textbf {\bibinfo {volume} {2014}},\
  \bibinfo {pages} {082A01} (\bibinfo {year} {2014})}\BibitemShut {NoStop}%
\bibitem [{\citenamefont {Hasegawa}\ \emph {et~al.}(2003)\citenamefont
  {Hasegawa}, \citenamefont {Loidl}, \citenamefont {Badurek}, \citenamefont
  {Baron},\ and\ \citenamefont {Rauch}}]{Hasegawa_2003}%
  \BibitemOpen
  \bibfield  {author} {\bibinfo {author} {\bibfnamefont {Y.}~\bibnamefont
  {Hasegawa}}, \bibinfo {author} {\bibfnamefont {R.}~\bibnamefont {Loidl}},
  \bibinfo {author} {\bibfnamefont {G.}~\bibnamefont {Badurek}}, \bibinfo
  {author} {\bibfnamefont {M.}~\bibnamefont {Baron}},\ and\ \bibinfo {author}
  {\bibfnamefont {H.}~\bibnamefont {Rauch}},\ }\href
  {https://doi.org/10.1038/nature01881} {\bibfield  {journal} {\bibinfo
  {journal} {Nature}\ }\textbf {\bibinfo {volume} {425}},\ \bibinfo {pages}
  {45} (\bibinfo {year} {2003})}\BibitemShut {NoStop}%
\bibitem [{\citenamefont {Rekveldt}(1996)}]{Rekveldt_1996}%
  \BibitemOpen
  \bibfield  {author} {\bibinfo {author} {\bibfnamefont {M.}~\bibnamefont
  {Rekveldt}},\ }\href {https://doi.org/10.1016/0168-583X(96)00213-3}
  {\bibfield  {journal} {\bibinfo  {journal} {Nucl. Instrum. Methods Phys. Res.
  B}\ }\textbf {\bibinfo {volume} {114}},\ \bibinfo {pages} {366–370}
  (\bibinfo {year} {1996})}\BibitemShut {NoStop}%
\bibitem [{\citenamefont {Rekveldt}\ \emph {et~al.}(2005)\citenamefont
  {Rekveldt}, \citenamefont {Plomp}, \citenamefont {Bouwman}, \citenamefont
  {Kraan}, \citenamefont {Grigoriev},\ and\ \citenamefont
  {Blaauw}}]{Rekveldt_2005}%
  \BibitemOpen
  \bibfield  {author} {\bibinfo {author} {\bibfnamefont {M.~T.}\ \bibnamefont
  {Rekveldt}}, \bibinfo {author} {\bibfnamefont {J.}~\bibnamefont {Plomp}},
  \bibinfo {author} {\bibfnamefont {W.~G.}\ \bibnamefont {Bouwman}}, \bibinfo
  {author} {\bibfnamefont {W.~H.}\ \bibnamefont {Kraan}}, \bibinfo {author}
  {\bibfnamefont {S.}~\bibnamefont {Grigoriev}},\ and\ \bibinfo {author}
  {\bibfnamefont {M.}~\bibnamefont {Blaauw}},\ }\href
  {https://doi.org/10.1063/1.1858579} {\bibfield  {journal} {\bibinfo
  {journal} {Rev. Sci. Instrum.}\ }\textbf {\bibinfo {volume} {76}},\ \bibinfo
  {pages} {033901} (\bibinfo {year} {2005})}\BibitemShut {NoStop}%
\bibitem [{\citenamefont {Born}\ and\ \citenamefont {Wolf}(2002)}]{BornWolf}%
  \BibitemOpen
  \bibfield  {author} {\bibinfo {author} {\bibfnamefont {M.}~\bibnamefont
  {Born}}\ and\ \bibinfo {author} {\bibfnamefont {E.}~\bibnamefont {Wolf}},\
  }\href@noop {} {\emph {\bibinfo {title} {Principles of Optics}}},\ \bibinfo
  {edition} {7th}\ ed.\ (\bibinfo  {publisher} {Cambridge University Press,
  Cambridge},\ \bibinfo {year} {2002})\BibitemShut {NoStop}%
\bibitem [{\citenamefont {Li}\ \emph {et~al.}(2016)\citenamefont {Li},
  \citenamefont {Parnell}, \citenamefont {Bai}, \citenamefont {Yang},
  \citenamefont {Hamilton}, \citenamefont {Maranville}, \citenamefont {Ashkar},
  \citenamefont {Baxter}, \citenamefont {Cremer},\ and\ \citenamefont
  {Pynn}}]{Li_2016}%
  \BibitemOpen
  \bibfield  {author} {\bibinfo {author} {\bibfnamefont {F.}~\bibnamefont
  {Li}}, \bibinfo {author} {\bibfnamefont {S.~R.}\ \bibnamefont {Parnell}},
  \bibinfo {author} {\bibfnamefont {H.}~\bibnamefont {Bai}}, \bibinfo {author}
  {\bibfnamefont {W.}~\bibnamefont {Yang}}, \bibinfo {author} {\bibfnamefont
  {W.~A.}\ \bibnamefont {Hamilton}}, \bibinfo {author} {\bibfnamefont {B.~B.}\
  \bibnamefont {Maranville}}, \bibinfo {author} {\bibfnamefont
  {R.}~\bibnamefont {Ashkar}}, \bibinfo {author} {\bibfnamefont {D.~V.}\
  \bibnamefont {Baxter}}, \bibinfo {author} {\bibfnamefont {J.~T.}\
  \bibnamefont {Cremer}},\ and\ \bibinfo {author} {\bibfnamefont
  {R.}~\bibnamefont {Pynn}},\ }\href
  {https://doi.org/10.1107/S1600576715021573} {\bibfield  {journal} {\bibinfo
  {journal} {J. Appl. Crystallogr.}\ }\textbf {\bibinfo {volume} {49}},\
  \bibinfo {pages} {55} (\bibinfo {year} {2016})}\BibitemShut {NoStop}%
\bibitem [{\citenamefont {Andersson}\ \emph {et~al.}(2008)\citenamefont
  {Andersson}, \citenamefont {van Heijkamp}, \citenamefont {de~Schepper},\ and\
  \citenamefont {Bouwman}}]{Andersson_2008}%
  \BibitemOpen
  \bibfield  {author} {\bibinfo {author} {\bibfnamefont {R.}~\bibnamefont
  {Andersson}}, \bibinfo {author} {\bibfnamefont {L.~F.}\ \bibnamefont {van
  Heijkamp}}, \bibinfo {author} {\bibfnamefont {I.~M.}\ \bibnamefont
  {de~Schepper}},\ and\ \bibinfo {author} {\bibfnamefont {W.~G.}\ \bibnamefont
  {Bouwman}},\ }\href {https://doi.org/10.1107/S0021889808026770} {\bibfield
  {journal} {\bibinfo  {journal} {Journal of Applied Crystallography}\ }\textbf
  {\bibinfo {volume} {41}},\ \bibinfo {pages} {868–885} (\bibinfo {year}
  {2008})}\BibitemShut {NoStop}%
\bibitem [{\citenamefont {Strobl}\ \emph {et~al.}(2012)\citenamefont {Strobl},
  \citenamefont {Tremsin}, \citenamefont {Hilger}, \citenamefont {Wieder},
  \citenamefont {Kardjilov}, \citenamefont {Manke}, \citenamefont {Bouwman},\
  and\ \citenamefont {Plomp}}]{Strobl_2012}%
  \BibitemOpen
  \bibfield  {author} {\bibinfo {author} {\bibfnamefont {M.}~\bibnamefont
  {Strobl}}, \bibinfo {author} {\bibfnamefont {A.~S.}\ \bibnamefont {Tremsin}},
  \bibinfo {author} {\bibfnamefont {A.}~\bibnamefont {Hilger}}, \bibinfo
  {author} {\bibfnamefont {F.}~\bibnamefont {Wieder}}, \bibinfo {author}
  {\bibfnamefont {N.}~\bibnamefont {Kardjilov}}, \bibinfo {author}
  {\bibfnamefont {I.}~\bibnamefont {Manke}}, \bibinfo {author} {\bibfnamefont
  {W.~G.}\ \bibnamefont {Bouwman}},\ and\ \bibinfo {author} {\bibfnamefont
  {J.}~\bibnamefont {Plomp}},\ }\href {https://doi.org/10.1063/1.4730775}
  {\bibfield  {journal} {\bibinfo  {journal} {Journal of Applied Physics}\
  }\textbf {\bibinfo {volume} {112}},\ \bibinfo {pages} {014503} (\bibinfo
  {year} {2012})}\BibitemShut {NoStop}%
\bibitem [{\citenamefont {Li}\ \emph {et~al.}(2019{\natexlab{a}})\citenamefont
  {Li}, \citenamefont {Parnell}, \citenamefont {Dalgliesh}, \citenamefont
  {Washington}, \citenamefont {Plomp},\ and\ \citenamefont
  {Pynn}}]{Li_Parnell_Dalgliesh_2019}%
  \BibitemOpen
  \bibfield  {author} {\bibinfo {author} {\bibfnamefont {F.}~\bibnamefont
  {Li}}, \bibinfo {author} {\bibfnamefont {S.~R.}\ \bibnamefont {Parnell}},
  \bibinfo {author} {\bibfnamefont {R.}~\bibnamefont {Dalgliesh}}, \bibinfo
  {author} {\bibfnamefont {A.}~\bibnamefont {Washington}}, \bibinfo {author}
  {\bibfnamefont {J.}~\bibnamefont {Plomp}},\ and\ \bibinfo {author}
  {\bibfnamefont {R.}~\bibnamefont {Pynn}},\ }\href
  {https://doi.org/10.1038/s41598-019-44493-9} {\bibfield  {journal} {\bibinfo
  {journal} {Scientific Reports}\ }\textbf {\bibinfo {volume} {9}},\ \bibinfo
  {pages} {8563} (\bibinfo {year} {2019}{\natexlab{a}})}\BibitemShut {NoStop}%
\bibitem [{\citenamefont {Li}\ \emph {et~al.}(2021)\citenamefont {Li},
  \citenamefont {Steinke}, \citenamefont {Dalgliesh}, \citenamefont
  {Washington}, \citenamefont {Shen}, \citenamefont {Pynn},\ and\ \citenamefont
  {Parnell}}]{Li_Parnell_2021}%
  \BibitemOpen
  \bibfield  {author} {\bibinfo {author} {\bibfnamefont {F.}~\bibnamefont
  {Li}}, \bibinfo {author} {\bibfnamefont {N.~J.}\ \bibnamefont {Steinke}},
  \bibinfo {author} {\bibfnamefont {R.~M.}\ \bibnamefont {Dalgliesh}}, \bibinfo
  {author} {\bibfnamefont {A.~L.}\ \bibnamefont {Washington}}, \bibinfo
  {author} {\bibfnamefont {J.}~\bibnamefont {Shen}}, \bibinfo {author}
  {\bibfnamefont {R.}~\bibnamefont {Pynn}},\ and\ \bibinfo {author}
  {\bibfnamefont {S.~R.}\ \bibnamefont {Parnell}},\ }\href
  {https://doi.org/10.1016/j.nima.2021.165705} {\bibfield  {journal} {\bibinfo
  {journal} {Nuclear Instruments and Methods in Physics Research Section A:
  Accelerators, Spectrometers, Detectors and Associated Equipment}\ }\textbf
  {\bibinfo {volume} {1014}},\ \bibinfo {pages} {165705} (\bibinfo {year}
  {2021})}\BibitemShut {NoStop}%
\bibitem [{\citenamefont {Pfeiffer}\ \emph {et~al.}(2006)\citenamefont
  {Pfeiffer}, \citenamefont {Weitkamp}, \citenamefont {Bunk},\ and\
  \citenamefont {David}}]{Pfeiffer_2006}%
  \BibitemOpen
  \bibfield  {author} {\bibinfo {author} {\bibfnamefont {F.}~\bibnamefont
  {Pfeiffer}}, \bibinfo {author} {\bibfnamefont {T.}~\bibnamefont {Weitkamp}},
  \bibinfo {author} {\bibfnamefont {O.}~\bibnamefont {Bunk}},\ and\ \bibinfo
  {author} {\bibfnamefont {C.}~\bibnamefont {David}},\ }\href
  {https://doi.org/10.1038/nphys265} {\bibfield  {journal} {\bibinfo  {journal}
  {Nat. Phys.}\ }\textbf {\bibinfo {volume} {2}},\ \bibinfo {pages} {258}
  (\bibinfo {year} {2006})}\BibitemShut {NoStop}%
\bibitem [{\citenamefont {Brezger}\ \emph {et~al.}(2002)\citenamefont
  {Brezger}, \citenamefont {Hackermüller}, \citenamefont {Uttenthaler},
  \citenamefont {Petschinka}, \citenamefont {Arndt},\ and\ \citenamefont
  {Zeilinger}}]{Brezger_2002}%
  \BibitemOpen
  \bibfield  {author} {\bibinfo {author} {\bibfnamefont {B.}~\bibnamefont
  {Brezger}}, \bibinfo {author} {\bibfnamefont {L.}~\bibnamefont
  {Hackermüller}}, \bibinfo {author} {\bibfnamefont {S.}~\bibnamefont
  {Uttenthaler}}, \bibinfo {author} {\bibfnamefont {J.}~\bibnamefont
  {Petschinka}}, \bibinfo {author} {\bibfnamefont {M.}~\bibnamefont {Arndt}},\
  and\ \bibinfo {author} {\bibfnamefont {A.}~\bibnamefont {Zeilinger}},\ }\href
  {https://doi.org/10.1103/PhysRevLett.88.100404} {\bibfield  {journal}
  {\bibinfo  {journal} {Phys. Rev. Lett.}\ }\textbf {\bibinfo {volume} {88}},\
  \bibinfo {pages} {100404} (\bibinfo {year} {2002})}\BibitemShut {NoStop}%
\bibitem [{\citenamefont {{P. Storey}}\ and\ \citenamefont {{C.
  Cohen-Tannoudji}}(1994)}]{cohen-tannoudji}%
  \BibitemOpen
  \bibfield  {author} {\bibinfo {author} {\bibnamefont {{P. Storey}}}\ and\
  \bibinfo {author} {\bibnamefont {{C. Cohen-Tannoudji}}},\ }\href
  {https://doi.org/10.1051/jp2:1994103} {\bibfield  {journal} {\bibinfo
  {journal} {J. Phys. II France}\ }\textbf {\bibinfo {volume} {4}},\ \bibinfo
  {pages} {1999} (\bibinfo {year} {1994})}\BibitemShut {NoStop}%
\bibitem [{\citenamefont {Shull}(1969)}]{Shull_1969}%
  \BibitemOpen
  \bibfield  {author} {\bibinfo {author} {\bibfnamefont {C.~G.}\ \bibnamefont
  {Shull}},\ }\href {https://doi.org/10.1103/PhysRev.179.752} {\bibfield
  {journal} {\bibinfo  {journal} {Phys. Rev.}\ }\textbf {\bibinfo {volume}
  {179}},\ \bibinfo {pages} {752} (\bibinfo {year} {1969})}\BibitemShut
  {NoStop}%
\bibitem [{\citenamefont {Majkrzak}\ \emph {et~al.}(2022)\citenamefont
  {Majkrzak}, \citenamefont {Berk}, \citenamefont {Maranville}, \citenamefont
  {Dura},\ and\ \citenamefont
  {Jach}}]{Majkrzak_Berk_Maranville_Dura_Jach_2022}%
  \BibitemOpen
  \bibfield  {author} {\bibinfo {author} {\bibfnamefont {C.~F.}\ \bibnamefont
  {Majkrzak}}, \bibinfo {author} {\bibfnamefont {N.~F.}\ \bibnamefont {Berk}},
  \bibinfo {author} {\bibfnamefont {B.~B.}\ \bibnamefont {Maranville}},
  \bibinfo {author} {\bibfnamefont {J.~A.}\ \bibnamefont {Dura}},\ and\
  \bibinfo {author} {\bibfnamefont {T.}~\bibnamefont {Jach}},\ }\href
  {https://doi.org/10.1107/S160057672200440X} {\bibfield  {journal} {\bibinfo
  {journal} {J. Appl. Cryst.}\ }\textbf {\bibinfo {volume} {55}},\ \bibinfo
  {pages} {787} (\bibinfo {year} {2022})}\BibitemShut {NoStop}%
\bibitem [{\citenamefont {Majkrzak}\ \emph {et~al.}(2014)\citenamefont
  {Majkrzak}, \citenamefont {Metting}, \citenamefont {Maranville},
  \citenamefont {Dura}, \citenamefont {Satija}, \citenamefont {Udovic},\ and\
  \citenamefont {Berk}}]{Majkrzak_2014}%
  \BibitemOpen
  \bibfield  {author} {\bibinfo {author} {\bibfnamefont {C.~F.}\ \bibnamefont
  {Majkrzak}}, \bibinfo {author} {\bibfnamefont {C.}~\bibnamefont {Metting}},
  \bibinfo {author} {\bibfnamefont {B.~B.}\ \bibnamefont {Maranville}},
  \bibinfo {author} {\bibfnamefont {J.~A.}\ \bibnamefont {Dura}}, \bibinfo
  {author} {\bibfnamefont {S.}~\bibnamefont {Satija}}, \bibinfo {author}
  {\bibfnamefont {T.}~\bibnamefont {Udovic}},\ and\ \bibinfo {author}
  {\bibfnamefont {N.~F.}\ \bibnamefont {Berk}},\ }\href
  {https://doi.org/10.1103/PhysRevA.89.033851} {\bibfield  {journal} {\bibinfo
  {journal} {Phys. Rev. A}\ }\textbf {\bibinfo {volume} {89}},\ \bibinfo
  {pages} {033851} (\bibinfo {year} {2014})}\BibitemShut {NoStop}%
\bibitem [{\citenamefont {Treimer}\ \emph {et~al.}(2006)\citenamefont
  {Treimer}, \citenamefont {Hilger},\ and\ \citenamefont
  {Strobl}}]{Treimer_Hilger_Strobl_2006}%
  \BibitemOpen
  \bibfield  {author} {\bibinfo {author} {\bibfnamefont {W.}~\bibnamefont
  {Treimer}}, \bibinfo {author} {\bibfnamefont {A.}~\bibnamefont {Hilger}},\
  and\ \bibinfo {author} {\bibfnamefont {M.}~\bibnamefont {Strobl}},\ }\href
  {https://doi.org/10.1016/j.physb.2006.05.205} {\bibfield  {journal} {\bibinfo
   {journal} {Physica B Condens. Matter}\ }\textbf {\bibinfo {volume}
  {385–386}},\ \bibinfo {pages} {1388} (\bibinfo {year} {2006})}\BibitemShut
  {NoStop}%
\bibitem [{\citenamefont {Wagh}\ \emph {et~al.}(2011)\citenamefont {Wagh},
  \citenamefont {Abbas},\ and\ \citenamefont
  {Treimer}}]{Wagh_Abbas_Treimer_2011}%
  \BibitemOpen
  \bibfield  {author} {\bibinfo {author} {\bibfnamefont {A.~G.}\ \bibnamefont
  {Wagh}}, \bibinfo {author} {\bibfnamefont {S.}~\bibnamefont {Abbas}},\ and\
  \bibinfo {author} {\bibfnamefont {W.}~\bibnamefont {Treimer}},\ }\href
  {https://doi.org/10.1016/j.nima.2010.06.269} {\bibfield  {journal} {\bibinfo
  {journal} {Nucl. Instrum. Methods Phys. Res. A: Accel. Spectrom. Detect.
  Assoc. Equip.}\ }\textbf {\bibinfo {volume} {634}},\ \bibinfo {pages} {S41}
  (\bibinfo {year} {2011})}\BibitemShut {NoStop}%
\bibitem [{\citenamefont {Altissimo}\ \emph {et~al.}(2008)\citenamefont
  {Altissimo}, \citenamefont {Petrillo}, \citenamefont {Sacchetti},
  \citenamefont {Sani},\ and\ \citenamefont {Stahn}}]{Altissimo_2008}%
  \BibitemOpen
  \bibfield  {author} {\bibinfo {author} {\bibfnamefont {M.}~\bibnamefont
  {Altissimo}}, \bibinfo {author} {\bibfnamefont {C.}~\bibnamefont {Petrillo}},
  \bibinfo {author} {\bibfnamefont {F.}~\bibnamefont {Sacchetti}}, \bibinfo
  {author} {\bibfnamefont {L.}~\bibnamefont {Sani}},\ and\ \bibinfo {author}
  {\bibfnamefont {J.}~\bibnamefont {Stahn}},\ }\href
  {https://doi.org/10.1016/j.nima.2007.11.063} {\bibfield  {journal} {\bibinfo
  {journal} {Nucl. Instrum. Methods Phys. Res. A: Accel. Spectrom. Detect.
  Assoc. Equip.}\ }\textbf {\bibinfo {volume} {586}},\ \bibinfo {pages} {68}
  (\bibinfo {year} {2008})}\BibitemShut {NoStop}%
\bibitem [{\citenamefont {Rauch}\ \emph {et~al.}(1996)\citenamefont {Rauch},
  \citenamefont {Wölwitsch}, \citenamefont {Kaiser}, \citenamefont
  {Clothier},\ and\ \citenamefont {Werner}}]{Rauch_1996}%
  \BibitemOpen
  \bibfield  {author} {\bibinfo {author} {\bibfnamefont {H.}~\bibnamefont
  {Rauch}}, \bibinfo {author} {\bibfnamefont {H.}~\bibnamefont {Wölwitsch}},
  \bibinfo {author} {\bibfnamefont {H.}~\bibnamefont {Kaiser}}, \bibinfo
  {author} {\bibfnamefont {R.}~\bibnamefont {Clothier}},\ and\ \bibinfo
  {author} {\bibfnamefont {S.~A.}\ \bibnamefont {Werner}},\ }\href
  {https://doi.org/10.1103/PhysRevA.53.902} {\bibfield  {journal} {\bibinfo
  {journal} {Phys. Rev. A}\ }\textbf {\bibinfo {volume} {53}},\ \bibinfo
  {pages} {902} (\bibinfo {year} {1996})}\BibitemShut {NoStop}%
\bibitem [{\citenamefont {Pushin}\ \emph {et~al.}(2008)\citenamefont {Pushin},
  \citenamefont {Arif}, \citenamefont {Huber},\ and\ \citenamefont
  {Cory}}]{Pushin_2008}%
  \BibitemOpen
  \bibfield  {author} {\bibinfo {author} {\bibfnamefont {D.~A.}\ \bibnamefont
  {Pushin}}, \bibinfo {author} {\bibfnamefont {M.}~\bibnamefont {Arif}},
  \bibinfo {author} {\bibfnamefont {M.~G.}\ \bibnamefont {Huber}},\ and\
  \bibinfo {author} {\bibfnamefont {D.~G.}\ \bibnamefont {Cory}},\ }\href
  {https://doi.org/10.1103/PhysRevLett.100.250404} {\bibfield  {journal}
  {\bibinfo  {journal} {Phys. Rev. Lett.}\ }\textbf {\bibinfo {volume} {100}},\
  \bibinfo {pages} {250404} (\bibinfo {year} {2008})}\BibitemShut {NoStop}%
\bibitem [{\citenamefont {Mandel}\ and\ \citenamefont
  {Wolf}(1995)}]{Mandel_Wolf_1995}%
  \BibitemOpen
  \bibfield  {author} {\bibinfo {author} {\bibfnamefont {L.}~\bibnamefont
  {Mandel}}\ and\ \bibinfo {author} {\bibfnamefont {E.}~\bibnamefont {Wolf}},\
  }\href@noop {} {\emph {\bibinfo {title} {Optical coherence and quantum
  optics}}}\ (\bibinfo  {publisher} {Cambridge University Press},\ \bibinfo
  {address} {Cambridge; New York},\ \bibinfo {year} {1995})\BibitemShut
  {NoStop}%
\bibitem [{\citenamefont {Zarubin}(1993)}]{Zarubin_1993}%
  \BibitemOpen
  \bibfield  {author} {\bibinfo {author} {\bibfnamefont {A.~M.}\ \bibnamefont
  {Zarubin}},\ }\href {https://doi.org/10.1016/0030-4018(93)90251-Y} {\bibfield
   {journal} {\bibinfo  {journal} {Opt. Commun.}\ }\textbf {\bibinfo {volume}
  {100}},\ \bibinfo {pages} {491} (\bibinfo {year} {1993})}\BibitemShut
  {NoStop}%
\bibitem [{\citenamefont {Taylor}\ \emph {et~al.}(1994)\citenamefont {Taylor},
  \citenamefont {Schernthanner}, \citenamefont {Lenz},\ and\ \citenamefont
  {Meystre}}]{Taylor_1994}%
  \BibitemOpen
  \bibfield  {author} {\bibinfo {author} {\bibfnamefont {B.}~\bibnamefont
  {Taylor}}, \bibinfo {author} {\bibfnamefont {K.}~\bibnamefont
  {Schernthanner}}, \bibinfo {author} {\bibfnamefont {G.}~\bibnamefont
  {Lenz}},\ and\ \bibinfo {author} {\bibfnamefont {P.}~\bibnamefont
  {Meystre}},\ }\href {https://doi.org/10.1016/0030-4018(94)90252-6} {\bibfield
   {journal} {\bibinfo  {journal} {Opt. Commun.}\ }\textbf {\bibinfo {volume}
  {110}},\ \bibinfo {pages} {569} (\bibinfo {year} {1994})}\BibitemShut
  {NoStop}%
\bibitem [{\citenamefont {Keller}\ \emph {et~al.}(1997)\citenamefont {Keller},
  \citenamefont {Besenböck}, \citenamefont {Feller}, \citenamefont {Gähler},
  \citenamefont {Golub}, \citenamefont {Hank},\ and\ \citenamefont
  {Köppe}}]{Keller_1997}%
  \BibitemOpen
  \bibfield  {author} {\bibinfo {author} {\bibfnamefont {T.}~\bibnamefont
  {Keller}}, \bibinfo {author} {\bibfnamefont {W.}~\bibnamefont {Besenböck}},
  \bibinfo {author} {\bibfnamefont {J.}~\bibnamefont {Feller}}, \bibinfo
  {author} {\bibfnamefont {R.}~\bibnamefont {Gähler}}, \bibinfo {author}
  {\bibfnamefont {R.}~\bibnamefont {Golub}}, \bibinfo {author} {\bibfnamefont
  {P.}~\bibnamefont {Hank}},\ and\ \bibinfo {author} {\bibfnamefont
  {M.}~\bibnamefont {Köppe}},\ }\href
  {https://doi.org/10.1016/S0921-4526(97)00130-0} {\bibfield  {journal}
  {\bibinfo  {journal} {Physica B Condens. Matter}\ }\textbf {\bibinfo {volume}
  {234–236}},\ \bibinfo {pages} {1120} (\bibinfo {year} {1997})}\BibitemShut
  {NoStop}%
\bibitem [{\citenamefont {Felber}\ \emph {et~al.}(1998)\citenamefont {Felber},
  \citenamefont {Gähler}, \citenamefont {Golub},\ and\ \citenamefont
  {Prechtel}}]{Felber_1998}%
  \BibitemOpen
  \bibfield  {author} {\bibinfo {author} {\bibfnamefont {J.}~\bibnamefont
  {Felber}}, \bibinfo {author} {\bibfnamefont {R.}~\bibnamefont {Gähler}},
  \bibinfo {author} {\bibfnamefont {R.}~\bibnamefont {Golub}},\ and\ \bibinfo
  {author} {\bibfnamefont {K.}~\bibnamefont {Prechtel}},\ }\href
  {https://doi.org/10.1016/S0921-4526(97)00999-X} {\bibfield  {journal}
  {\bibinfo  {journal} {Physica B Condens. Matter}\ }\textbf {\bibinfo {volume}
  {252}},\ \bibinfo {pages} {34} (\bibinfo {year} {1998})}\BibitemShut
  {NoStop}%
\bibitem [{\citenamefont {Barrachina}\ \emph {et~al.}(2019)\citenamefont
  {Barrachina}, \citenamefont {Navarrete},\ and\ \citenamefont
  {Ciappina}}]{Barrachina_Navarrete_Ciappina_2019}%
  \BibitemOpen
  \bibfield  {author} {\bibinfo {author} {\bibfnamefont {R.~O.}\ \bibnamefont
  {Barrachina}}, \bibinfo {author} {\bibfnamefont {F.}~\bibnamefont
  {Navarrete}},\ and\ \bibinfo {author} {\bibfnamefont {M.~F.}\ \bibnamefont
  {Ciappina}},\ }\href {https://doi.org/10.1088/1361-6404/ab3b6f} {\bibfield
  {journal} {\bibinfo  {journal} {Eur. J. Phys.}\ }\textbf {\bibinfo {volume}
  {40}},\ \bibinfo {pages} {065402} (\bibinfo {year} {2019})}\BibitemShut
  {NoStop}%
\bibitem [{\citenamefont {Barrachina}\ \emph {et~al.}(2020)\citenamefont
  {Barrachina}, \citenamefont {Navarrete},\ and\ \citenamefont
  {Ciappina}}]{Barrachina_Navarrete_Ciappina_2020}%
  \BibitemOpen
  \bibfield  {author} {\bibinfo {author} {\bibfnamefont {R.~O.}\ \bibnamefont
  {Barrachina}}, \bibinfo {author} {\bibfnamefont {F.}~\bibnamefont
  {Navarrete}},\ and\ \bibinfo {author} {\bibfnamefont {M.~F.}\ \bibnamefont
  {Ciappina}},\ }\href {https://doi.org/10.1103/PhysRevResearch.2.043353}
  {\bibfield  {journal} {\bibinfo  {journal} {Phys. Rev. Research}\ }\textbf
  {\bibinfo {volume} {2}},\ \bibinfo {pages} {043353} (\bibinfo {year}
  {2020})}\BibitemShut {NoStop}%
\bibitem [{\citenamefont {Fabre}\ \emph {et~al.}(2018)\citenamefont {Fabre},
  \citenamefont {Navarrete}, \citenamefont {Sarkadi},\ and\ \citenamefont
  {Barrachina}}]{Fabre_Navarrete_Sarkadi_Barrachina_2018}%
  \BibitemOpen
  \bibfield  {author} {\bibinfo {author} {\bibfnamefont {I.}~\bibnamefont
  {Fabre}}, \bibinfo {author} {\bibfnamefont {F.}~\bibnamefont {Navarrete}},
  \bibinfo {author} {\bibfnamefont {L.}~\bibnamefont {Sarkadi}},\ and\ \bibinfo
  {author} {\bibfnamefont {R.~O.}\ \bibnamefont {Barrachina}},\ }\href
  {https://doi.org/10.1088/1361-6404/aa8e74} {\bibfield  {journal} {\bibinfo
  {journal} {European Journal of Physics}\ }\textbf {\bibinfo {volume} {39}},\
  \bibinfo {pages} {015401} (\bibinfo {year} {2018})}\BibitemShut {NoStop}%
\bibitem [{\citenamefont {Mariano}(2002)}]{Mariano_2002}%
  \BibitemOpen
  \bibfield  {author} {\bibinfo {author} {\bibfnamefont {A.}~\bibnamefont
  {Mariano}},\ }in\ \href {https://doi.org/10.1117/12.475886} {\emph {\bibinfo
  {booktitle} {First International Workshop on Classical and Quantum
  Interference}}},\ Vol.\ \bibinfo {volume} {4888},\ \bibinfo {editor} {edited
  by\ \bibinfo {editor} {\bibfnamefont {J.}~\bibnamefont {Perina}}, \bibinfo
  {editor} {\bibfnamefont {M.}~\bibnamefont {Hrabovsky}},\ and\ \bibinfo
  {editor} {\bibfnamefont {J.}~\bibnamefont {Krepelka}}},\ \bibinfo
  {organization} {International Society for Optics and Photonics}\ (\bibinfo
  {publisher} {SPIE},\ \bibinfo {year} {2002})\ p.~\bibinfo {pages}
  {75}\BibitemShut {NoStop}%
\bibitem [{\citenamefont {Stodolsky}(1998)}]{Stodolsky_1998}%
  \BibitemOpen
  \bibfield  {author} {\bibinfo {author} {\bibfnamefont {L.}~\bibnamefont
  {Stodolsky}},\ }\href {https://doi.org/10.1103/PhysRevD.58.036006} {\bibfield
   {journal} {\bibinfo  {journal} {Phys. Rev. D}\ }\textbf {\bibinfo {volume}
  {58}},\ \bibinfo {pages} {036006} (\bibinfo {year} {1998})}\BibitemShut
  {NoStop}%
\bibitem [{\citenamefont {Kiers}\ \emph {et~al.}(1996)\citenamefont {Kiers},
  \citenamefont {Nussinov},\ and\ \citenamefont {Weiss}}]{Kiers_1996}%
  \BibitemOpen
  \bibfield  {author} {\bibinfo {author} {\bibfnamefont {K.}~\bibnamefont
  {Kiers}}, \bibinfo {author} {\bibfnamefont {S.}~\bibnamefont {Nussinov}},\
  and\ \bibinfo {author} {\bibfnamefont {N.}~\bibnamefont {Weiss}},\ }\href
  {https://doi.org/10.1103/PhysRevD.53.537} {\bibfield  {journal} {\bibinfo
  {journal} {Phys. Rev. D}\ }\textbf {\bibinfo {volume} {53}},\ \bibinfo
  {pages} {537} (\bibinfo {year} {1996})}\BibitemShut {NoStop}%
\bibitem [{\citenamefont {Golub}\ and\ \citenamefont
  {Lamoreaux}(1992)}]{Golub_Lamoreaux_1992}%
  \BibitemOpen
  \bibfield  {author} {\bibinfo {author} {\bibfnamefont {R.}~\bibnamefont
  {Golub}}\ and\ \bibinfo {author} {\bibfnamefont {S.}~\bibnamefont
  {Lamoreaux}},\ }\href {https://doi.org/10.1016/0375-9601(92)90987-W}
  {\bibfield  {journal} {\bibinfo  {journal} {Phys. Lett. A}\ }\textbf
  {\bibinfo {volume} {162}},\ \bibinfo {pages} {122} (\bibinfo {year}
  {1992})}\BibitemShut {NoStop}%
\bibitem [{\citenamefont {Berk}(2014)}]{Berk_2014}%
  \BibitemOpen
  \bibfield  {author} {\bibinfo {author} {\bibfnamefont {N.~F.}\ \bibnamefont
  {Berk}},\ }\href {https://doi.org/10.1103/PhysRevA.89.033852} {\bibfield
  {journal} {\bibinfo  {journal} {Phys. Rev. A}\ }\textbf {\bibinfo {volume}
  {89}},\ \bibinfo {pages} {033852} (\bibinfo {year} {2014})}\BibitemShut
  {NoStop}%
\bibitem [{\citenamefont {Keller}\ \emph {et~al.}(2002)\citenamefont {Keller},
  \citenamefont {Golub},\ and\ \citenamefont {Gähler}}]{keller_book}%
  \BibitemOpen
  \bibfield  {author} {\bibinfo {author} {\bibfnamefont {T.}~\bibnamefont
  {Keller}}, \bibinfo {author} {\bibfnamefont {R.}~\bibnamefont {Golub}},\ and\
  \bibinfo {author} {\bibfnamefont {R.}~\bibnamefont {Gähler}},\ }\href@noop
  {} {\emph {\bibinfo {title} {Scattering and Inverse Scattering in Pure and
  Applied Science}}}\ (\bibinfo  {publisher} {Academic Press},\ \bibinfo {year}
  {2002})\ Chap.\ \bibinfo {chapter} {2.8.6: Neutron Spin Echo—A Technique
  for High-Resolution Neutron Scattering}, p.\ \bibinfo {pages}
  {1264}\BibitemShut {NoStop}%
\bibitem [{\citenamefont {Klein}\ and\ \citenamefont
  {Werner}(1983)}]{Klein_Werner_1983}%
  \BibitemOpen
  \bibfield  {author} {\bibinfo {author} {\bibfnamefont {A.~G.}\ \bibnamefont
  {Klein}}\ and\ \bibinfo {author} {\bibfnamefont {S.~A.}\ \bibnamefont
  {Werner}},\ }\href {https://doi.org/10.1088/0034-4885/46/3/001} {\bibfield
  {journal} {\bibinfo  {journal} {Rep. Prog. Phys.}\ }\textbf {\bibinfo
  {volume} {46}},\ \bibinfo {pages} {259} (\bibinfo {year} {1983})}\BibitemShut
  {NoStop}%
\bibitem [{\citenamefont {Goldstein}(1950)}]{Goldstein_1950}%
  \BibitemOpen
  \bibfield  {author} {\bibinfo {author} {\bibfnamefont {H.}~\bibnamefont
  {Goldstein}},\ }\href@noop {} {\emph {\bibinfo {title} {Classical
  Mechanics}}},\ \bibinfo {edition} {1st}\ ed.\ (\bibinfo  {publisher}
  {Addison-Wesley Longman},\ \bibinfo {year} {1950})\BibitemShut {NoStop}%
\bibitem [{\citenamefont {Rauch}\ and\ \citenamefont
  {Werner}(2015)}]{rauch-werner2015}%
  \BibitemOpen
  \bibfield  {author} {\bibinfo {author} {\bibfnamefont {H.}~\bibnamefont
  {Rauch}}\ and\ \bibinfo {author} {\bibfnamefont {S.}~\bibnamefont {Werner}},\
  }\href {https://books.google.com/books?id=S7XTBQAAQBAJ} {\emph {\bibinfo
  {title} {Neutron Interferometry: Lessons in Experimental Quantum Mechanics,
  Wave-particle Duality, and Entanglement}}},\ Oxford series in neutron
  scattering in condensed matter\ (\bibinfo  {publisher} {Oxford University
  Press},\ \bibinfo {year} {2015})\BibitemShut {NoStop}%
\bibitem [{\citenamefont {Dadisman}\ \emph {et~al.}(2019)\citenamefont
  {Dadisman}, \citenamefont {Shen}, \citenamefont {Feng}, \citenamefont {Crow},
  \citenamefont {Jiang}, \citenamefont {Wang}, \citenamefont {Zhang},
  \citenamefont {Bilheux}, \citenamefont {Parnell}, \citenamefont {Pynn},\ and\
  \citenamefont {Li}}]{Dadisman_2019}%
  \BibitemOpen
  \bibfield  {author} {\bibinfo {author} {\bibfnamefont {R.}~\bibnamefont
  {Dadisman}}, \bibinfo {author} {\bibfnamefont {J.}~\bibnamefont {Shen}},
  \bibinfo {author} {\bibfnamefont {H.}~\bibnamefont {Feng}}, \bibinfo {author}
  {\bibfnamefont {L.}~\bibnamefont {Crow}}, \bibinfo {author} {\bibfnamefont
  {C.}~\bibnamefont {Jiang}}, \bibinfo {author} {\bibfnamefont
  {T.}~\bibnamefont {Wang}}, \bibinfo {author} {\bibfnamefont {Y.}~\bibnamefont
  {Zhang}}, \bibinfo {author} {\bibfnamefont {H.}~\bibnamefont {Bilheux}},
  \bibinfo {author} {\bibfnamefont {S.~R.}\ \bibnamefont {Parnell}}, \bibinfo
  {author} {\bibfnamefont {R.}~\bibnamefont {Pynn}},\ and\ \bibinfo {author}
  {\bibfnamefont {F.}~\bibnamefont {Li}},\ }\href
  {https://doi.org/10.1016/j.nima.2019.05.092} {\bibfield  {journal} {\bibinfo
  {journal} {Nuclear Instruments and Methods in Physics Research Section A:
  Accelerators, Spectrometers, Detectors and Associated Equipment}\ }\textbf
  {\bibinfo {volume} {940}},\ \bibinfo {pages} {174–180} (\bibinfo {year}
  {2019})}\BibitemShut {NoStop}%
\bibitem [{\citenamefont {Li}\ \emph {et~al.}(2019{\natexlab{b}})\citenamefont
  {Li}, \citenamefont {Zhang}, \citenamefont {Li}, \citenamefont {Venero},
  \citenamefont {White}, \citenamefont {Cubitt}, \citenamefont {Huang},
  \citenamefont {Chen}, \citenamefont {He}, \citenamefont {van~der Laan},
  \citenamefont {Wang}, \citenamefont {Hesjedal},\ and\ \citenamefont
  {Wang}}]{li2019}%
  \BibitemOpen
  \bibfield  {author} {\bibinfo {author} {\bibfnamefont {X.}~\bibnamefont
  {Li}}, \bibinfo {author} {\bibfnamefont {S.}~\bibnamefont {Zhang}}, \bibinfo
  {author} {\bibfnamefont {H.}~\bibnamefont {Li}}, \bibinfo {author}
  {\bibfnamefont {D.~A.}\ \bibnamefont {Venero}}, \bibinfo {author}
  {\bibfnamefont {J.~S.}\ \bibnamefont {White}}, \bibinfo {author}
  {\bibfnamefont {R.}~\bibnamefont {Cubitt}}, \bibinfo {author} {\bibfnamefont
  {Q.}~\bibnamefont {Huang}}, \bibinfo {author} {\bibfnamefont
  {J.}~\bibnamefont {Chen}}, \bibinfo {author} {\bibfnamefont {L.}~\bibnamefont
  {He}}, \bibinfo {author} {\bibfnamefont {G.}~\bibnamefont {van~der Laan}},
  \bibinfo {author} {\bibfnamefont {W.}~\bibnamefont {Wang}}, \bibinfo {author}
  {\bibfnamefont {T.}~\bibnamefont {Hesjedal}},\ and\ \bibinfo {author}
  {\bibfnamefont {F.}~\bibnamefont {Wang}},\ }\href
  {https://doi.org/https://doi.org/10.1002/adma.201900264} {\bibfield
  {journal} {\bibinfo  {journal} {Adv. Mater.}\ }\textbf {\bibinfo {volume}
  {31}},\ \bibinfo {pages} {1900264} (\bibinfo {year}
  {2019}{\natexlab{b}})}\BibitemShut {NoStop}%
\bibitem [{\citenamefont {Nguyen}\ \emph {et~al.}(2022)\citenamefont {Nguyen},
  \citenamefont {Tsurimaki}, \citenamefont {Pablo-Pedro}, \citenamefont
  {Bednik}, \citenamefont {Liu}, \citenamefont {Apte}, \citenamefont
  {Andrejevic},\ and\ \citenamefont {Li}}]{thanh2022}%
  \BibitemOpen
  \bibfield  {author} {\bibinfo {author} {\bibfnamefont {T.}~\bibnamefont
  {Nguyen}}, \bibinfo {author} {\bibfnamefont {Y.}~\bibnamefont {Tsurimaki}},
  \bibinfo {author} {\bibfnamefont {R.}~\bibnamefont {Pablo-Pedro}}, \bibinfo
  {author} {\bibfnamefont {G.}~\bibnamefont {Bednik}}, \bibinfo {author}
  {\bibfnamefont {T.}~\bibnamefont {Liu}}, \bibinfo {author} {\bibfnamefont
  {A.}~\bibnamefont {Apte}}, \bibinfo {author} {\bibfnamefont {N.}~\bibnamefont
  {Andrejevic}},\ and\ \bibinfo {author} {\bibfnamefont {M.}~\bibnamefont
  {Li}},\ }\href {https://doi.org/10.1088/1367-2630/ac45cb} {\bibfield
  {journal} {\bibinfo  {journal} {New J. Phys.}\ }\textbf {\bibinfo {volume}
  {24}},\ \bibinfo {pages} {013016} (\bibinfo {year} {2022})}\BibitemShut
  {NoStop}%
\bibitem [{\citenamefont {Snow}\ \emph {et~al.}(2022)\citenamefont {Snow},
  \citenamefont {Haddock},\ and\ \citenamefont {Heacock}}]{snow2022}%
  \BibitemOpen
  \bibfield  {author} {\bibinfo {author} {\bibfnamefont {W.~M.}\ \bibnamefont
  {Snow}}, \bibinfo {author} {\bibfnamefont {C.}~\bibnamefont {Haddock}},\ and\
  \bibinfo {author} {\bibfnamefont {B.}~\bibnamefont {Heacock}},\ }\href
  {https://www.mdpi.com/2073-8994/14/1/10} {\bibfield  {journal} {\bibinfo
  {journal} {Symmetry}\ }\textbf {\bibinfo {volume} {14}} (\bibinfo {year}
  {2022})}\BibitemShut {NoStop}%
\bibitem [{\citenamefont {Shinichiro}\ and\ \citenamefont
  {Masahito}(2016)}]{seki}%
  \BibitemOpen
  \bibfield  {author} {\bibinfo {author} {\bibfnamefont {S.}~\bibnamefont
  {Shinichiro}}\ and\ \bibinfo {author} {\bibfnamefont {M.}~\bibnamefont
  {Masahito}},\ }\href
  {https://doi.org/https://doi.org/10.1007/978-3-319-24651-2} {\emph {\bibinfo
  {title} {Skyrmions in Magnetic Materials}}}\ (\bibinfo  {publisher} {Springer
  Cham},\ \bibinfo {year} {2016})\BibitemShut {NoStop}%
\bibitem [{\citenamefont {Tkachov}(2022)}]{tkachov}%
  \BibitemOpen
  \bibfield  {author} {\bibinfo {author} {\bibfnamefont {G.}~\bibnamefont
  {Tkachov}},\ }\href
  {https://doi.org/https://doi.org/10.1007/978-3-319-24651-2} {\emph {\bibinfo
  {title} {Topological Quantum Materials Concepts, Models, and Phenomena}}}\
  (\bibinfo  {publisher} {Jenny Stanford Publishing},\ \bibinfo {year}
  {2022})\BibitemShut {NoStop}%
\bibitem [{\citenamefont {Rojo}\ and\ \citenamefont {Bloch}(2018)}]{Rojo}%
  \BibitemOpen
  \bibfield  {author} {\bibinfo {author} {\bibfnamefont {A.}~\bibnamefont
  {Rojo}}\ and\ \bibinfo {author} {\bibfnamefont {A.}~\bibnamefont {Bloch}},\
  }\href {https://doi.org/10.1017/9781139021029} {\emph {\bibinfo {title} {The
  Principle of Least Action: History and Physics}}}\ (\bibinfo  {publisher}
  {Cambridge University Press},\ \bibinfo {year} {2018})\BibitemShut {NoStop}%
\end{thebibliography}%
\end{document}